\documentclass{elsart}
\usepackage{graphicx}
\usepackage{amssymb,amsmath}
\usepackage{hyperref}

\usepackage{ifpdf}
\ifpdf
    \DeclareGraphicsExtensions{.pdf}
\else
    \DeclareGraphicsExtensions{.eps}
\fi

\usepackage{ring}
\def\dot{\ring}

\hypersetup{
pdfauthor = {Marc Wouts},
pdftitle={A Coarse graining for the Fortuin-Kasteleyn measure in random media},
pdfsubject = {},
pdfkeywords = {Coarse Graining, Multiscale Analysis, Random Media, Fortuin-Kasteleyn Measure, Dilute Ising Model},
pdfcreator = {LaTeX},
}

\theoremstyle{plain}
\newtheorem{theorem}{Theorem}[section]
\newtheorem{proposition}[theorem]{Proposition}
\newtheorem{corollary}[theorem]{Corollary}
\newtheorem{lemma}[theorem]{Lemma}

\theoremstyle{remark}

\newtheorem{definition}[theorem]{Definition}

\newcommand{\tmop}[1]{\ensuremath{\operatorname{#1}}}
\newcommand{\tmscript}[1]{\mbox{\scriptsize{$#1$}}}

\def\Top{{\mathrm{Top}}}
\def\Bottom{{\mathrm{Bottom}}}
\def\BrLi{{\dot{B}^{L}_i}}
\def\BrL0{{\dot{B}^{L}_0}}
\def\BrLck{{\dot{B}^{L}_{c_k}}}
\def\BrLik{{\dot{B}^{L}_{i_k}}}
\def\ErLi{{\dot{E}^{L}_i}}
\def\ErL0{{\dot{E}^{L}_0}}
\def\oext{{\omega_{\tmop{ext}}}}

\def\theenumi{\roman{enumi}}

\begin{document}

\begin{frontmatter}



\title{A Coarse graining for the Fortuin-Kasteleyn measure in random media}
\author{Marc Wouts}


\address{Universit\'e Paris~7 - Laboratoire de Probabilit\'es et Mod\`eles Al\'eatoires. Site Chevaleret, Case 7012, 75205 Paris Cedex 13, France}
\address{Presently at Universit\'e Paris~10 - Modal'X. B\^{a}timent G, 200 avenue de la R\'epublique, 92001 Nanterre Cedex, France}
\ead{wouts@math.jussieu.fr}

\begin{abstract}
By means of a multi-scale analysis we describe the typical geometrical structure of the clusters under the FK measure in random media. Our result holds in any dimension $d \geqslant 2$ provided that slab percolation occurs under the averaged measure, which should be the case in the whole supercritical phase. This work extends the one of Pisztora~\cite{N09} and provides an essential tool for the analysis of the supercritical regime in disordered FK models and in the corresponding disordered Ising and Potts models.
\end{abstract}

\begin{keyword}
Coarse Graining \sep Multiscale Analysis \sep Random Media \sep Fortuin-Kasteleyn Measure \sep Dilute Ising Model

{\par\leavevmode\hbox {\it 2000 MSC:\ }} 60K35 \sep 82B44 \sep 82B28
\end{keyword}

\end{frontmatter}

\section{Introduction}

The introduction of disorder in the Ising model leads to major changes in the behavior of the system. Several types of disorder have been studied, including random field (in that case, the phase transition disappears if and only if the dimension is less or equal to~$2$~\cite{N234,N22,N230}) and random couplings.

In this article our interest goes to the case of random but still \emph{ferromagnetic} and independent couplings. One such model is the \emph{dilute} Ising model in which the interactions between adjacent spins equal $\beta$ or $0$ independently, with respective probabilities~$p$ and $1-p$. 
The ferromagnetic media randomness is responsible for a new region in the phase diagram: the {\emph{Griffiths phase}} $p<1$ and $\beta_c<\beta<\beta_c(p)$. Indeed, on the one hand the phase transition occurs at $\beta_c(p)>\beta_c$
for any $p<1$ that exceeds the percolation threshold $p_c$, and does not occur (i.e.~$\beta_c(p)=\infty$) if $p\leqslant p_c$,
$\beta_c=\beta_c(1)$ being the critical inverse temperature in absence of dilution~\cite{N21,N27}. Yet, on the second hand, for any $p<1$ and $\beta>\beta_c$, the magnetization is a \textit{non-analytic} function of the external field at $h=0$~\cite{N228}. See also the reviews~\cite{N14,N231}.

The \textit{paramagnetic phase} $p\leqslant 1$ and $\beta < \beta_c$ is well understood as the spin correlations are not larger than in the corresponding undiluted model, and the Glauber dynamics have then a positive spectral gap~\cite{N15}. The study of the Griffiths phase is already more challenging and other phenomena than the break in the analyticity betray the presence of the Griffiths phase, as the sub-exponential relaxation under the Glauber dynamics~\cite{N227}. In the present article we focus on the domain of phase transition $p>p_c$ and $\beta > \beta_c (p)$ and on the elaboration of a coarse graining.

A coarse graining consists in a renormalized description of the microscopic spin
system. It permits to define precisely the notion of local phase and constitutes therefore
a fundamental tool for the study of the phase coexistence phenomenon.
In the case of percolation, Ising and Potts models with uniform couplings,
such a coarse graining was established by Pisztora~\cite{N09} and among the applications
stands the study of the {\nobreak $L^1$-phase} coexistence by 
Bodineau et al.~\cite{N06,N07} and Cerf, Pisztora~\cite{N11,N12,N13}, see also Cerf's lecture notes
\cite{N70}.

In the case of random media there are numerous motivations for the construction of a coarse graining.
 Just as for the uniform case, the coarse graining is a major step towards the
 $L^1$-description of the equilibrium phase coexistence phenomenon -- the second important step being
the analysis of surface tension and its fluctuations~\cite{M02}. But our motivations do not stop there
as the coarse graining also permits the study of the dynamics of the corresponding systems, which are modified in a definite way by the introduction of media randomness. We confirm in~\cite{M03} the prediction of Fisher and Huse~\cite{N05} that the dilution dramatically slows down the dynamics,
proving that the average spin autocorrelation, under the Glauber dynamics, decays not quicker than a negative power of time.

Let us conclude with a few words on the technical aspects of the present work. First, 
the construction of the coarse graining is done under the random media FK model which constitutes 
a convenient mathematical framework, while the
adaptation of the coarse graining to the Ising and Potts models is straightforward, cf.~Section~\ref{sec-cg-Ising}.
Second, instead of the assumption of phase transition we require \emph{percolation in slabs} as in~\cite{N09} (under the averaged measure),
yet we believe that the two notions correspond to the same threshold~$\beta_c(p)$.
At last, there is a major difference between the present work and~\cite{N09}: 
on the contrary to the uniform FK measure, the averaged random media FK measure does not satisfy the DLR equation.
 This ruins all expectancies for a simple adaptation of the original proof, and it was indeed a challenging task to design an alternative proof.

\section{The model and our results}

\subsection{The random media FK model}

\label{sec-def-RCRM}
\subsubsection{Geometry, configurations sets}
We define the FK model on finite subsets of the standard lattice $\mathbb{Z}^d$ for $d \in \{1,2,\ldots\}$. Domains that often appear in this work include the box $\Lambda_N =\{1, \ldots, N - 1\}^d$, its symmetric version $\hat{\Lambda}_N =\{- N, \ldots, N\}^d$ and the slab $S_{N, H} =\{1, \ldots, N - 1\}^{d- 1} \times \{1, \ldots, H - 1\}$ for any $N, H \in \mathbb{N}^{\star}$, $d \geqslant 2$.

Let us consider the norms
\begin{equation*}
\|x\|_2 = \left(\sum_{i=1}^d x_i^2\right)^{1/2}\mbox{ \  and \  } \|x\|_\infty = \max_{i=1\ldots d} |x_i| \mbox{ \  , \  } \forall x\in\mathbb{Z}^d
\end{equation*}
and denote $(\mathbf{e}_i)_{i=1\ldots d}$ the canonical basis of $\mathbb{Z}^d$.
We say that $x,y\in \mathbb{Z}^d$ are \textit{nearest neighbors} if $\|x - y\|_2 = 1$ and denote this as $x \sim y$. Given any $\Lambda \subset
\mathbb{Z}^d$, we define its exterior boundary
\begin{equation}
  \partial \Lambda = \left\{ x \in \mathbb{Z}^d \setminus \Lambda : \exists y
  \in \Lambda, x \sim y \right\} \label{eq-def-dL}
\end{equation}
and to $\Lambda$ we associate the edge sets
\begin{eqnarray}
  E^w (\Lambda) & = & \{\{x, y\}: x \in \Lambda, y \in \mathbb{Z}^d \mbox{ and } x \sim  y \} \label{eq-def-EL} \\
\mbox{and \ } E^f (\Lambda) & = & \{\{x, y\}: x,y \in \Lambda \mbox{ and } x \sim  y \}. \label{eq-def-ELint}
\end{eqnarray} In other words, $E^w(\Lambda)$ is the set of edges that touch $\Lambda$ while $E^f (\Lambda)$ is
the set of edges between two adjacent points of $\Lambda$. Note that the set of points attained by $E^w (\Lambda)$ equals, thus, $\Lambda \cup
\partial \Lambda$. We also denote $E(\mathbb{Z}^d) = E^w(\mathbb{Z}^d) = E^f(\mathbb{Z}^d) $.

The set of \textit{cluster} configurations and that of \textit{media} configurations are respectively
\begin{equation*}  \Omega = \left\{ \omega : E \left( \mathbb{Z}^d \right)  \rightarrow \{0, 1\} \right\}
\mbox{ \ and \ } \mathcal{J} = \left\{ J : E \left( \mathbb{Z}^d \right)  \rightarrow [0, 1] \right\}.
\end{equation*}
Given any $E \subset E(\mathbb{Z}^d) $ we denote by $\omega_{|E}$ (resp. $J_{|E}$)
the restriction of $\omega\in\Omega$ (resp. $J\in{\mathcal{J}}$) to $E$, that is the configuration that
coincides with $\omega$ on $E$ and equals $0$ on $E^c$. We consider then
\begin{equation*} \Omega_E = \left\{ \omega_{|E} , \omega\in\Omega \right\}
\mbox{ \ and \ } \mathcal{J}_E = \left\{ J_{|E} , J\in\mathcal{J} \right\} 
\end{equation*}
the set of configurations that equal $0$ outside $E$.
Given $\omega\in\Omega$, we say that 
 an edge $e \in E(\mathbb{Z}^d) $ is 
\emph{open} for $\omega$ if $\omega_e = 1$, \emph{closed}
otherwise. A \emph{cluster} for $\omega$ is a connected
component of the graph $(\mathbb{Z}^d, \mathcal{O}(\omega))$ where
$\mathcal{O}(\omega) \subset E(\mathbb{Z}^d) $ is the set of open edges for
$\omega$. At last, given $x, y \in \mathbb{Z}^d$ we say that $x$ and $y$ are
connected by $\omega$ (and denote it as $x \stackrel{\omega}{\leftrightarrow}
y$) if they belong to the same $\omega$-cluster.

\subsubsection{FK measure under frozen disorder}
We now define the FK measure under \textit{frozen disorder} $J\in\mathcal{J}$ in function of two parameters $p$ and $q$. The first one 
$p:[0,1] \rightarrow [0,1]$ is an increasing function such that $p(0)=0$, $p(x)>0$
if $x>0$ and $p(1)< 1$, that quantifies the strength of interactions in function of the media. The second one $q\geqslant 1$
corresponds to the spin multiplicity.

Given $E \subset E(\mathbb{Z}^d) $ finite, $J \in \mathcal{J}$ a realization of the media
and $\pi \in \Omega_{E^c}$ a boundary condition, we define the measure $\Phi^{J, p, q, \pi}_E$ by its weight on each $\omega \in \Omega_E$:
\begin{equation}
  \Phi^{J, p, q, \pi}_E (\{\omega\}) = \frac{1}{Z^{J, p, q, \pi}_E} \prod_{e
  \in E} \left( p (J_e) \right)^{\omega_e} (1 - p (J_e))^{1 - \omega_e} \times
  q^{C^{\pi}_E (\omega)} \label{eq-def-Phi}
\end{equation}
where $C^{\pi}_E (\omega)$ is the number of $\omega$-clusters touching $E$ under the configuration $\omega \vee \pi$ defined by
\begin{equation*}
  \left( \omega \vee \pi \right)_e = \left\{ \begin{array}{ll}
   \omega_e & \mbox{if } e \in E\\
   \pi_e & \mbox{else}
  \end{array} \right. \end{equation*}
and $Z^{J, p, q, \pi}_E$ is the partition function
\begin{equation}
  Z^{J, p, q, \pi}_E = \sum_{\omega \in \Omega_E} \prod_{e \in E} \left( p
  (J_e) \right)^{\omega_e} (1 - p (J_e))^{1 - \omega_e} \times q^{C^{\pi}_E(\omega)}.
\end{equation}
Note that we often use a simpler form for $\Phi^{J, p, q, \pi}_E$: if the parameters $p$ and $q$ are clear from the context, we omit them, and if $E$ is of the form $E^w (\Lambda)$ for some $\Lambda \subset\mathbb{Z}^d$ we simply write $\Phi^{J, \pi}_{\Lambda}$ instead of $\Phi^{J,\pi}_{E^w (\Lambda)}$.
For convenience we use the same notation for the probability measure $\Phi^{J, \pi}_E$ and for its expectation. 
Let us at last denote $f,w$ the two extremal boundary conditions: $f\in\Omega_{E^c}$ with $f_e=0, \forall e\in E^c$ is 
the \emph{free} boundary condition while $w\in\Omega_{E^c}$ with $w_e=1, \forall e\in E^c$ is
the \emph{wired} boundary condition.

When $q=2$ and $p(J)=1-\exp(-2\beta J)$, the measure $\Phi^{J, p, q, w}_{\Lambda}$ is the random cluster
representation of the Ising model with couplings $J$, and when $q\in\{2,3\ldots\}$ and $p(J)=1-\exp(-\beta J)$
it is the random cluster
representation of the $q$-Potts model with couplings $J$, see Section~\ref{sec-cg-Ising} and~\cite{N20}.
Yet, most of the results we present here are independent of this particular form for $p$.

Let us recall the most important properties of the FK measure $\Phi^{J, \pi}_E$.  Given $\omega, \omega' \in \Omega$ we write $\omega \leqslant
\omega'$ if and only if $\omega_e \leqslant \omega_e', \forall e \in E(\mathbb{Z}^d)$. A function $f : \Omega \rightarrow \mathbb{R}$ is said
increasing if for any $\omega, \omega' \in \Omega$ we have $\omega \leqslant
\omega' \Rightarrow f (\omega) \leqslant f (\omega')$.
For any finite
$E \subset E(\mathbb{Z}^d) $, for any $J \in \mathcal{J}$, $\pi \in
\Omega_{E^c}$, the following holds:
\begin{description}
\item[The DLR equation] For any function $h : \Omega \rightarrow
\mathbb{R}$, any $E'\subset E$,
\begin{equation}
  \Phi^{J, \pi}_E \left( h \left( \omega \right) \right) = \Phi^{J, \pi}_E
  \left[ \Phi^{J, (\omega \vee \pi)_{| (E')^c}}_{E'} h \left( \omega_{|
  (E')^c} \vee \omega' \right) \right] \label{eq-DLR-Phi}
\end{equation}
where $\omega'$ denotes the variable associated to the measure $\Phi^{J,
(\omega \vee \pi)_{|E^c}}_{E'}$.

\item[The FKG inequality] If $f, g : \Omega \rightarrow \mathbb{R}^+$ are positive increasing functions, then
\begin{equation}
  \Phi^{J, \pi}_E (f g) \geqslant \Phi^{J, \pi}_E (f) \Phi^{J, \pi}_E (g) .
  \label{eq-FKG-phi}
\end{equation}
\item[Monotonicity along $\pi$ and $p$] If $f : \Omega \rightarrow
\mathbb{R}^+$ is a positive increasing function and if $\pi, \pi' \in
\Omega_{E^c}$, $p, p' : [0, 1] \rightarrow [0, 1]$ satisfy $\pi \leqslant
\pi'$ and $p (J_e) \leqslant p' (J_e)$ for all $e \in E$, then
\begin{equation}
  \Phi^{J, p, q, \pi}_E (f) \leqslant \Phi^{J, p', q, \pi'}_E (f) .
  \label{eq-mon-Phi}
\end{equation}
\item[Comparison with percolation] If $\tilde{p} = p / (p + q (1 - p))$, for
any positive increasing function $f : \Omega \rightarrow \mathbb{R}^+$ we
have
\begin{equation}
  \Phi^{J, \tilde{p}, 1, f}_E (f) \leqslant \Phi^{J, p, q, \pi}_E (f)
  \leqslant \Phi^{J, p, 1, f}_E (f) \label{eq-comp-phi-P} .
\end{equation}
\end{description}
The proofs of these statements can be found in~\cite{N210} or in
the reference book~\cite{N222} (yet for uniform $J$).
Let us mention that the assumption $q \geqslant 1$ is fundamental for (\ref{eq-FKG-phi}).

\subsubsection{Random media}
We continue with the description of the law on the random media. Given a Borel probability distribution $\rho$ on $[0,1]$, we call~$\mathbb{P}$ the product measure
on $J\in\mathcal{J}$ that makes the $J_e$ i.i.d.\ variables with marginal law $\rho$, and denote~$\mathbb{E}$ the expectation associated to~$\mathbb{P}$. We also denote~$\mathcal{B}_E$ the $\sigma$-algebra generated by $J_{|E}$, for any $E\subset E(\mathbb{Z}^d) $.

We now turn towards the properties of $\Phi^{J, \pi}_E$ as a function of
$J$. Given $E, E' \subset E(\mathbb{Z}^d) $ with $E$ finite and a
function $h : \mathcal{J} \times \Omega \rightarrow \mathbb{R}^+$ such that
$h (., \omega)$ is $\mathcal{B}_{E'}$-measurable for each $\omega \in
\Omega_E$, the following holds:

\begin{description}
\item[Measurability] The function $J \rightarrow \Phi^{J, \pi}_E (\{\omega_0\})$ is  $\mathcal{B}_E$-measurable
while 
\begin{equation}
J \rightarrow \Phi^{J, \pi}_E (h (J,\omega)) \mbox{ \ and \ }
J \rightarrow \sup_{\pi \in \Omega_{E^c}} \Phi^{J, \pi}_E (h (J, \omega))
  \label{eq-meas-phiJ}
\end{equation}
are $\mathcal{B}_{E \cup E'}$-measurable, for all
$\omega_0 \in \Omega_E$ and $\pi \in \Omega_{E^c}$.

\item[Worst boundary condition] There exists a $\mathcal{B}_{E \cup
E'}$-measurable function $\tilde{\pi} : \mathcal{J} \mapsto \Omega_{E^c}$ such
that, for all $J \in \mathcal{J}$,
\begin{equation}
  \Phi^{J, \tilde{\pi} (J)}_E (h (J, \omega)) = \sup_{\pi \in \Omega_{E^c}}
  \Phi^{J, \pi}_E (h (J, \omega)) . \label{eq-meas-pit}
\end{equation}
\end{description}

The first point is a consequence of the fact that $\Phi^{J, \pi}_E
(\{\omega\})$ is a continuous function of the $p (J_e)$ and of the remark that 
\begin{equation}
\Phi^{J, \pi}_E (h (J, \omega)) =
\sum_{\omega \in \Omega_E} \Phi^{J, \pi}_E \left( \{\omega\} \right) h (J,
\omega). \notag
\end{equation}
For proving the existence of $\tilde{\pi}$ in (\ref{eq-meas-pit}) we
partition the set of possible boundary conditions $\Omega_{E^c}$ into finitely
many classes according to the equivalence relation
\begin{equation*} \pi \sim \pi' \Leftrightarrow \forall \omega \in \Omega_E, C^{\pi}_E
   (\omega) = C^{\pi'}_E (\omega) . \end{equation*}
A geometrical interpretation for this condition is the following: $\pi$ and
$\pi'$ are equivalent if they partition the interior boundary of the set of
vertices of $E$ in the same way. Consider now $\pi_1, \pi_2, \ldots, \pi_n \in
\Omega_{E^c}$ in each of the $n$ classes and define:
\begin{equation*} k (J) = \inf \left\{ k \in \{1, \ldots, n\}: \Phi^{J, \pi_k}_E (h (J,
   \omega)) = \sup_{\pi \in \Omega_{E^c}} \Phi^{J, \pi}_E (h (J, \omega))
   \right\}, \end{equation*}
it is a finite, $\mathcal{B}_{E \cup E'}$-measurable function and $\tilde{\pi} =
\pi_{k (J)}$ is a solution to \eqref{eq-meas-pit}.

\subsubsection{Quenched, averaged and averaged worst FK measures}
A consequence of (\ref{eq-meas-phiJ}) is that one can consider the joint law $\mathbb{E} \Phi^{J, \pi}_E$ on $(J,\omega)$.
We will be interested in the behavior of $\omega$ under both $\Phi^{J, \pi}_E$ for frozen $J\in\mathcal{J}$ -- we call $\Phi^{J, \pi}_E$
the {\emph{quenched}} measure -- and under the joint random media FK measure $\mathbb{E} \Phi^{J, \pi}_E$ 
-- we will refer to the marginal distribution of $\omega$ under $\mathbb{E} \Phi^{J, \pi}_E$  as the {\emph{averaged}} measure.
In view of Markov's inequality the \textit{averaged worst measure} constitutes a convenient way of
controlling both the $\mathbb{P}$ and the $\sup_{\pi} \Phi^{J,\pi}_E$-probabilities of \textit{rare} events 
(yet it is not a measure):
for any $\mathcal{A} \subset \Omega_E$ and
$C > 0$, 
\begin{align}
  \mathbb{E} \sup_{\pi \in \Omega_E^c} \Phi^{J, \pi}_E \left( \mathcal{A}
  \right) \leqslant \exp (- 2 C) & \Rightarrow \mathbb{P} \left( \sup_{\pi
  \in \Omega_E^c} \Phi^{J, \pi}_E \left( \mathcal{A} \right) \geqslant \exp (-
  C) \right) \leqslant \exp (- C) \notag\\
  & \Rightarrow \mathbb{E} \sup_{\pi \in \Omega_E^c} \Phi^{J, \pi}_E
  \left( \mathcal{A} \right) \leqslant 2 \exp (- C).  \label{eq-EPPhi}
\end{align}

\subsubsection{Absence of DLR equation for the averaged measure} 
Similarly to systems with quenched disorder that are non-Gibbsian~\cite{N236}, or to averaged laws of Markov chains in random media that are not Markov, the averaged FK measure lacks the DLR equation.
We present here a simple counterexample. Consider $\rho = \lambda
\delta_1 + (1 - \lambda) \delta_0$ for $\lambda \in (0, 1)$, $q > 1$ and $p(J_e) = p J_e$ with $p \in (0, 1)$. Let $E =\{e, f\}$ where $e
=\{x, y\}$ and $f =\{y, z\}$ with $z \neq x$ and 
$\pi$ a boundary condition that connects $x$ to $z$ but not to $y$. Then,
\begin{align*}
  \mathbb{E} \Phi_E^{\pi} \left( \omega_e = 1 \mbox{ and } \omega_f = 1
  \right) & =  \lambda^2 p \hat{p}\\
  \mbox{ } \mathbb{E} \Phi_E^{\pi} \left( \omega_e = 0 \mbox{ and } \omega_f
  = 1 \right) & = \lambda^2 (1 - p) \hat{p} + (1 - \lambda) \lambda
  \tilde{p}
\end{align*}
where
\begin{equation*} \tilde{p} =  \frac{p}{1 + (1 - p) (q - 1)} \mbox{ \ \ and \ \ }  \hat{p} = \frac{p}{1 + (1 - p)^2 (q - 1)} \end{equation*} 
and it follows that the conditional expectation of $\omega_e$ knowing
$\omega_f = 1$ equals
\begin{equation*} \left( \mathbb{E} \Phi_E^{\pi} \right) \left( \omega_e | \omega_f = 1
   \right) = \frac{\lambda p}{\lambda + (1 - \lambda) \tilde{p} / \hat{p}} >
   \lambda p \end{equation*}
since $\tilde{p} < \hat{p}$. As $\mathbb{E} \sup_{\pi'}
\Phi_{\{e\}}^{\pi'} (\omega_e) =\mathbb{E} \Phi_{\{e\}}^w (\omega_e) =
\lambda p$ we have proved that the averaged measure conditioned on the event $\omega_e=1$ strictly dominates any
averaged FK measure on $\{e\}$ with the same parameters, hence the DLR equation cannot hold.

\subsection{Slab percolation}
\label{sec-sp-def}The regime of \emph{percolation} under the averaged measure is characterized by
\begin{equation}
   \lim_{N \rightarrow \infty} \mathbb{E} \Phi^{J, f}_{\hat{\Lambda}_N} \left(
  0 \stackrel{\omega}{\leftrightarrow} \partial \hat{\Lambda}_N \right) > 0 \tag{\textbf{P}} \label{eq-def-P}
\end{equation}
yet we could not elaborate a coarse graining under the only assumption of percolation. As in \cite{N09} our work
relies on the stronger requirement of {\emph{slab percolation}} under the averaged measure, that is:
\begin{gather}
\exists H \in \mathbb{N}^{\star}\mbox{, \ }
  \inf_{N \in \mathbb{N}^{\star}} \inf_{x, y \in S_{N, H} \cup \partial S_{N, H}} \mathbb{E} \Phi^{J, f}_{S_{N, H}} \left( x
  \stackrel{\omega}{\leftrightarrow} y \right) > 0 \label{eq-def-SP3} \tag{\textbf{SP}, $d\geqslant3$} \\
\lim_{N \rightarrow \infty} \mathbb{E} \Phi^{J, f}_{S_{N, \kappa (N)}}  \left( \exists \mbox{ an horizontal crossing for $\omega$} \right) > 0
\label{eq-def-SP2} \tag{\textbf{SP}, $d=2$}
\end{gather}
for some function $\kappa: \mathbb{N}^\star\mapsto\mathbb{N}^\star$ with $\lim_{N \rightarrow\infty}{\kappa(N)/N}= 0$,
where an \emph{horizontal crossing} for $\omega$ means an $\omega$-cluster that connects the two vertical faces of $\partial S_{N, \kappa (N)}$.

The choice of the averaged measure for defining \eqref{eq-def-SP3} is not arbitrary and one should note that slab percolation does not occur in general under the quenched measure, even for high values of $\beta$ when $p(J_e)=1-\exp(-\beta J_e)$: as soon as $\mathbb{P}( J_e = 0 ) > 0$, the $\mathbb{P}$-probability that some vertex in the slab is $J$-disconnected goes to $1$ as $N \rightarrow\infty$, hence
\begin{equation*} \forall H\in\mathbb{N}^\star, \lim_{N\rightarrow\infty}
 \mathbb{P} \left( \inf_{x, y \in S_{N, H} \cup \partial S_{N, H}} \Phi^{J,
  f}_{S_{N, H}} \left( x \stackrel{\omega}{\leftrightarrow} y \right) = 0
  \right) = 1. \end{equation*}
This fact makes the construction of the coarse graining difficult. Indeed, the averaged measure lacks some mathematical properties with respect to the quenched measure -- notably the DLR equation -- and this impedes the generalization of Pisztora's construction~\cite{N09}, while under the quenched measure
the assumption of percolation in slabs is not relevant.

Let us discuss the generality of assumption (\textbf{SP}). It is remarkable that (\textbf{SP}) is \textit{equivalent} to the \textit{coarse graining} described by Theorem~\ref{thm-strong-cg} (the converse of Theorem~\ref{thm-strong-cg} is an easy exercise in view of the renormalization methods developed in Section~\ref{sec-renorm}). Yet, the fundamental question is whether \eqref{eq-def-P} and (\textbf{SP}) are
equivalent.

In the uniform case, when $d \geqslant 3$ it has been proved that the thresholds
for percolation and slab percolation coincide in the case of 
percolation ($q = 1$) by Grimmett and Marstrand~\cite{N170} and for the Ising model ($q = 2$) by Bodineau~\cite{N46}.
It is generally believed that they coincide for all $q\geqslant 1$. In the two dimensional case,
the threshold for \eqref{eq-def-SP2}
coincides again with the threshold for percolation $p_c$ when $q=1$, as $p_c$ coincides with
the threshold for exponential decay of connectivities in the dual lattice~\cite{N232,N233}.

In the random case the equality of thresholds holds when $q=1$ as the averaged measure 
boils down to a simple independent bond percolation process of intensity $\mathbb{E}(p (J_e))$. For larger $q$
we have no clue for a rigorous proof, yet we believe that the equality of thresholds should hold.
The argument of Aizenman et~al.~\cite{N21} provides efficient necessary and sufficient conditions for assumption (\textbf{SP}). 
Indeed, the averaged FK measure can be compared to 
independent bond percolation processes of respective intensities $\mathbb{E} \left( p (J_e)
/ (p (J_e) + q (1 - p (J_e))) \right)$ and $\mathbb{E}(p (J_e))$ (see also (\ref{eq-comp-phi-P})), which implies that 
\begin{equation}
\forall d \geqslant 2, \; \;\; \mathbb{E} \left( \frac{p (J_e)}{p (J_e) + q (1 - p (J_e))} \right) > p_c
  (d) \Rightarrow \mbox{(\textbf{SP})} \Rightarrow \mathbb{E}(p (J_e))
  \geqslant p_c (d)  \end{equation}
according to the equality
of thresholds for \eqref{eq-def-P} and (\textbf{SP}) for (non-random)
percolation. If we consider $p(J)=1-\exp(-\beta J)$, then 
(\textbf{SP}) occurs for $\beta$ large when $\mathbb{P}(J_e>0)>p_c$.

\subsection{Our results}

The most striking result we obtain is a generalization of the coarse graining
of Pisztora~\cite{N09}. Given $\omega \in \Omega_{E^w (\Lambda_N)}$, we say that a cluster
$\mathcal{C}$ for $\omega$ is a {\emph{crossing cluster}} if it touches every
face of $\partial \Lambda_N$.

\begin{theorem}
  \label{thm-strong-cg}Assumption {\rm(\textbf{SP})} implies the
  existence of $c > 0$ and $\kappa < \infty$ such that, for any $N \in
  \mathbb{N}^{\star}$ large enough and for all  $l \in [\kappa \log N, N]$,
\begin{equation*} \mathbb{E} \inf_{\pi} \Phi^{J, \pi}_{\Lambda_N} \left( \begin{array}{l}
       \mbox{There exists a crossing $\omega$-cluster $\mathcal{C}^{\star}$ in
       $\Lambda_N$}\\
       \mbox{and it is the unique cluster of diameter} \geqslant l
     \end{array} \right) \geqslant 1 - \exp \left( - cl \right) \end{equation*}
  where the infimum $\inf_{\pi}$ is taken over all boundary conditions $\pi
  \in \Omega_{E(\mathbb{Z}^d)  \setminus E^w (\Lambda_N)}$.
\end{theorem}

This result is completed by the following controls on the density of the main
cluster: if
\begin{equation}
  \theta^f = \lim_{N \rightarrow \infty} \mathbb{E} \Phi^{J,
  f}_{\hat{\Lambda}_N} \left( 0 \stackrel{\omega}{\leftrightarrow} \partial
  \hat{\Lambda}_N \right) \mbox{ \ and \ } \theta^w = \lim_{N \rightarrow
  \infty} \mathbb{E} \Phi^{J, w}_{\hat{\Lambda}_N} \left( 0
  \stackrel{\omega}{\leftrightarrow} \partial \hat{\Lambda}_N \right) \label{eq-def-thetafw}
\end{equation}
are the limit probabilities for percolation under the averaged measure with
{\emph{free}} and {\emph{wired}} boundary conditions, and if we define the
density of a cluster in $\Lambda_N$ as the ratio of its cardinal over $|
\Lambda_N |$, we have:

\begin{proposition}
  \label{prop-cg-dens}For any $\varepsilon > 0$ and $d \geqslant 1$,
\begin{equation}
 \limsup_N  \frac{1}{N^d} \log \mathbb{E} \sup_{\pi} \Phi_{\Lambda_N}^{J,
     \pi} \left( \begin{array}{l}
       \mbox{Some crossing cluster $\mathcal{C}^{\star}$ has}\\
       \mbox{a density larger than $\theta^w + \varepsilon$}
     \end{array} \right) < 0 \label{eq-dens-up}
\end{equation} 
  while assumption {\rm{(\textbf{SP})}} implies, for
  any $\varepsilon > 0$ and $d \geqslant 2$:
\begin{equation}
 \limsup_N  \frac{1}{N^{d - 1}} \log \mathbb{E} \sup_{\pi}
     \Phi_{\Lambda_N}^{J, \pi} \left( \begin{array}{l}
       \mbox{There is no crossing cluster $\mathcal{C}^{\star}$}\\
       \mbox{of density larger than $\theta^f - \varepsilon$}
     \end{array}  \right) < 0. \label{eq-dens-low}
\end{equation} 
\end{proposition}

In other words, the density of the crossing cluster determined by Theorem
\ref{thm-strong-cg} lies between $\theta^f$ and $\theta^w$. Yet in most cases
these two quantities coincide thanks to our last result, which generalizes
those of Lebowitz~\cite{N33} and Grimmett~\cite{N03}:

\begin{theorem}
  \label{thm-unique-infvol}If the interaction equals $p (J_e) =
1 - \exp (- \beta J_e)$, for any Borel probability measure $\rho$ on $[0, 1]$, any $q \geqslant 1$
and any dimension $d \geqslant 1$, the set
\begin{equation*} \mathcal{D}_{\rho, q, d} = \left\{ \beta \geqslant 0 : \lim_{N \rightarrow
   \infty} \mathbb{E} \Phi^{J, f}_{\hat{\Lambda}_N} \neq \lim_{N \rightarrow
   \infty} \mathbb{E} \Phi^{J, w}_{\hat{\Lambda}_N} \right\} \end{equation*}
is at most countable.
\end{theorem}
We also give an application of the coarse graining for the FK measure to the Ising model with ferromagnetic random interactions, see Theorem~\ref{thm-cg-Ising}.

\subsection{Overview of the paper}

\label{sec-overview}

A significant part of the paper is dedicated to the proof of the coarse graining -- Theorem~\ref{thm-strong-cg} -- under the assumption of slab percolation under the averaged measure \eqref{eq-def-SP3}. Let us recall that no simple adaptation of the original proof for the uniform media~\cite{N09} is possible as, on the one hand, the averaged measure does not satisfy the DLR equation while, on the second hand, slab percolation does not occur under the quenched measure.

In Section~\ref{sec-dense-cl} we prove the existence of a crossing cluster in a large box, with large probability under the averaged measure. We provide as well a much finer result: a stochastic comparison between the joint measure
and a product of local joint measures, that permits to describe some aspects of the structure of $(J,\omega)$ under the joint measure.

In Section~\ref{sec-cg-unique} we complete the difficult part of the coarse graining: we prove the uniqueness of large clusters with large probability. In order to achieve such a result we establish first a \textit{quenched} and \textit{uniform} characterization of \eqref{eq-def-SP3} that we call \eqref{eq-def-USP}: for $\varepsilon >0$ small enough and $L$ large enough, with a $\mathbb{P}$-probability at least $\varepsilon$ for $n$ large enough, each $x$ in the bottom of a slab $S$ of length $n L$, height $L \log n$ is, with a $\Phi^J_S$-probability at least $\varepsilon$, either connected to the origin of $S$, or disconnected from the top of the slab. For proving the (nontrivial) implication \eqref{eq-def-SP3}$\Rightarrow$\eqref{eq-def-USP} we describe first the typical structure of $(J,\omega)$ under the joint measure (Section \ref{sec-typ-struct}), then we introduce the notion of \textit{first pivotal bond} (Section \ref{sec-fpb}) that enables to make \textit{recognizable} local modifications for turning bad configurations (in terms of \eqref{eq-def-USP}) into appropriate ones. Finally, in Section~\ref{sec-weak-cg} we prove a first version of the coarse graining, while in Section~\ref{sec-dim-two} we give the same conclusion for the two dimensional case using a much simpler argument.

The objective of Section \ref{sec-renorm} is to present the adaptation of the renormalization techniques to the random media case. As a first application we state the final form of the coarse graining -- Theorem~\ref{thm-strong-cg} -- and complete it with estimates on the density of the crossing cluster -- Proposition \ref{prop-cg-dens}. We generalize then the results of \cite{N33,N03} on the uniqueness of the infinite volume measure -- see Theorem~\ref{thm-unique-infvol}. We conclude the article with an adaptation of the coarse graining to the Ising model with ferromagnetic disorder and discuss the structure of the local phase profile in Theorem~\ref{thm-cg-Ising}.

\section{Existence of a dense cluster}

\label{sec-dense-cl}In this Section we concentrate on the proof of existence
of a {\emph{dense}} $\omega$-cluster in a large box. As our proof is based on a
multi-scale analysis we begin with a few notations for the decomposition of
the domain into $L$-blocks: given $L \in \mathbb{N}^{\star}$, we say that a
domain $\Lambda \subset \mathbb{Z}^d$ is $L$-{\emph{admissible}} if it is of
the form $\Lambda = \prod_{i = 1}^d \{1, \ldots, a_i L - 1\}$ with $a_1,
\ldots, a_d \in \{2, 3, \ldots\}$. Such a domain can be decomposed into blocks
(and edge blocks) of side-length $L$ as follows: we let
\begin{equation}
  \BrLi = \left\{ 1, \ldots, L - 1 \right\}^d + Li \mbox{ \ \ and \ \ }
  B^L_i =\{0, \ldots, L\}^d + Li \label{eq-def-BLi}
\end{equation}
and denote
\begin{equation}
  \ErLi = E^w \left( \BrLi\right) \mbox{ \ \ and \ \ } E_i^L = E^f
  (B^L_i) \cap E^w (\Lambda) \label{eq-def-ELi}.
\end{equation}
We recall that $E^f (\Lambda')$ was defined at (\ref{eq-def-ELint}), 
it is the set of \emph{interior edges} of $\Lambda' \subset \mathbb{Z}^d$,
in opposition with $E^w (\Lambda')$ defined at (\ref{eq-def-EL}) that includes
the edges from $\Lambda'$ to the exterior. We call at last
\begin{equation}
  I_{\Lambda, L} = \left\{ i \in \mathbb{Z}^d : \BrLi \subset \Lambda
  \right\} . \label{eq-def-ILL}
\end{equation}
Remark that the $\ErLi$ are disjoint with $\ErLi
\subset E^L_i$. The edge set  $E^L_i$ includes the edges on the faces of $B^L_i$,
which makes $E^L_i$ and $E^L_j$ disjoint if and only if $i, j \in
I_{\Lambda, L}$ satisfy $\|i - j\|_2 > 1$. See also Figure
\ref{fig-decL-BLi} for an illustration.

In order to describe the structure of configurations $\omega \in \Omega_{E^w
(\Lambda)}$ we say that $(\mathcal{E}_i)_{i \in I_{\Lambda, L}}$ is a
$L$-{\emph{connecting family}} for $\Lambda$ if $\mathcal{E}_i$ is $\omega_{|E^L_i}$-measurable, $\forall i \in I_{\Lambda, L}$ and if it has the following
property: given any connected path $c_1, \ldots, c_n$ in $I_{\Lambda, L}$ (that is : we assume that
$\|c_{i+1}-c_i\|_2=1$, for all $i\in\{1,\ldots,n-1\}$ ), for
any $\omega \in \bigcap_{k = 1}^n \mathcal{E}_{c_k}$ there exists an
$\omega$-cluster in $\bigcup_{k = 1}^n E^L_{c_k}$ that connects all
faces of all~$\partial\BrLck$, for $k = 1 \ldots n$.

\medskip
Let us present the main result of this Section:

\begin{theorem}
  \label{thm-meas-Psi}Given any $L \in \mathbb{N}^{\star}$ and $\Lambda
  \subset \mathbb{Z}^d$ a $L$-admissible domain, there exist a measure
  $\Psi^L_{\Lambda}$ on $\mathcal{J}_{E^w (\Lambda)} \times \Omega_{E^w
  (\Lambda)}$ and a $L$-connecting family $(\mathcal{E}_i)_{i \in I_{\Lambda,
  L}}$ such that
  \begin{enumerate}
    \item the measure $\Psi^L_{\Lambda}$ is stochastically smaller than
    $\mathbb{E} \Phi^{J, f}_{\Lambda}$,
    
    \item under $\Psi^L_{\Lambda}$, each $\mathcal{E}_i$ is independent of the
    collection $(\mathcal{E}_j)_{j \in I_{\Lambda, L} : \|j - i\|_2 > 1}$,
    
    \item there exists $\rho_L \in [0, 1]$ independent of the choice of
    $\Lambda$ such that
    \begin{equation*} \inf_{i \in I_{\Lambda, L}} \Psi^L_{\Lambda} \left( \mathcal{E}_i
       \right) \geqslant \rho_L \end{equation*}
    with furthermore $\rho_L \underset{L \rightarrow \infty}{\longrightarrow}
    1$ if \eqref{eq-def-SP3}.
  \end{enumerate}
\end{theorem}

An immediate consequence of this Theorem is that \eqref{eq-def-SP3} implies the existence of a crossing cluster in the box
$\Lambda_{LN}$ for $L, N \in \mathbb{N}^{\star}$ large with large probability
under the averaged measure $\mathbb{E} \Phi^{J, f}_{\Lambda_{LN}}$, cf.
Corollary~\ref{corollary-exists-cl}. Yet, the information provided by Theorem
\ref{thm-meas-Psi} goes much further than Corollary~\ref{corollary-exists-cl}
and we will see in Section~\ref{sec-cg-unique} that it is also the basis for
the proof of the uniform slab percolation criterion \eqref{eq-def-USP}.

\subsection{The measure \texorpdfstring{$\Psi^L_{\Lambda}$}{Psi}}

\label{sec-meas-PsiLL} The absence of DLR equation
for the averaged FK measure makes impossible an immediate adaptation of
Pisztora's argument for the coarse graining~\cite{N09}. As an alternative to
the DLR equation one can however consider product measures and compare them to the
joint measure.

\begin{figure}[!h]
\begin{center}
\includegraphics[width=9cm]{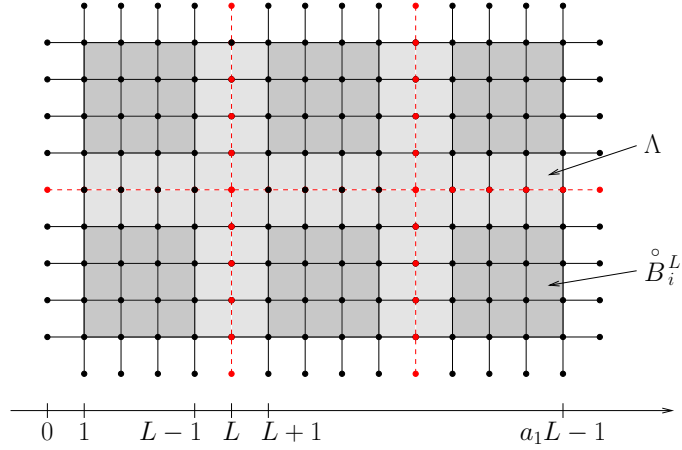}
\end{center}
\caption{\label{fig-decL-BLi}$E^w(\Lambda)$ partitioned into the $\ErLi$ and 
the lateral (dashed) edges.}
\end{figure}

Assuming that $\Lambda \subset \mathbb{Z}^d$ is a $L$-admissible domain, we
begin with the description of a partition of $E^w (\Lambda) = \bigsqcup_{k =
1}^n E_k$. On the one hand, we take all the $\ErLi$ with $i \in
I_{\Lambda, L}$ and then separately all the remaining edges, namely the
lateral edges of the $B^L_i$ (see (\ref{eq-def-ELi}) for the definition of
$\ErLi$ and Figure~\ref{fig-decL-BLi} for an illustration of the
partition). This can be written down as
\begin{equation}
  E^w (\Lambda) = \bigsqcup_{k = 1}^n E_k = \left( \bigsqcup_{i \in I_{\Lambda,
  L}} \ErLi \right) \bigsqcup \left( \bigsqcup_{e \in E_{\tmop{lat}}^L(\Lambda)} \{e\} \right) \label{eq-def-part-EL}
\end{equation}
where $E_{\tmop{lat}}^L (\Lambda) = E^w (\Lambda) \setminus \bigcup_{i \in
I_{\Lambda, L}} \ErLi = \bigcup_{i \in I_{\Lambda, L}} (E^L_i \setminus
\ErLi)$. We consider then for $\Psi^L_{\Lambda}$ the measure on
$\mathcal{J}_{E^w (\Lambda)} \times \Omega_{E^w (\Lambda)}$ defined by
\begin{equation}
  \Psi^L_{\Lambda} (h (J, \omega)) = \left[ \bigotimes_{k = 1}^n
  \mathbb{E}_{E_k} \Phi_{E_k}^{J_k, f} \right] \left( h (J_1 \vee \ldots \vee
  J_n, \omega_1 \vee \ldots \vee \omega_n) \right) \label{eq-def-PsiLL}
\end{equation}
for any $h : \mathcal{J} \times \Omega \rightarrow \mathbb{R}^+$ such that
$h (., \omega)$ is $\mathcal{B}_{E^w(\Lambda)}$-measurable for each $\omega \in \Omega_{E^w(\Lambda)}$, where
$J_1 \vee \ldots \vee J_n \in \mathcal{J}_{E^w(\Lambda)}$ (resp. $\omega_1 \vee \ldots \vee
\omega_n \in \Omega_{E^w(\Lambda)}$) stands for the configuration which restriction to
$E_k$ equals $J_k$ (resp. $\omega_k$), for all $k$.

The first crucial feature of $\Psi^L_{\Lambda}$ is its product structure:
under $\Psi^L_{\Lambda}$ the restriction of $(J, \omega)$ to any $\ErLi$
with $i \in I_{\Lambda, L}$ or to $\{e\}$ with $e \in E_{\tmop{lat}}^L
(\Lambda)$ is independent of the rest of the configuration, so that in
particular the restriction of $\omega$ to $E^L_i$ is independent of 
its restriction to 
$\bigcup_{j \in I_{\Lambda, L} : \|j - i\|_2 > 1} E^L_j$, and for any
$L$-connecting family point ({ii}) of Theorem~\ref{thm-meas-Psi} is
verified.

The second essential property of $\Psi^L_{\Lambda}$ is that it is
stochastically smaller than the averaged measure on $\Lambda$ with free
boundary condition, namely point ({i}) of Theorem~\ref{thm-meas-Psi} is
true. This is an immediate consequence of the following Proposition:

\begin{proposition}
  \label{prop-comp-averaged-prod}Consider a finite edge set $E$ and a
  partition $(E_i)_{i = 1 \ldots n}$ of $E$. Assume that $h : \mathcal{J}
  \times \Omega \rightarrow \mathbb{R}$ is $\mathcal{B}_E$-measurable in the first variable and
  that for every $J$, $h (J, .)$ is an increasing function. If we denote by
  $(J_i, \omega_i) \in \mathcal{J}_{E_i} \times \Omega_{E_i}$ the variables
  associated to the measure $\mathbb{E} \Phi_{E_i}^{J_i, f}$ (resp. $\mathbb{E}
  \Phi_{E_i}^{J_i, w}$), we have:
  \begin{align}
    \mathbb{E} \Phi^{J, w}_E h & \leqslant \left[ \bigotimes_{i = 1}^n
    \mathbb{E}_{E_i} \Phi_{E_i}^{J_i, w} \right] \left( h (J_1 \vee \ldots
    \vee J_n, \omega_1 \vee \ldots \vee \omega_n) \right) \\
    \mbox{resp. \ \ \ } \mathbb{E} \Phi^{J, f}_E h & \geqslant \left[
    \bigotimes_{i = 1}^n \mathbb{E}_{E_i} \Phi_{E_i}^{J_i, f} \right] \left(
    h (J_1 \vee \ldots \vee J_n, \omega_1 \vee \ldots \vee \omega_n) \right) .
    \label{ineq-EPhi-prod}
  \end{align}
\end{proposition}

\begin{pf}
  We focus on the proof of the second inequality since both proofs are
  similar. We begin with the case $n = 2$. Applying twice the DLR equation
  (\ref{eq-DLR-Phi}) for $\Phi$ we get, for any $J \in \mathcal{J}$:
  \begin{equation*}
    \Phi^{J, f}_E  \left( h (J, \omega) \right) = \Phi^{J, f}_E \left[
    \Phi^{J, \omega_{|E_2}}_{E_1} \left[ \Phi^{J, \omega_1}_{E_2}  \left( h
    \left( J, \omega_1 \vee \omega_2 \right) \right) \right] \right] \end{equation*}
  where $\omega$ is the variable for $\Phi^{J, f}_E$, $\omega_1$ that for
  $\Phi^{J, \omega_{|E_2}}_{E_1}$ and $\omega_2$ that for $\Phi^{J,
  \omega_1}_{E_2}$. Since $h \left( J, \omega_1 \vee \omega_2 \right)$ is an
  increasing function of $\omega_1$ and $\omega_2$, it is enough to use the
  monotonicity (\ref{eq-mon-Phi}) of $\Phi^{J, \pi}_E$ along $\pi$ to conclude
  that
  \begin{equation*}
    \Phi^{J, f}_E  \left( h (J, \omega) \right) \geqslant \Phi^{J, f}_{E_1}
    \left[ \Phi^{J, f}_{E_2}  \left( h \left( J, \omega_1 \vee \omega_2
    \right) \right) \right] . \end{equation*}
  The same question on the $J$-variable is trivial since $\mathbb{P}$ is a
  product measure, namely if $J, J_1, J_2$ are the variables corresponding to
  $\mathbb{E}_E, \mathbb{E}_{E_1}$ and $\mathbb{E}_{E_2}$:
  \begin{equation*}
    \mathbb{E}_E \Phi^{J, f}_E  \left( h (J, \omega) \right) \geqslant
    \mathbb{E}_{E_1} \mathbb{E}_{E_2} \Phi^{J_1, f}_{E_1} \left[ \Phi^{J_2,
    f}_{E_2}  \left( h \left( J_1 \vee J_2, \omega_1 \vee \omega_2 \right)
    \right) \right] . \end{equation*}
  It is clear that $\mathbb{E}_{E_2}$ and $\Phi^{J_1, f}_{E_1}$ commute, and
  that $\mathbb{E}_{E_1} \Phi^{J_1, f}_{E_1}$ and $\mathbb{E}_{E_2}
  \Phi^{J_2, f}_{E_2}$ also commute, hence the claim is proved for $n = 2$. We
  end the proof with the induction step, assuming that (\ref{ineq-EPhi-prod})
  holds for $n$ and that $E$ is partitioned into $(E_i)_{i = 1 \ldots n + 1}$.
  Applying the inductive hypothesis at rank $2$ to $E_1$ and $E'_2 = \cup_{i
  \geqslant 2} E_i$ we see that $\mathbb{E} \Phi^{J, f}_E h (J, \omega)
  \geqslant \mathbb{E}_{E_1} \Phi^{J_1, f}_{E_1} \mathbb{E}_{E_2'}
  \Phi^{J_2', f}_{E_2'} h \left( J_1 \vee J_2', \omega_1 \vee \omega'_2
  \right)$. Remarking that for any fixed $(J_1, \omega_1)$ the function
  $\left( J, \omega \right) \in \mathcal{J} \times \Omega \mapsto h
  (J_1 \vee J_{|E'_2}, \omega_1 \vee \omega)$ is $\mathcal{B}_{E'_2}$-measurable in $J$ and
  increases with $\omega$ we can apply the inductive hypothesis at rank $n$ in
  order to expand further on $J'_2$ and $\omega'_2$ and the proof is over.
\end{pf}

\subsection{The \texorpdfstring{$L$}{L}-connecting family \texorpdfstring{$\mathcal{E}^{L, H}_i$}{E}}

\label{sec-omega-struct}
\begin{figure}[!h]
\begin{center}
\includegraphics[width=10cm]{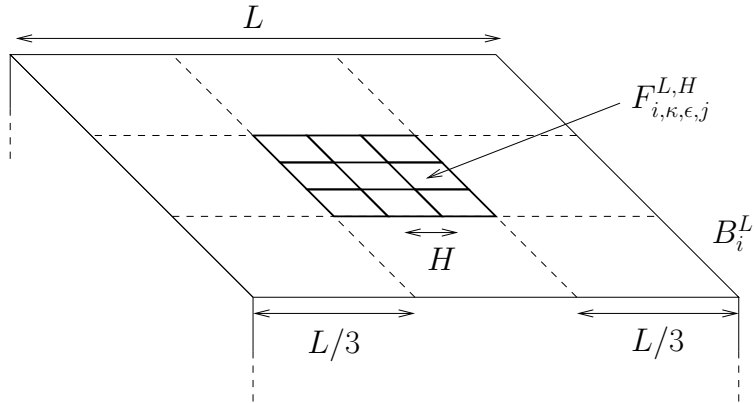}
\end{center}
\caption{\label{fig-BLi-FLH}The $d - 1$ dimensional facets $F^{L, H}_{i, \kappa, \varepsilon, j}$.}
\end{figure}

The second step towards the proof of Theorem
\ref{thm-meas-Psi} is the construction of a $L$-connecting family. The faces
of the blocks $B^L_i$ play an important role hence we continue with some more
notations. Remark that $(\kappa, \varepsilon) \in \{1, \ldots, d\} \times \{0,
1\}$ indexes conveniently the $2 d$ faces of $B^L_i$ if to $(\kappa,
\varepsilon)$ we associate the face $Li + L \varepsilon \mathbf{e}_{\kappa}
+\mathcal{F}^L_{\kappa}$ where
\begin{equation}
  \mathcal{F}^L_{\kappa} =\{0, \ldots, L\}^d \cap \{x \cdot
  \mathbf{e}_{\kappa} = 0\}.
\end{equation}
We decompose then each of these faces into smaller $d - 1$ dimensional
hypercubes and let
\begin{equation}
  \mathcal{H}^{L, H}_{\kappa} = \left\{ j \in \mathbb{Z}^d : 
\begin{array}{l}
j \cdot \mathbf{e}_{\kappa} = 0 \mbox{ and }\forall k \in \{1, \ldots, d\} \setminus \{\kappa\},
   \\
 L/({3 H}) \leqslant j \cdot
  \mathbf{e}_k \leqslant {2 L}/({3 H}) - 1
\end{array}     \right\} \label{eq-cond-j-seed}
\end{equation}
and for any $j \in \mathcal{H}^{L, H}_{\kappa}$ we denote
\begin{equation}
  F^{L, H}_{i, \kappa, \varepsilon, j} = Li + L \varepsilon
  \mathbf{e}_{\kappa} + H j +\mathcal{F}^{H - 1}_{\kappa}
\end{equation}
so that $F^{L, H}_{i, \kappa, \varepsilon, j}$ is the translated of
$\mathcal{F}^{H - 1}_{\kappa}$ positioned at $H j$ on the face $(\kappa,
\varepsilon)$ of $B^L_i$, as illustrated on Figure~\ref{fig-BLi-FLH}.

The facets $F^{L, H}_{i, \kappa, \varepsilon, j}$ will play the role of seeds
for the $L$-connecting family. Given $\omega \in \Omega_{E^w (\Lambda)}$ and $i
\in I_{\Lambda, L}$, $(\kappa, \varepsilon) \in \{1, \ldots, d\} \times \{0,
1\}$ and $j_0 \in \mathcal{H}^{L, H}_{\kappa}$, we say that $F^{L, H}_{i,
\kappa, \varepsilon, j_0}$ is a {\emph{seed at scale}} $H$ for the face
$(\kappa, \varepsilon)$ of $B^L_i$ if $j_0$ is the smallest index in the
lexicographical order among the $j \in \mathcal{H}^{L, H}_{\kappa}$ such that
either $F^{L, H}_{i, \kappa, \varepsilon, j} \cap \Lambda = \emptyset$, or all the edges $e
\in E^f (F^{L, H}_{i, \kappa, \varepsilon, j})$ are open for
$\omega$ (we recall that $E^f (\Lambda')$ is the set of
edges between any two adjacent points of $\Lambda' \subset \mathbb{Z}^d$, cf.~(\ref{eq-def-ELint}))

The first condition is designed to handle the case when the face $(\kappa,
\varepsilon)$ of $B^L_i$ is not in $\Lambda$ (this happens if $\BrLi$
touches the border of $\Lambda$, cf.~Figure~\ref{fig-decL-BLi}): with our
conventions, there always exists a seed in that case and it is the $F^{L,
H}_{i, \kappa, \varepsilon, j}$ of smallest index $j \in \mathcal{H}^{L, H}_{\kappa}$.

Then, we let
\begin{equation}
  \mathcal{E}^{L, H}_i = \left\{ \omega \in \Omega_{E^w (\Lambda)} :
  \mbox{\begin{tabular}{l}
    Each face of $B^L_i$ owns a seed and\\
    these are connected under $\omega_{|E^L_i}$
  \end{tabular}} \right\} \label{eq-def-E}, \forall i \in I_{\Lambda, L}
\end{equation}
which is clearly a $L$-connecting family since, on the one hand,
$\mathcal{E}^{L, H}_i$ depends on $\omega_{|E^L_i}$ only and, on the second
hand, the seed on the face $(\kappa, 1)$ of $B^L_i$ corresponds {\emph{by construction}} to that on the
face $(\kappa, 0)$ of $B^L_{i +\mathbf{e}_{\kappa}}$, for any $i, i
+\mathbf{e}_{\kappa} \in I_{\Lambda, L}$. Hence we
are left with the proof of part ({iii}) of Theorem~\ref{thm-meas-Psi}.

\subsection{Large probability for \texorpdfstring{$\mathcal{E}^L_i$}{E} under \texorpdfstring{$\Psi^L_{\Lambda}$}{Psi}}

In this Section we conclude the proof of Theorem~\ref{thm-meas-Psi} with an
estimate over the $\Psi^L_{\Lambda}$-probability of
\begin{equation}
  \mathcal{E}^L_i =\mathcal{E}^{L, [ \sqrt[d]{\delta \log L}]}_i
  \label{eq-def-calEL}
\end{equation}
for $\delta > 0$ small enough, and show as required that $\Psi^L_{\Lambda}
(\mathcal{E}^L_i) \rightarrow 1$ as $L \rightarrow \infty$ assuming \eqref{eq-def-SP3}, uniformly over $\Lambda$
and $i \in I_{\Lambda, L}$. Our proof is made of the two Lemmas below:
 first we prove the existence of seeds with large probability
 and then we estimate the
conditional probability for connecting them.

\begin{lemma}
  \label{lemma-exists-seed}Assume that $c =\mathbb{E}_{} \Phi^{J, f}_{\{e\}}
  (\omega_e) > 0$ and let $\delta < - 1 / \log c$. Then there exists
  $(\rho_L)_{L \geqslant 3}$ with $\lim_{L \rightarrow \infty} \rho_L = 1$
  such that, for every $L$-admissible $\Lambda$ and every $i \in I_{\Lambda,
  L}$,
  \begin{equation*}
    \Psi^L_{\Lambda} \left( \left\{ \mbox{Each face of $B^L_i$ bears a seed at
    scale } H_L = [ \sqrt[d]{\delta \log L}] \right\} \right) \geqslant
    \rho_L . \end{equation*}
\end{lemma}

\begin{lemma}
  \label{lemma-cond-ELH}Assume \eqref{eq-def-SP3}.
  Then, there exists $(\rho'_H)_{H \in \mathbb{N}^{\star}}$ with $\rho'_H \rightarrow 1$ as $H
  \rightarrow \infty$ such that, for any $L \in
  \mathbb{N}^{\star}$ such that $\mathcal{H}^{L, H}_1 \neq \emptyset$, any
  $L$-admissible domain $\Lambda \subset \mathbb{Z}^d$ and any $i \in
  I_{\Lambda, L}$,
  \begin{equation}
    \Psi^L_{\Lambda} \left( \mathcal{E}_i^{L, H} \left| \left\{
    \mbox{\begin{tabular}{l}
      Each face of $B^L_i$ bears\\
      a seed at scale $H$
    \end{tabular}} \right. \right\} \right) \geqslant \rho'_H . \label{eq-PsiL-cond}
  \end{equation}
\end{lemma}

Before proving Lemmas~\ref{lemma-exists-seed} and
\ref{lemma-cond-ELH} we state an important warning: the fact that
$\Psi^L_{\Lambda} (\mathcal{E}^L_i) \rightarrow 1$ as $L \rightarrow \infty$
does not give any information on the probability of $\mathcal{E}_i^L$ under
the averaged measure $\mathbb{E} \Phi^J_{\Lambda}$ as $\mathcal{E}^L_i$ is \emph{not} an increasing event !

\begin{pf}
  (Lemma~\ref{lemma-exists-seed}). The $\Psi^L_{\Lambda}$-probability for any lateral edge of $B^L_i$ to be open
  equals~$c$, hence a facet $F^{L, H_L}_{i, \kappa, \varepsilon, j}
  \subset \Lambda$ is entirely open with a probability
  \begin{equation*} c^{(d - 1) (H_L - 1) H_L^{d - 2}} \geqslant c^{H_L^d} \end{equation*}
  for $L$ large enough. Consequently, the probability that there is a seed at
  scale $H_L$ on each face of $B^L_i$ is at least
  \begin{align}
 \rho_L &= 1 - 2 d \left( 1 - c^{H_L^d} \right)^{\left[ \frac{L}{3 H_L} - 2 \right]^{d - 1}} \notag \\ 
& \geqslant 1 - 2 d \exp \left( - c^{H_L^d} \times \left[ \frac{L}{3 H_L} - 2 \right]^{d -
    1} \right)  \label{eq-ub-noseed}
  \end{align}
  using the inequality $1 - u \leqslant \exp \left( - u \right)$. We remark at
  last that for $L$ large,
  \begin{align*}
    \log \left( c^{H_L^d} \times \left[ \frac{L}{3 H_L} - 2
    \right]^{d-1} \right) & \geqslant H_L^d \log c + (d-1)\left( \log L -  \log (4 H_L) \right) \\
    & \geqslant \left( 1 + \delta \log c \right) \log L - \log (4
    H_L)
  \end{align*}
  with $1 + \delta \log c > 0$ thanks to the assumption on $\delta$, hence the
  term in the exponential in (\ref{eq-ub-noseed}) goes to $- \infty$ as $L
  \rightarrow \infty$ and we have proved that $\lim_{L \rightarrow \infty}
  \rho_L = 1$.
\end{pf}

\begin{figure}[!h]
\begin{center}
\includegraphics[width=10cm]{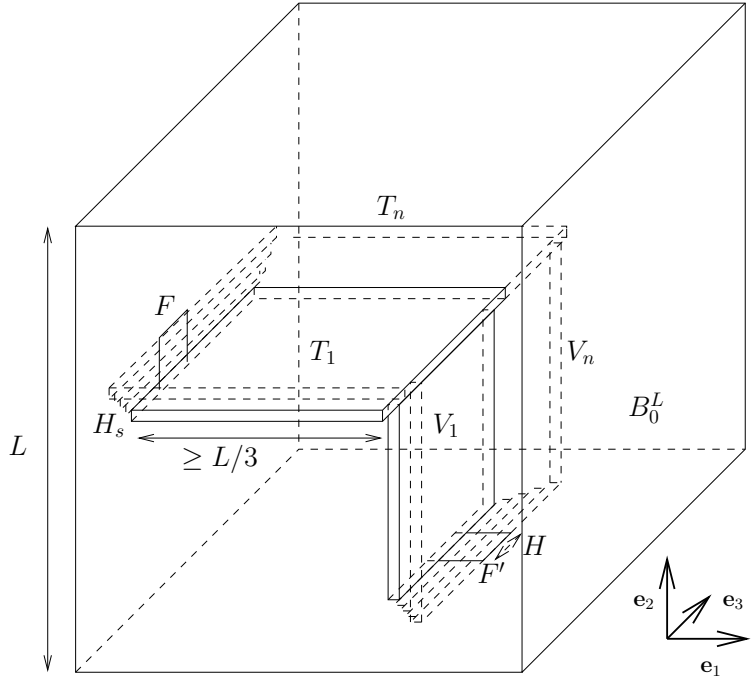}
\end{center}
  \caption{\label{fig-slab-dec-E}The slabs $V_n$ and $T_n$ in the proof of
  Lemma~\ref{lemma-cond-ELH}.}
\end{figure}

\begin{pf}
  (Lemma~\ref{lemma-cond-ELH}). We fix a realization
  $\oext \in \Omega_{E^w (\Lambda) \setminus \ErLi}$ such
  that each face of $B^L_i$ bears a seed under $\oext$. Thanks
  to the product structure of $\Psi^L_{\Lambda}$, the restriction to
  $\ErLi$ of the conditional measure $\Psi^L_{\Lambda} \left( . | \omega
  = \oext \right)$ equals $\mathbb{E} \Phi^{J,
  f}_{\ErLi}$, hence the probability for connecting all seeds together is
  \begin{equation*} \mathbb{E} \Phi^{J, f}_{\ErLi} \left( \omega \vee
     \oext \in \mathcal{E}^{L, H}_i \right). \end{equation*}
We will prove below that with large probability 
one can connect a seed to the seed in any adjacent face, and this will be enough for concluding
the proof.
Indeed, denote $s_1, \ldots, s_{2 d}$ the seeds of $\oext$. Thanks to the requirement $a_i \geqslant 2$ in the definition of $L$-admissible
  sets, we can assume that $s_1$ and $s_2$ are on adjacent faces, both of them {\emph{inside}} $\Lambda$ so that in fact $s_1$ and $s_2$ are entirely open for $\oext$. If we connect $s_1$ to each of
  the seeds $s_2, \ldots, s_{2 d - 1}$ in the adjacent faces of $B^L_i$, and then in turn connect $s_2$ to $s_{2 d}$ we have connected all seeds together. As a consequence one can take
  \begin{equation} \rho'_H = 1 - (2 d - 1) (1 - \inf_{L : \mathcal{H}^{L, H}_1 \neq
     \emptyset} \rho''_{H, L}) 
\label{eq-rhop-H}
\end{equation}
  as a lower bound in (\ref{eq-PsiL-cond}), where $\rho''_{H, L}$ is the least
  probability under $\mathbb{E} \Phi^{J, f}_{\ErLi}$ for connecting two
  facets $F^{L, H}_{i, \kappa, \varepsilon, j}$ and $F^{L, H}_{i, \kappa',
  \varepsilon', j'}$ in adjacent faces of $B^L_i$.
  
  For the sake of simplicity we let $i = 0$, $(\kappa, \varepsilon) = (1, 0)$
  and $(\kappa', \varepsilon') = (2, 0)$. Our objective is to
  connect any two facets $F^{L, H}_{0, 1, 0, j}$ and $F^{L, H}_{0, 2, 0, j'}$
  ($j \in \mathcal{H}^{L, H}_1$ and $j' \in \mathcal{H}^{L, H}_2$) with large
  probability under $\mathbb{E} \Phi^{J, f}_{\BrL0}$, and we achieve this placing slabs in $B^L_0$.
  Thanks to assumption \eqref{eq-def-SP3} there exist $\alpha > 0$ and $H_s
  \in \mathbb{N}^{\star}$ such that any two points in $S \cup \partial S$ are
  connected by $\omega$ with probability at least $\alpha$ under $\mathbb{E}
  \Phi^{J, f}_S$, provided that $S$ is of the form
  $ S =\{1, \ldots, N - 1\}^{d - 1} \times \{1, H_s - 1\} $
  with $N \in \mathbb{N}$ large enough. We describe now two sequences of
  slabs of height $H_s$ linking the seeds $F^{L, H}_{0, 1, 0, j}$ and $F^{L, H}_{0, 2, 0, j'}$
  to each other. Let first, for $l \in \mathbb{N}$ and $\kappa \in \{1, 2\}$:
  \begin{equation}
    S (l, \kappa) =\{1, \ldots, l - 1\}^d \cap \{x : 1 \leqslant x \cdot
    \mathbf{e}_{\kappa} \leqslant H_s - 1\}
  \end{equation}
  and then
  \begin{equation}
    U (l, h, \kappa) = S (l, \kappa) + h\mathbf{e}_{\kappa} + \sum_{k
    \geqslant 3} \left[ \frac{L - l}{2} \right] \mathbf{e}_k
  \end{equation}
  for $l \in \{1, \ldots, L\}$, $h \in \{0, \ldots, L - H_s \}$ and $\kappa
  \in \{1, 2\}$. The slab $U (l, h, \kappa)$ is normal to
  $\mathbf{e}_{\kappa}$ and the $\mathbf{e}_{\kappa}$-coordinates of its points remain in $\{h + 1, \ldots, h + H_s - 1\}$, it is in
  contact with the face $(\kappa', 0)$ of $\BrL0$ where
  $\{\kappa' \}=\{1, 2\} \setminus \{\kappa\}$ and it is positioned roughly at
  the center of $\BrLi$ in every other direction $\mathbf{e}_k$ for $k
  \geqslant 3$. We conclude these geometrical definitions letting
  \begin{equation}
    V_n = U \left( j \cdot \mathbf{e}_2 H + (n - 1) H_s, j' \cdot
    \mathbf{e}_1 H + (n - 1) H_s, 1 \right)
  \end{equation}
  which are vertical slabs and
  \begin{equation}
    T_n = U \left( j' \cdot \mathbf{e}_1 H + n H_s, j \cdot \mathbf{e}_2 H
    + (n - 1) H_s, 2 \right)
  \end{equation}
  which are horizontal slabs, for any $n \in \{1, \ldots, \lceil H / H_s
  \rceil\}$. As illustrated on Figure~\ref{fig-slab-dec-E}, for any $n \in
  \{1, \ldots, \lceil H / H_s \rceil\}$, $F^{L, H}_{0, 1, 0, j}$ is in contact
  with $E^w (T_n)$ since the largest dimension of the slab is at least $L / 3$,
  while $E^w (V_n)$ touches $F^{L, H}_{0, 2, 0, j'}$, and by construction $E^w
  (V_n)$ and $E^w (T_n)$ touch each other. Furthermore the edge sets $E^w (V_n)$ and
  $E^w (T_n)$ are all disjoint, and all included in $E^w ( \BrL0)$. Consider
  now the product measure
  \begin{equation}
    \Theta = \bigotimes_{n = 1}^{\lceil H / H_s \rceil} \left( \mathbb{E}
    \Phi^{J, f}_{V_n} \otimes \mathbb{E} \Phi^{J, f}_{T_n} \right) .
  \end{equation}
  Under the measure $\Theta$, the probability that there is a $\omega$-open
  path in $E^w (V_n) \cup E^w (T_n)$ between the two seeds $F^{L, H}_{0, 1, 0, j}$
  and $F^{L, H}_{0, 2, 0, j'}$ is at least $\alpha^2$ thanks to \eqref{eq-def-SP3}. By independence of the restrictions of $\omega$ to the
unions of slabs
  $\left( E^w (V_n) \cup E^w (T_n) \right)_{n = 1 \ldots \lceil H / H_s \rceil}$,
  it follows that the $\Theta$-probability that $\omega$ does not
  connect $F^{L, H}_{0, 1, 0, j}$ to $F^{L, H}_{0, 2, 0, j'}$ in $E^w (\BrL0)$ is not larger than $\left( 1 - \alpha^2 \right)^{\lceil H /
  H_s \rceil}$. Thanks to the stochastic domination $\Theta
  \underset{stoch.}{\leqslant} \mathbb{E} \Phi^{J, f}_{\BrL0}$
  seen in Proposition~\ref{prop-comp-averaged-prod}, the same control holds
  for the measure $\mathbb{E} \Phi^{J, f}_{\BrL0}$ and we have proved
  that
  \begin{equation*} \rho''_{H, L} \geqslant \left( 1 - \alpha^2 \right)^{\lceil H / H_s
     \rceil} \end{equation*}
  for any $L$ such that $\mathcal{H}^{L, H}_1 \neq \emptyset$. In view of (\ref{eq-rhop-H}) this yields $\lim_{H \rightarrow \infty} \rho'_H = 1$.
\end{pf}

\subsection{Existence of a crossing cluster}

\label{sec-exists-cc}An easy consequence of Theorem
\ref{thm-meas-Psi} is the following:
\begin{corollary}
  \label{corollary-exists-cl}If \eqref{eq-def-SP3}, for any $L \in \mathbb{N}^{\star}$ large enough one has
  \begin{equation*} \lim_N \mathbb{E} \Phi^{J, f}_{\Lambda_{LN}} \left( \mbox{There exists a
     crossing cluster for } \omega \mbox{ in } \Lambda_{LN} \right) = 1. \end{equation*}
\end{corollary}

\begin{pf}
  The existence of a crossing cluster is an increasing event hence it is
  enough to prove the estimate under the stochastically smaller measure
  $\Psi^L_{\Lambda_{LN}}$. Under $\Psi^L_{\Lambda_{LN}}$ the events
  $\mathcal{E}_i^L$ are only $1$-dependent thus for $L$ large enough the
  collection $(\mathbf{1}_{\{\mathcal{E}_i^L \}})_{i \in I_{\Lambda, L}}$
  stochastically dominates a site percolation process with high density~\cite{N43}.
 In particular, the coarse graining~\cite{N09} yields the
  existence of a crossing cluster for $(\mathbf{1}_{\{\mathcal{E}_i^L
  \}})_{i \in I_{\Lambda, L}}$ in $I_{\Lambda, L}$ with large probability as
  $N \rightarrow \infty$, and the latter event implies the existence of a
  crossing cluster for $\omega$ in $\Lambda_{LN}$ as $(\mathcal{E}_i^L)_{i \in
  I_{\Lambda, L}}$ is a $L$-connecting family.
\end{pf}

\section{Uniqueness of large clusters}

\label{sec-cg-unique}In the previous Section we established Theorem
\ref{thm-meas-Psi}, that gives a first description of the behavior of clusters in a
large box. Our present objective is to use that information in order to infer from the slab
percolation assumption \eqref{eq-def-SP3} a uniform slab percolation
criterion \eqref{eq-def-USP}. 

Given $L, n \in \mathbb{N}^{\star}$ with $n
\geqslant 3$ we let
\begin{equation}
  \Lambda_{n, L}^{\log} =\{1, Ln - 1\}^{d - 1} \times \{1, L \lceil \log n
  \rceil - 1\}, \label{eq-def-LlognL}
\end{equation}
call $\Bottom(\Lambda^{\log}_{n, L}) = \{1, \ldots, L n - 1\}^{d - 1} \times \{0\}$ and $\Top(\Lambda^{\log}_{n, L}) =  \{1, \ldots, L n - 1\}^{d - 1} \times \{L \lceil \log n \rceil\}$
the horizontal faces of $\partial \Lambda_{n, L}^{\log}$, consider $o = (1, \ldots, 1, 1) \in \mathbb{Z}^d$ a reference point in $\Lambda_{n,
L}^{\log}$ and $\mathbb{H}^-$ the discrete lower half space
\begin{equation}
  \mathbb{H}^- =\{x \in \mathbb{Z}^d : x \cdot \mathbf{e}_d \leqslant 0\}.
  \label{eq-def-Hminus}
\end{equation}
as well as $E^- = E^f (\mathbb{H}^-)$ the set of edges with all
extremities in $\mathbb{H}^-$. Then, we define \eqref{eq-def-USP} as follows:
\begin{align}
&\exists L \in \mathbb{N}^{\star}, \exists \varepsilon > 0 \mbox{ such that for
any $n$ large enough,} \notag \\
 & \mathbb{P} \left( \begin{array}{l}
    \forall x \in \Bottom(\Lambda^{\log}_{n, L}), \forall \pi \in
    \Omega_{E^w (\Lambda^{\log}_{n, L})^c}, \forall \xi \in \Omega_{E^-} :\\
    \Phi_{\Lambda^{\log}_{n, L}}^{J, \pi} \left( x \stackrel{\omega \vee
    \xi}{\leftrightarrow} o \mbox{ or } x \stackrel{\omega \vee
    \xi}{\nleftrightarrow} \Top(\Lambda^{\log}_{n, L}) \right)
    \geqslant \varepsilon
  \end{array} \right) \geqslant \varepsilon . \label{eq-def-USP} \tag{\textbf{USP}}
\end{align}
The implication \eqref{eq-def-SP3}$\Rightarrow$\eqref{eq-def-USP} will be
finally proved in Proposition \ref{prop-unif-conn}, and its consequence -- the
uniqueness of large clusters -- detailed in Proposition~\ref{prop-weak-cg}.

\subsection{Typical structure in slabs of logarithmic height}
\label{sec-typ-struct}
As a first step towards the proof of the implication \eqref{eq-def-SP3}$\Rightarrow$\eqref{eq-def-USP}
we work on the proof of Proposition \ref{prop-block-int-final} below. We need still a few more definitions. On the one hand, given
 $\omega \in \Omega$ and $x\in\mathbb{Z}^d$ we say that $o$ and $x$ are doubly connected under $\omega$ if there
exist two $\omega$-open paths from $o$ to $x$ made of disjoint edges, and consider
\begin{equation}
  \mathcal{C}^2_o \left( \omega \right) =\{x \in \mathbb{Z}^d : x \mbox{ is
  doubly connected to } o \mbox{ under } \omega\}.
\end{equation}
On the second hand we describe the typical $J$-structure in order to permit local surgery on $\omega$ later
on. Given a rectangular parallelepiped $\Lambda \subset \mathbb{Z}^d$ that is $L$-admissible,
we generalize the notation $B^L_i$ defining
\begin{equation}
  B^{L, n}_i = \left( L i +\{- n L + 1, \ldots, (n + 1) L - 1\}^d \right) \cap
  \Lambda \label{eq-def-BLin}
\end{equation}
for $n \in \mathbb{N}$; note that $\BrLi = B_i^{L, 0}$ if $i \in I_{\Lambda,L}$ (see \eqref{eq-def-ILL}).
Given $J\in\mathcal{J}$ and $e\in E(\mathbb{Z}^d)$, we say that $e$ is $J$-open if $J_e>0$.
For all $i \in I_{\Lambda,L}$ we denote
\begin{equation}
  \mathcal{G}_i^L = \left\{ \begin{array}{l}
    \mbox{There exists a unique $J$-open cluster in}\\
    E^w (B^{L, 1}_i) \mbox{ of diameter larger or equal to $L$}\\
    \mbox{and } \forall e \in E^w (B_i^{L, 3}), J_e = 0 \mbox{ or } J_e
    \geqslant \varepsilon_L
  \end{array} \right\} \label{eq-def-G}
\end{equation}
where $\varepsilon_L > 0$ is a cutoff that satisfies $\mathbb{P} \left( 0 <
J_e < \varepsilon_L \right) \leqslant e^{- L}$. 
Given a finite rectangular parallelepiped $R \subset \mathbb{Z}^d$ and $I \subset \mathbb{Z}^d$
we say that $I$ presents an horizontal interface in $R$ if there exists
no $\ast$-connected path $c_1, \ldots, c_n$ (i.e.~$\|c_{i+1}-c_i\|_\infty=1$, $\forall i=1\ldots n-1$) in  $R \setminus I$
with $c_1 \cdot \mathbf{e}_d = \min_{x \in R} x \cdot \mathbf{e}_d$ and $c_n \cdot \mathbf{e}_d = \max_{x \in R} x \cdot \mathbf{e}_d$.
We consider at last the event
  \begin{equation}
    \mathcal{L}= \left\{ (J, \omega) : \begin{array}{l}
      \mbox{There exists an horizontal interface } \mathcal{I} \mbox{ in}\\
        \{0, \ldots, n - 1\}^{d - 1} \times \left\{ 1, \ldots, \lceil \log n \rceil - 1 \right\} \mbox{ such }\\
      \mbox{that: }\forall i \in \mathcal{I}\mbox{, }\mathcal{C}^2_o \left( \omega
      \right) \cap B^L_i \neq \emptyset \mbox{ and } J\in\mathcal{G}^L_i
    \end{array} \right\} \label{eq-def-L}
  \end{equation}
and claim:
\begin{proposition}
  \label{prop-block-int-final} \eqref{eq-def-SP3} implies the existence of $L \in \mathbb{N}^{\star}$ such that
  \begin{equation*} \liminf_{n \rightarrow \infty} \mathbb{E} \inf_{\pi} \Phi_{\Lambda_{n,
     L}^{\log}}^{J, \pi} \left( \mathcal{L} \right) > 0. \end{equation*}
\end{proposition}
The proof of this Proposition is not straightforward and we achieve first several intermediary estimates under the product measure 
$\Psi^{L / 3}_{\Lambda^{\log}_{n,L}}$.
\subsubsection{Double connections}

\label{sec-dbl-cnx}The event $\mathcal{E}^L_i$ introduced in the former
Section efficiently describes connections between sub-blocks in the domain~$\Lambda$. However, as it will appear in the proof of Proposition
\ref{prop-unif-conn}, the information provided by $\mathcal{E}^L_i$ is not
enough to be able to proceed to local modifications on $\omega$ in a {\emph{recognizable}} way
and this is the motivation for introducing the notion of double connections.
Assuming that~$L$ is a multiple of~$3$, that $\Lambda \subset \mathbb{Z}^d$
is a $L$-admissible domain and that $i \in I_{\Lambda, L}$ we define
\begin{equation}
  \mathcal{D}^L_i = \bigcap_{j \in 3 i +\{0, 1, 2\}^d} \mathcal{E}_j^{L / 3}
  \label{eq-def-D} .
\end{equation}
Note that $\mathcal{D}^L_i$ depends on $\omega$ in $E^L_i$, a box of side-length $L$, while the measure $\Psi^{L / 3}_{\Lambda}$ associated to the $\mathcal{E}_j^{L / 3}$ has a decorrelation length $L/3$.

\begin{figure}[!h]
\begin{center}
 \includegraphics[width=7cm]{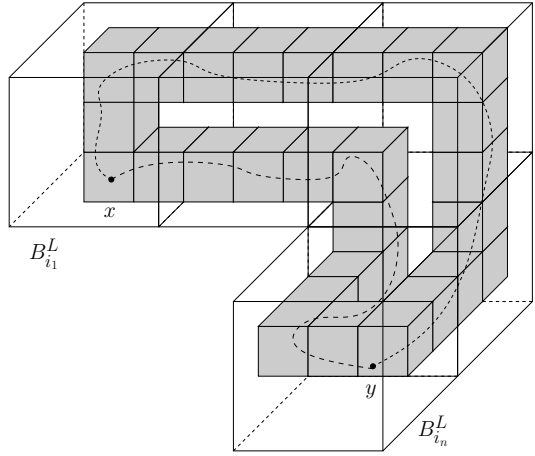}
\end{center}
  \caption{\label{fig-dbl-cnx-ens} A loop made of good $\mathcal{E}^{L / 3}$-blocks in a path of good $\mathcal{D}^L$-blocks.}
\end{figure}

An immediate consequence of Theorem~\ref{thm-meas-Psi} is the following fact:
\begin{lemma}
  \label{lemma-probaD}Assumption \eqref{eq-def-SP3} implies:
  \begin{equation*} \lim_{L \rightarrow \infty, 3| L} \inf_{\tmscript{\begin{array}{c}
       \Lambda \subset \mathbb{Z}^d L \mbox{-admissible}\\
       i \in I_{\Lambda, L}
     \end{array}}} \Psi^{L / 3}_{\Lambda} \left( \mathcal{D}^L_i \right) = 1.
  \end{equation*}
  Moreover, the event $\mathcal{D}^L_i$ depends only on $\omega_{|E^L_i}$. For
  any $i, j \in I_{\Lambda, L}$ with $\|i - j\|_2 > 1$ the events
  $\mathcal{D}^L_i$ and $\mathcal{D}^L_j$ are independent under $\Psi^{L /
  3}_{\Lambda}$.
\end{lemma}
The relation between $\mathcal{D}^L_i$ and the notion of double connections appears below:
\begin{lemma}
  \label{lemma-D-dbl-cnx}If $(i_1, \ldots, i_n)$ is a path in $\mathbb{Z}^d$
  such that $\BrLik \subset \Lambda, \forall k = 1 \ldots n$ and if
  $\omega \in \mathcal{D}_{i_1}^L \cap \ldots \cap \mathcal{D}_{i_n}^L$, then
  there exist $x \in B^L_{i_1}$ and $y \in B^L_{i_n}$ and two $\omega$-open
  paths from $x$ to $y$ in $\bigcup_{k = 1}^n E^L_{i_k}$ made of distinct
  edges.
\end{lemma}
This fact is an immediate consequence of the properties of $\mathcal{E}_j^{L / 3}$, see Figure~\ref{fig-dbl-cnx-ens}. Note that the factor $3$ in $\mathcal{E}_j^{L / 3}$ is necessary as the $\omega$-open clusters described by $\mathcal{E}_j^{L / 3}$ may use the edges on the faces of $B_j^{L / 3}$.

\subsubsection{Local \texorpdfstring{$J$}{J}-structure}

\label{sec-struct-J}We describe now the typical
$J$-structure with the help of the event $\mathcal{G}_i^L$ (see \eqref{eq-def-G}).

\begin{lemma}
  \label{lemma-probaG}The event $\mathcal{G}_i^L$ depends only on $J_{|E^w
  (B^{L, 3}_i)}$, and \eqref{eq-def-SP3} implies
  \begin{equation*} \lim_{L \rightarrow \infty} \inf_{\tmscript{\begin{array}{c}
       \Lambda \subset \mathbb{Z}^d L \mbox{-admissible}\\
       i \in I_{\Lambda, L}
     \end{array}}} \Psi^{L / 3}_{\Lambda} \left( \mathcal{G}^L_i \right) = 1.
  \end{equation*}
\end{lemma}

\begin{pf}
  The domain of dependence of $\mathcal{G}_i^L$ is trivial. Concerning the
  estimate on its probability, we remark that the marginal on $J$ of $\Psi^{L /
  3}_{\Lambda}$ equals $\mathbb{P}$, while \eqref{eq-def-SP3} ensures that
  percolation in slabs holds for the variable $\mathbf{1}_{\{J_e > 0\}}$
  under $\mathbb{P}$. Hence the condition on the structure of the $J$-open
  clusters holds with a probability larger than $1 - e^{- c L}$ for some $c >
  0$ according to~\cite{N09}. The condition on the value of
  $J_e$ also has a very large probability thanks to the choice of
  $\varepsilon_L$: remark that $| E^w ( B_i^{L, 3} )|
  \leqslant d (7 L)^d$, hence
  \begin{equation*} \mathbb{P} \left( \exists e \in E^w \left( B_i^{L, 3} \right) : J_e \in
     (0, \varepsilon_L) \right) \leqslant d (7 L)^d e^{- L} \end{equation*}
  which goes to $0$ as $L \rightarrow \infty$.
\end{pf}

\subsubsection{Typical structure in logarithmic slabs}

\label{sec-omega-stcl}We proceed now with Peierls estimates in order to infer
some controls on the global structure of $(J, \omega)$ in slabs of logarithmic
height. We define
\begin{equation*} \mathcal{T}_i^L =\mathcal{D}^L_i \cap \mathcal{G}^L_i \end{equation*}
where $\mathcal{D}^L_i$ and $\mathcal{G}^L_i$ are the events defined at
(\ref{eq-def-D}) and (\ref{eq-def-G}) (see also (\ref{eq-def-calEL}) for the
definition of $\mathcal{E}^L_i$). An immediate consequence of Lemmas~\ref{lemma-probaD} and
\ref{lemma-probaG} is that
\begin{equation*} \lim_{L \rightarrow \infty, 3| L} \inf_{\tmscript{\begin{array}{c}
     \Lambda \subset \mathbb{Z}^d L \mbox{-admissible}\\
     i \in I_{\Lambda, L}
   \end{array}}} \Psi^{L / 3}_{\Lambda} \left( \mathcal{T}^L_i \right) = 1 \end{equation*}
if \eqref{eq-def-SP3}, together with the independence of
$\mathcal{T}^L_i$ and $\mathcal{T}^L_j$ under $\Psi^{L / 3}_{\Lambda}$ if \linebreak $\|i
- j\|_{\infty} \geqslant 7$.
We recall the notation $\Lambda_{n, L}^{\log} =\{1, Ln - 1\}^{d - 1} \times
\{1, L \lceil \log n \rceil - 1\}$  (\ref{eq-def-LlognL}) and claim:
\begin{lemma}
  \label{lemma-block-int}Assume \eqref{eq-def-SP3}. For any $\varepsilon > 0$, $L$ large enough multiple of $3$,
  \begin{equation*} \liminf_n \Psi_{\Lambda_{n, L}^{\log}}^{L / 3} \left( \begin{array}{l}
       \mbox{The cluster of $\mathcal{T}_i^L$-good blocks issued}\\
       \mbox{from $0$ presents an horizontal interface}\\
       \mbox{in } \{0, n - 1\}^d \times \{1, \lceil \log n \rceil - 1\}
     \end{array} \right) \geqslant 1 - \varepsilon . \end{equation*}
\end{lemma}

Remark that the cluster of $\mathcal{T}_i^L$-good blocks issued
  from $0$ lives in $\{0, n - 1\}^d \times \{\mathbf{0}, \lceil \log n
  \rceil - 1\}$, hence we require here (as in the definition of $\mathcal{L}$ at (\ref{eq-def-L})) that the interface does not use the
  first layer of blocks. This is done in prevision for the proof of Lemma~\ref{lemma-mod-fpb}.

\begin{pf}
  The proof is made of two Peierls estimates. A first estimate that we do not expand here permits to prove that some $\left( \mathcal{T}_i^L
  \right)$-cluster forms an horizontal interface with large probability in the
  desired region $\{0, \ldots, n - 1\}^{d - 1} \times \{1, \ldots, \lceil \log n
  \rceil - 1\}$ if $L$ is large enough. The second estimate concerns
  the probability that there exists a $\mathcal{T}$-open path from $0$ to the
  top of the region.
  
  If the $\mathcal{T}$-cluster issued from $0$ does
  not touch the top of the region, there exists a $\ast$-connected, self
  avoiding path of $\mathcal{T}$-closed sites in the vertical section $\left\{
  0, \ldots, n - 1 \right\} \times \{0\}^{d - 2} \times \left\{ 0, \ldots,
  \lceil \log n \rceil - 1 \right\}$ separating $0$ from the top of the
  region. We call this event $\mathcal{C}$ and enumerate the possible paths
  according to their first coordinate on the left side $h \geqslant 0$ and
  their length $l \geqslant h + 1$: there are not more than $7^l$ such paths.
  On the other hand, in any path of length $l$ we can select at least $\lceil
  l / 13^2 \rceil$ positions at $\|.\|_{\infty}$-distance at least $7$ from
  any other. As the corresponding $\mathcal{T}$-events are independent under $\Psi_{\Lambda_{n, L}^{\log}}^{L / 3}$,
  \begin{equation*} \Psi_{\Lambda_{n, L}^{\log}}^{L / 3} \left( \mathcal{C} \right) \leqslant
     \sum_{h \geqslant 0} \sum_{l \geqslant h + 1} 7^l  \left( 1 - \rho_L
     \right)^{l / 13^2} \end{equation*}
  where $\rho_L = \inf_{i \in I_{\Lambda, L}} \Psi^{L / 3}_{\Lambda} \left(
  \mathcal{T}^L_i \right)$. This is not larger than $a_L / (1 - a_L)^2$ if
  $a_L = 7 \left( 1 - \rho_L \right)^{1 / 13^2} < 1$, and since $\lim_L a_L =
  0$ the claim follows.
\end{pf}

\subsubsection{Proof of Proposition \ref{prop-block-int-final}}
We conclude these intermediary estimates with the proof of Proposition \ref{prop-block-int-final}. 
\begin{pf}(Proposition \ref{prop-block-int-final}).
  As the event $\mathcal{L}$ is increasing in $\omega$, thanks to Proposition~\ref{prop-comp-averaged-prod} it is enough to estimate its probability under
  the product measure $\Psi^{L / 3}_{\Lambda_{n, L}^{\log}}$. We consider the
  following events on $(J, \omega)$:
  \begin{align*}
    A & = \left\{ \mbox{\begin{tabular}{l}
      $(J, \omega)$: there exists a modification of $(J, \omega)$ on $E^w (
      \BrL0)$\\
      such that the $\mathcal{T}$-cluster issued from $0$ forms an\\
      horizontal interface in $\{0, \ldots, n - 1\}^{d - 1} \times \left\{ 1,
      \ldots, \lceil \log n \rceil - 1 \right\}$
    \end{tabular}} \right\} \\
    B & = \left\{ \forall e \in E^w ( \BrL0), J_e
    \geqslant \varepsilon_L \mbox{ and } \omega_e = 1 \right\} .
  \end{align*}
  By a modification of $(J, \omega)$ on $E^w ( \BrL0)$ we mean a
  configuration $(J', \omega') \in \mathcal{J} \times \Omega$ that coincides
  with $(J, \omega)$ outside $E^w ( \BrL0)$. Clearly, the event $A$ does
  not depend on $(J, \omega)_{|E^w ( \BrL0)}$, whereas $B$ depends
  uniquely on $(J, \omega)_{|E^w ( \BrL0)}$. According to the product
  structure of $\Psi^{L / 3}_{\Lambda^{\log}_{n, L}}$, we have
  \begin{equation*} \Psi_{\Lambda_{n, L}^{\log}}^{L / 3} \left( A \cap B \right) =
     \Psi_{\Lambda_{n, L}^{\log}}^{L / 3} \left( A \right) \times \Psi_{
     \BrL0}^{L / 3} \left( B \right) . \end{equation*}
  In view of Lemma~\ref{lemma-block-int}, $\liminf_n \Psi_{\Lambda_{n,
  L}^{\log}}^{L / 3} \left( A \right) \geqslant 1 / 2$ for $L$ large enough
  multiple of $3$, whereas $\Psi_{\BrL0}^{L / 3} \left( B \right) > 0$
  for any $L$ large enough (we just need $\mathbb{P} \left( J_e \geqslant
  \varepsilon_L \right) \geqslant \mathbb{P} \left( J_e > 0 \right) - e^{- L}
  > 0$). This proves that $\liminf_n \Psi_{\Lambda_{n, L}^{\log}}^{L / 3}
  \left( A \cap B \right) > 0$ for $L$ large enough. We prove at last that $A
  \cap B$ is a subset of $\mathcal{L}$ and consider $(J, \omega) \in A \cap
  B$. From the definition of $A$ we know that there exists a modification
  $(J', \omega')$ of $(J, \omega)$ on $E^w ( \BrL0)$ such that the
  $\mathcal{T}$-cluster for $(J', \omega')$ issued from $0$ forms an
  horizontal interface in $\{0, \ldots, n - 1\}^{d - 1} \times \left\{ 1,
  \ldots, \lceil \log n \rceil - 1 \right\}$. Let us call $\mathcal{C}$ that
  $\mathcal{T}$-cluster and
  \begin{equation*} \mathcal{I}=\mathcal{C} \cap \{0, \ldots, n - 1\}^{d - 1} \times \left\{
     1, \ldots, \lceil \log n \rceil - 1 \right\} . \end{equation*}
  From its definition it is clear that $\mathcal{I}$ contains an horizontal
  interface in $\{0, \ldots, n - 1\}^{d - 1} \times \left\{ 1, \ldots, \lceil
  \log n \rceil - 1 \right\}$; we must check now that $\forall i \in
  \mathcal{I}, \mathcal{C}^2_o \left( \omega \right) \cap B^L_i \neq
  \emptyset$ and $J \in \mathcal{G}^L_i$. We begin with the proof that
  $\mathcal{C}^2_o \left( \omega \right) \cap B^L_i \neq \emptyset$, for every
  $i \in \mathcal{I}$: since $i \in \mathcal{C}$, Lemma~\ref{lemma-D-dbl-cnx}
  tells us that there exist $x \in \BrL0$ and $y \in \BrLi$ which
  are doubly connected under $\omega'$. Since the corresponding paths enter at
  distinct positions in $E^w ( \BrL0)$, $y$ is also doubly connected to
  $o$ under $\omega$ which has all edges open in $E^w ( \BrL0)$. As for
  the $J$-structure, for every $i \in \mathcal{I}$ we have $(J', \omega') \in
  \mathcal{T}^L_i$, hence $J' \in \mathcal{G}^L_i$ and $J \in \mathcal{G}^L_i$
  for every $i \in \mathcal{I}$ such that $\BrL0 \cap B^{L, 1}_i = \emptyset$. We
  conclude with the remark that the replacement of $J'$ by $J$ in $E^w (
  \BrL0)$ just enlarges an already large $J'$-cluster (no new large
  cluster is created, hence $J' \in \mathcal{G}^L_i \Rightarrow J \in
  \mathcal{G}^L_i$): the inclusion $(J', \omega') \in \mathcal{T}^L_0$ implies
  the existence of a $\omega'$-open path of length $L$ in $E^w ( \BrL0)$,
  and this path is necessarily also $J'$-open, hence $J \in \mathcal{G}^L_i$ for all $i\in\mathcal{I}$
such that $\BrL0 \cap B^{L, 1}_i \neq \emptyset$, and this ends the proof that $A
  \cap B$ is a subset of $\mathcal{L}$.
\end{pf}

\subsection{First pivotal bond and local modifications}

\label{sec-fpb}We introduce here the notion of first pivotal
bond: given a configuration $\omega \in \Omega$, we call
  $\mathcal{C}^2_x (\omega)$ the set of points doubly connected to $x$ under
  $\omega$. Given $e \in E(\mathbb{Z}^d) $ we say that $e$ is a \emph{pivotal bond}
  between $x$ and $y$ under $\omega$ if $x \stackrel{\omega}{\leftrightarrow}
  y$ in $\omega$ and $x \stackrel{\omega_{|\{e\}^c}}{\nleftrightarrow} y$. At
  last we say that $e$ is the \emph{first pivotal bond} from $x$ to $y$ under
  $\omega$ if it is a pivotal bond between $x$ and $y$ under $\omega$ and if
  it touches~$\mathcal{C}^2_x \left( \omega \right)$.

There does not always exist a first pivotal bond between two connected
points: it requires in particular the existence of a pivotal bond between these two points.
When a first pivotal bond from $x$ to $y$ exists, it is unique.
Indeed, assume by contradiction that $e \neq e'$ are pivotal bonds under
$\omega$ between $x$ and $y$ and that both of them touch $\mathcal{C}^2_x
\left( \omega \right)$. If $c$ is an $\omega$-open path from $x$ to $y$, it
must contain both $e$ and $e'$. Assume that $c$ passes through $e$ before
passing through $e'$, then removing $e$ in $\omega$ we do not disconnect $x$
from $y$ since $e'$ touches $\mathcal{C}^2_x \left( \omega \right) \supset
\mathcal{C}_x \left( \omega_{|\{e\}^c} \right)$, and this contradicts the
assumption that $e$ is a pivotal bond.

In the following geometrical Lemma we relate the event $\mathcal{L}$ defined
at (\ref{eq-def-L}) to the notion of first pivotal bond. We recall the
notations $\Bottom(\Lambda^{\log}_{n, L}) =\{1, \ldots, L n - 1\}^{d -
1} \times \{0\}$ and $\Top(\Lambda^{\log}_{n, L}) =\{1, \ldots, L n -
1\}^{d - 1} \times \{L \lceil \log n \rceil\}$, as well as $E^-$ for the set
of edges in the discrete lower half space $\mathbb{H}^-$ (see
(\ref{eq-def-Hminus})). We say that $\omega \in \Omega_E$ is compatible with
$J \in \mathcal{J}_E$ if, for every $e \in E, J_e = 0 \Rightarrow \omega_e =
0$.

\begin{lemma}
  \label{lemma-mod-fpb}Consider $x \in \Bottom(\Lambda^{\log}_{n, L})$,
  $\xi \in \Omega_{E^-}$ and $(J, \omega) \in \mathcal{L}$ with $\omega$ such
  that
  \begin{equation*} \mbox{$x \stackrel{\omega \vee \xi}{\leftrightarrow} \Top(\Lambda^{\log}_{n, L}) \mbox{ and } x \stackrel{\omega \vee
     \xi}{\nleftrightarrow} o$} . \end{equation*}
  Then, there exists $i \in \{0, \ldots, n - 1\}^{d - 1} \times \{1, \ldots,
  \lceil \log n \rceil - 1\}$ such that $J \in \mathcal{G}^L_i$ and there
  exists a modification $\omega'$ of $\omega$ on $E^w (B^{L, 1}_i)$ compatible
  with $J$, such that the first pivotal bond from $o$ to $x$ under $\omega'
  \vee \xi$ exists and belongs to $E^w (B^{L, 1}_i)$.
\end{lemma}

The variable $\xi$ corresponds to the configuration below the slab
  $\Lambda^{\log}_{n, L}$. The point in introducing $\xi$ here (and in the formulation of \eqref{eq-def-USP}) is the need for an estimate that holds uniformly over the configuration below the slab in the proof of Lemma~\ref{lemma-proba-vert-cl}.

\begin{figure}[!h]
\begin{center}
\includegraphics[width=10cm]{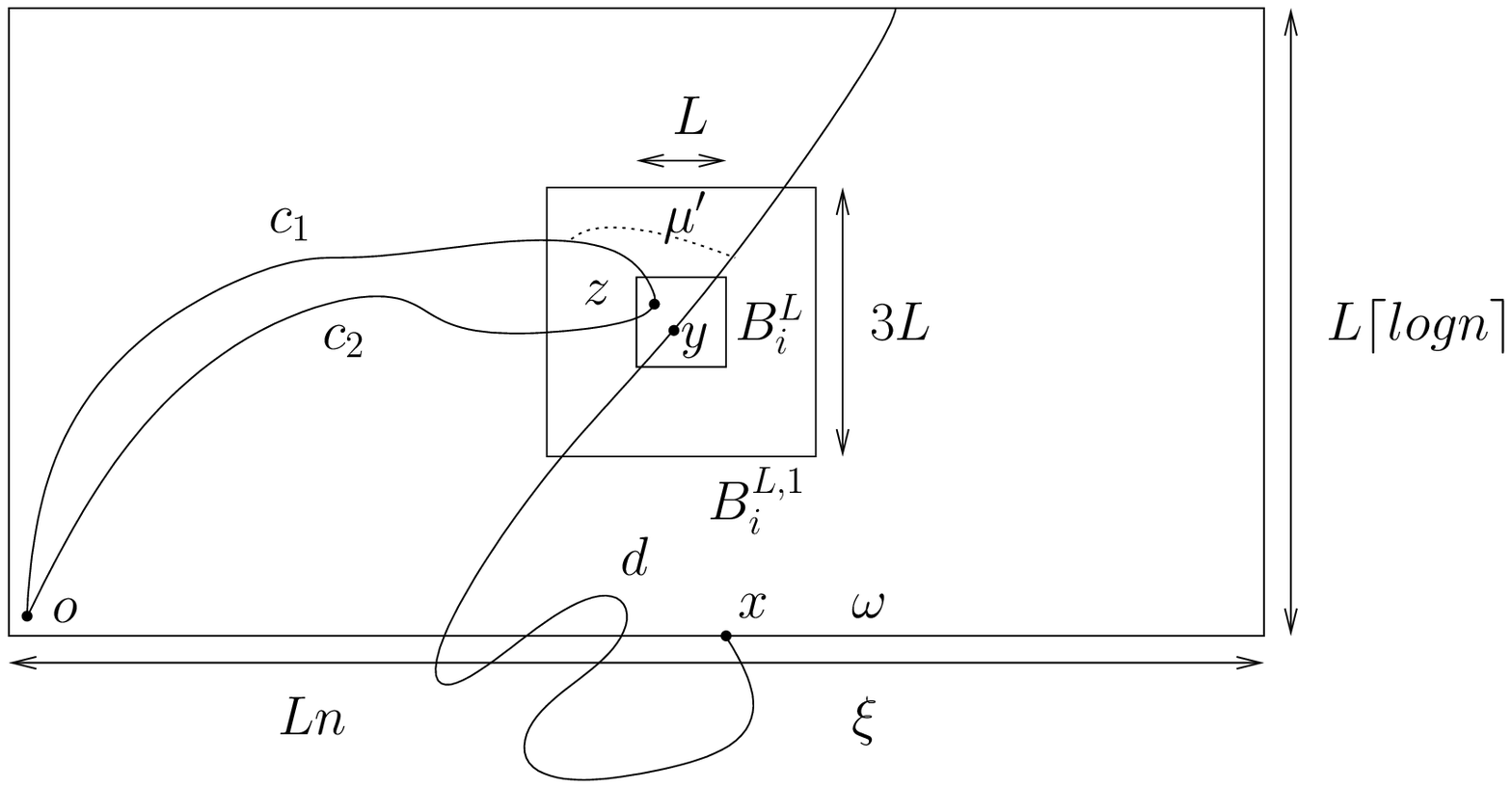}
\end{center}
  \caption{\label{fig-mod-fpb}Lemma~\ref{lemma-mod-fpb}: $c_1, c_2$ are $\omega$-open, $d$ is $\omega \vee \xi$-open and $\mu'$ is $J$-open.}
\end{figure}

\begin{pf}
  Note that Figure~\ref{fig-mod-fpb} provides an illustration for the
  objects considered in the proof. We build by hand the modification
  $\omega'$. Since $(J, \omega) \in \mathcal{L}$ there exists an horizontal
  interface $\mathcal{I}$ as in (\ref{eq-def-L}). Since on the other hand $x
  \stackrel{\omega \vee \xi}{\leftrightarrow} \Top(\Lambda^{\log}_{n,
  L})$, there exists $i \in \mathcal{I}$ such that $\mathcal{C}_x (\omega \vee
  \xi) \cap B^L_i \neq \emptyset$. Let us fix such an $i$: we clearly have $J
  \in \mathcal{G}^L_i$. From the definition of the event $\mathcal{L}$, we
  know that $\mathcal{C}^2_o (\omega) \cap B^L_i \neq \emptyset$. We fix $y
  \in \mathcal{C}_x (\omega \vee \xi) \cap B^L_i$ and $z \in \mathcal{C}^2_o(\omega)
  \cap B^L_i$. There exist two $\omega$-open paths $c_1, c_2$ in $E^w
  (\Lambda_{n, L}^{\log})$ made of disjoint edges, with no loop, that link $o$
  to $z$, as well as an $\omega \vee \xi$-open path $d$ in $E^w
  (\Lambda^{\log}_{n, L}) \cup E^-$, with no loop, that links $x$ to $y$. Of
  course, $d$ does not touch $c_1 \cup c_2$ since $x \nleftrightarrow o$ under
  $\omega \vee \xi$.
  
  Since $i$ is not in the first block layer (see the remark after Lemma
\ref{lemma-block-int}), $c_1 \cap E^w (B^{L, 1}_i)$ and $d \cap E^w (B^{L,
  1}_i)$ have a connected component of diameter larger or equal to $L$. Since
  $\mathbf{1}_{J_e > 0}$ is larger than $\omega$ these components are also
  $J$-open, and since $J \in \mathcal{G}^L_i$, this implies that there exists
  a $J$-open path $\mu$ in $E^w (B^{L, 1}_i)$, self-avoiding, joining $c_1$ to
  $d$. Noting $(\mu_t)_t$ the vertices of $\mu$, we call $v = \min \{t : \mu_t
  \cap d \neq \emptyset\}$, then $u = \max \{t \leqslant v : \mu_t \cap \{c_1
  \cup c_2 \} \neq \emptyset\}$, and $\mu'$ the portion of $\mu$ between $\mu_u$
  and $\mu_v$. Finally, we define the modified configuration as
  \begin{equation*} \omega'_e = \left\{ \begin{array}{ll}
       \omega_e & \mbox{if } e \notin E^w (B^{L, 1}_i)\\
       1 & \mbox{if } e \in E^w (B^{L, 1}_i) \cap \left\{ c_1 \cup c_2 \cup d
       \cup \mu' \right\}\\
       0 & \mbox{else}
     \end{array} \right. \end{equation*}
  and claim that $\{\mu_u, \mu_{u + 1} \}$ is the first pivotal bond from $o$ to
  $x$ under $\omega' \vee \xi$: first of all, there is actually a
  connection between $o$ and $x$ under $\omega' \vee \xi$ since $\mu'$ touches
  both $c_1 \cup c_2$ and $d$. Then, it is clear that $\mu_u$ is doubly
  connected to $o$, to prove this, if $\mu_u \in c_1$ for instance we just
  need to consider $c_1'$ the portion of $c_1$ from $o$ to $\mu_u$ and $c''_1$
  the rest of $c_1$; $c'_1$ is a path from $o$ to $x$, and a second path is
  made by $c''_1 \cup c_2$, which uses edges distinct from those of $c'_1$. At
  last, $\{\mu_u, \mu_{u + 1} \}$ is a pivotal bond between $o$ and $x$ (and
  more generally any edge of $\mu'$ is a pivotal bond) since $\mu'$ touches $c_1
  \cup c_2$ only at its first extremity.
\end{pf}

\subsection{The uniform estimate \texorpdfstring{\eqref{eq-def-USP}}{(USP)}}
\label{sec-USP}
We are now in a position to prove the uniform estimate \eqref{eq-def-USP} defined at the beginning of Section~\ref{sec-cg-unique}.
\begin{proposition}
  \label{prop-unif-conn} \eqref{eq-def-SP3} implies \eqref{eq-def-USP}.
\end{proposition}

\begin{pf}
  In view of Proposition~\ref{prop-block-int-final}, one can fix $L \in
  \mathbb{N}^{\star}$ and $\delta > 0$ such that
  \begin{equation*} \liminf_{n \rightarrow \infty} \mathbb{E} \inf_{\pi} \Phi_{\Lambda_{n,
     L}^{\log}}^{J, \pi} \left( (J, \omega) \in \mathcal{L} \right) \geqslant
     3 \delta . \end{equation*}
  According to Markov's inequality (\ref{eq-EPPhi}) we thus have
\begin{equation*}\mathbb{P}
  \left( \inf_{\pi} \Phi_{\Lambda^{\log}_{n, L}}^{J, \pi} \left( (J, \omega)
  \in \mathcal{L} \right) \geqslant \delta \right) \geqslant \delta\end{equation*} for any
  $n$ large enough. In the sequel we fix $J \in \mathcal{J}$ such that
  \begin{equation}
    \inf_{\pi} \Phi_{\Lambda^{\log}_{n, L}}^{J, \pi} \left( (J, \omega) \in
    \mathcal{L} \right) \geqslant \delta \label{eq-hypJL} .
  \end{equation}
  Consider $x \in \Bottom(\Lambda^{\log}_{n, L})$, $\pi \in \Omega_{E^w
  (\Lambda^{\log}_{n, L})^c}$ and $\xi \in \Omega_{E^-}$. One of the following
  cases must occur:
  \begin{enumerate}
    \item $\Phi_{\Lambda^{\log}_{n, L}}^{J, \pi} \left( x \stackrel{\omega \vee
    \xi}{\leftrightarrow} o \right) \geqslant \delta / 3$
    
    \item or $\Phi_{\Lambda^{\log}_{n, L}}^{J, \pi} \left( x \stackrel{\omega
    \vee \xi}{\nleftrightarrow} \Top(\Lambda^{\log}_{n, L}) \right)
    \geqslant \delta / 3$
    
    \item or $\Phi_{\Lambda^{\log}_{n, L}}^{J, \pi} \left( x \stackrel{\omega
    \vee \xi}{\nleftrightarrow} o \mbox{ and } x \stackrel{\omega \vee
    \xi}{\leftrightarrow} \Top(\Lambda^{\log}_{n, L}) \right) \geqslant
    1 - 2 \delta / 3$.
  \end{enumerate}
  The first two cases lead directly to the estimate 
\begin{equation*}\Phi_{\Lambda^{\log}_{n,
  L}}^{J, \pi} \left( x \stackrel{\omega \vee \xi}{\leftrightarrow} o \mbox{ or
  } x \stackrel{\omega \vee \xi}{\nleftrightarrow} \Top(\Lambda^{\log}_{n, L}) \right) \geqslant \delta / 3. \end{equation*} We focus hence on the
  third case. We let
  \begin{equation}
    \mathcal{L}_x = \left\{ \omega \in \Omega_{E^w (\Lambda^{\log}_{n, L})} :
    (J, \omega) \in \mathcal{L}, x \stackrel{\omega \vee \xi}{\nleftrightarrow}
    o \mbox{ and } x \stackrel{\omega \vee \xi}{\leftrightarrow} \Top(\Lambda^{\log}_{n, L}) \right\}, \label{eq-def-Lx}
  \end{equation}
  it follows from (iii) and (\ref{eq-hypJL}) that $\Phi_{E^w
  (\Lambda^{\log}_{n, L})}^{J, \pi} \left( \mathcal{L}_x \right) \geqslant
  \delta / 3$. Then, for $\omega \in \mathcal{L}_x$ we define the set of
  could-be first pivotal bond:
  \begin{equation*} F_x (\omega) = \left\{
     \mbox{\begin{tabular}{l}
        $e \in \mathcal{E}(\Lambda^{\log}_{n, L})$ : $\exists i \in \{0, \ldots, n - 1\}^{d - 1} \times \{1, \ldots, \lceil \log n \rceil - 1\}$ \\
with $J \in \mathcal{G}^L_i$ and a modification $\tilde{\omega}$ of $\omega$ on $E^w (B_i^{L, 1})$ \\
       compatible with $J$ such that $e \in E^w (B_i^{L, 1})$ is the \\
       first pivotal bond from $o$ to $x$ under $\tilde{\omega} \vee \xi$
     \end{tabular}} \right\} \end{equation*}
  where $\mathcal{E}(\Lambda^{\log}_{n, L}) = \bigcup_{i \in
  I_{\Lambda^{\log}_{n, L}, L}} \ErLi$. Lemma~\ref{lemma-mod-fpb} states
  that $F_x (\omega)$ is not empty whenever $\omega \in \mathcal{L}_x$. Hence,
  for all $\omega \in \mathcal{L}_x$ we can consider the edge $f_x (\omega) =
  \min F_x (\omega)$, where $\min$ refers to the lexicographical ordering of
  $\mathcal{E}(\Lambda^{\log}_{n, L})$. Given $e \in
  \mathcal{E}(\Lambda^{\log}_{n, L})$ we denote by $i (e)$ the unique index $i
  \in \mathbb{Z}^d$ such that $e \in \ErLi$. We prove now the existence
  of $c_L > 0$ such that
  \begin{equation}
    \forall \omega \in \mathcal{L}_x \cap \{\omega : f_x (\omega) = e\}, \
    \Phi_{E_i}^{J, \pi \vee \omega_{|E_i^c}} \left( \begin{array}{l}
      e \mbox{ first pivotal bond from $o$}\\
      \mbox{to } x \mbox{ under } \omega_{|E_i^c} \vee \omega' \vee \xi
    \end{array} \right) \geqslant c_L . \label{ineq-fpe}
  \end{equation}
  where $\omega'$ is the variable associated to $\Phi_{E_i}^{J, \pi \vee
  \omega_{|E_i^c}}$ and $E_i = E^w (B_{i (e)}^{L, 2})$. Let $\omega \in
  \mathcal{L}_x \cap \{\omega : f_x (\omega) = e\}$. According to the
  definition of $f_x$, there exists $i$ such that $J \in \mathcal{G}^L_i$, $e
  \in E^w (B^{L, 1}_i)$ and there exists a local modification $\tilde{\omega}$
  of $\omega$ on $E^w (B^{L, 1}_i)$, compatible with $J$ such that $e$ is the
  first pivotal bond from $o$ to $x$ under $\tilde{\omega} \vee \xi$. From the
  inclusion $E^w (B^{L, 1}_i) \subset E^w (B^{L, 2}_{i (e)})$ we deduce that
  $\tilde{\omega}$ is a modification of $\omega$ on the block $E^w(B^{L, 2}_{i (e)})$ that does not depend on $i$. On the other hand, $E^w (B^{L, 2}_{i (e)}) \subset E^w
  (B^{L, 3}_i)$ and in view of the definition of $\mathcal{G}^L_i$
  (\ref{eq-def-G}) this implies that for all $e \in E^w (B^{L, 2}_{i (e)})$,
  $J_e = 0 \mbox{ or } J_e \geqslant \varepsilon_L$. From the DLR equation (\ref{eq-DLR-Phi}) it follows that
  \begin{equation*}
    \Phi_{E_i}^{J, \pi \vee \omega_{|E_i^c}} (\{ \tilde{\omega} \}) 
    \geqslant \prod_{e \in E_i} \inf_{\pi} \Phi_{\{e\}}^{J, \pi} \left(
    \omega_e = \tilde{\omega}_e \right) 
  \end{equation*}
  and remarking that
  \begin{equation*} \forall J_e \in [0, 1] \mbox{, \ \ } \Phi_{\{e\}}^{J, \pi} \left(
     \omega_e = 0 \right) \geqslant \Phi_{\{e\}}^{J, w} \left( \omega_e = 0
     \right) = 1 - p (J_e) \geqslant 1 - p (1) > 0 \end{equation*}
  and
  \begin{equation*} \forall J_e \in [\varepsilon, 1] \mbox{, \ \ } \Phi_{\{e\}}^{J, \pi}
     \left( \omega_e = 1 \right) \geqslant \Phi_{\{e\}}^{J, f} \left( \omega_e
     = 1 \right) = \tilde{p} (J_e) \geqslant \frac{p (\varepsilon)}{p
     (\varepsilon) + q (1 - p (\varepsilon))} > 0 \end{equation*}
  thanks to the assumptions on $p$ stated before (\ref{eq-def-Phi}), we
  conclude that (\ref{ineq-fpe}) holds with
  \begin{equation*} c_L = \left[ \min \left( 1 - p (1), \frac{p (\varepsilon)}{p
     (\varepsilon) + q (1 - p (\varepsilon))} \right) \right]^{|E_i |} > 0. \end{equation*}
  Combining the DLR equation for $\Phi^J$ (\ref{eq-DLR-Phi}) with
  (\ref{ineq-fpe}), we obtain
  \begin{align}
    \lefteqn{\Phi_{\Lambda^{\log}_{n, L}}^{J, \pi} \left( \begin{array}{l}
      e \mbox{ first pivotal bond from}\\
      o \mbox{ to } x \mbox{ under } \omega \vee \xi
    \end{array} \right)} \notag \\ & =  \Phi_{\Lambda^{\log}_{n, L}}^{J, \pi} \left[
    \Phi_{E_i}^{J, \pi \vee \omega_{|E_i^c}} \left( \begin{array}{l}
      e \mbox{ first pivotal bond from $o$}\\
      \mbox{to } x \mbox{ under } \omega_{|E_i^c} \vee \omega' \vee \xi
    \end{array} \right) \right] \notag\\
    & \geqslant  c_L \Phi_{\Lambda^{\log}_{n, L}}^{J, \pi} \left(
    \mathcal{L}_x \cap \{f_x (\omega) = e\} \right) . 
  \end{align}
  If we now sum over $e \in \mathcal{E}(\Lambda^{\log}_{n, L})$ -- the events
  in the left-hand term are \emph{disjoint} for distinct edges $e$, and
  all included in $\{o \stackrel{\omega \vee \xi}{\leftrightarrow} x\}$ -- we
  obtain
  \begin{equation*}
    \Phi_{\Lambda^{\log}_{n, L}}^{J, \pi} \left( o \stackrel{\omega \vee
    \xi}{\leftrightarrow} x \right) \geqslant c_L \Phi_{\Lambda^{\log}_{n,
    L}}^{J, \pi} \left( \mathcal{L}_x \right)
  \end{equation*}
  which is larger than $c_L \delta / 3$ as seen after (\ref{eq-def-Lx}). To
  sum it up, under the assumption (\ref{eq-hypJL}) which holds with a
  $\mathbb{P}$-probability not smaller than $\delta$, we have shown that
  \begin{equation*} \Phi_{\Lambda^{\log}_{n, L}}^{J, \pi} \left( x \stackrel{\omega \vee
     \xi}{\leftrightarrow} o \mbox{ or } x \stackrel{\omega \vee
     \xi}{\nleftrightarrow} \Top(\Lambda^{\log}_{n, L}) \right)
     \geqslant \min (\delta / 3, c_L \delta / 3) \end{equation*}
  and the proof is over.
\end{pf}

\subsection{An intermediate coarse graining}

\label{sec-weak-cg}The strength of the criterion \eqref{eq-def-USP} lies in
the fact that it provides an estimate on the $\Phi^{J, \pi}$ connection
probabilities that is uniform over $x$, $\pi$ and $\xi$. This is a very strong
improvement compared to the original assumption of percolation in slabs
\eqref{eq-def-SP3}.

In this Section, we establish an intermediate formulation of the coarse
graining. We begin with an estimate on the probability of having two long
vertical and disjoint $\omega$-clusters in the domain
\begin{equation}
  \Lambda^{1 / 4}_N =\{1, N - 1\}^{d - 1} \times \{1, [N / 4] - 1\}.
\end{equation}
\begin{lemma}
  \label{lemma-proba-vert-cl}Assume \eqref{eq-def-SP3}. There exist $L \in \mathbb{N}^{\star}$ and $c > 0$ such
  that, for any $N \in \mathbb{N}^{\star}$ large enough multiple of $L$ and
  any $x, y \in \Bottom(\Lambda^{1 /  4}_N)$:
  \begin{equation}
    \mathbb{E} \inf_{\pi} \Phi^{J, \pi}_{\Lambda^{1 / 4}_N} \left( x
    \stackrel{\omega}{\leftrightarrow} \Top(\Lambda^{1 / 4}_N), y
    \stackrel{\omega}{\leftrightarrow} \Top(\Lambda^{1 / 4}_N) \mbox{
    and } x \stackrel{\omega}{\nleftrightarrow} y \right) \leqslant \exp \left(
    - c \frac{N}{\log N} \right) \label{eq-proba-vert-cl} .
  \end{equation}
\end{lemma}

\begin{pf}
  We fix some
  $L \in \mathbb{N}^{\star}$ and $\varepsilon > 0$ so that the uniform
  criterion \eqref{eq-def-USP} holds whenever $n = N / L$ is large enough. The
  domain $\Lambda^{1 / 4}_N$ contains all slabs
  \begin{equation*} S_h = \Lambda_{n, L}^{\log} + h L \lceil \log n \rceil \mathbf{e}_d, \ h
     \in \{0, \ldots, n / \left( 4 \lceil \log n \rceil \right) - 1\} \end{equation*}
  Given $J \in \mathcal{J}$, we say that $S_h$ is $J$-good if for all
$x \in \Bottom(S_h)$, $\pi \in \Omega_{E^w (S_h)^c}$ and $\xi \in \Omega_{E^-_h}$,
  \begin{equation}
    \Phi_{S_h}^{J, \pi} \left( x
    \stackrel{\omega \vee \xi}{\leftrightarrow} o_h \mbox{ or } x
    \stackrel{\omega \vee \xi}{\nleftrightarrow} \Top\left( S_h \right)
    \right) \geqslant \varepsilon \label{eq-def-Sh-good}
  \end{equation}
  where
  \begin{equation*} E^-_h = E^f \left( \mathbb{H}^- + h L \lceil \log n \rceil
     \mathbf{e}_d \right) \end{equation*}
  (cf.~(\ref{eq-def-Hminus})) and $o_h = o + h L \lceil \log n \rceil
  \mathbf{e}_d$. The event that $S_h$ is $J$-good depends only on $J_e$ for
  $e \in E^w (S_h)$, thus for distinct $h$ these events are independent. Since
  they all have the same probability larger than $\varepsilon$, Cram\'er's
  Theorem yields the existence of $c > 0$ such that
  \begin{equation}
    \mathbb{P} \left( \mbox{
\begin{tabular}{l}
There are at least $[ \varepsilon n/(8\log n)]$ \\
$J$-good slabs in $\Lambda^{1 / 4}_N$ \end{tabular} } \right) \geqslant
    1 - \exp \left( - c \frac{n}{\log n} \right) \label{eq-proba-Shgood}
  \end{equation}
  for any $n$ large enough.
  
  Let us denote $\kappa = \left[ \varepsilon n / \left( 8 \log n \right)
  \right]$ and fix $J \in \mathcal{J}$ such that there are at least $\kappa$
  $J$-good slabs. We denote by $h_1, \ldots, h_{\kappa}$ the positions (in
  increasing order) of the first $\kappa$ $J$-good slabs. Given some boundary
  condition $\pi$ and $x, y \in \Bottom(\Lambda^{1 / 4}_N)$, we pass to
  an inductive proof of the fact that, for all $k \in \{1, \ldots, \kappa\}$:
  \begin{equation}
    \Phi^{J, \pi}_{E^w (\Lambda_N^{1 / 4}) \cap E^-_{h_k + 1}} \left( x
    \stackrel{\omega}{\leftrightarrow} \Top(S_{h_k}), y
    \stackrel{\omega}{\leftrightarrow} \Top(S_{h_k}) \mbox{ and } x
    \stackrel{\omega}{\nleftrightarrow} y \right) \leqslant (1 - \varepsilon^2
    / 4)^k \label{eq-ind-hyp}.
  \end{equation}
  We assume that either $k = 1$ or that (\ref{eq-ind-hyp}) holds for $k - 1$
  and we let
  \begin{equation*} D_h = \left\{ \begin{array}{cl}
       \Omega & \mbox{if } h < h_1\\
       \left\{
\omega \in \Omega_{\Lambda_N^{1 / 4}} :  \begin{array}{l}
x \stackrel{\omega}{\leftrightarrow} \Top(S_h), y \stackrel{\omega}{\leftrightarrow} \Top(S_h) \\ 
\mbox{and } x \nleftrightarrow y \mbox{ under } \omega_{|E^-_{h + 1}}
\end{array} \right\} & \mbox{else.} \end{array} \right. \end{equation*}
  It is obvious that $D_h \subset D_{h - 1}$ for any $h \geqslant 1$. For any
  $k$ such that $h_k \geqslant 1$ and $\omega \in D_{h_k - 1}$, we define $x_k
  (\omega)$ as the first point (under the lexicographical order) of
  $\Bottom(S_{h_k}) = \Top\left( S_{h_k - 1} \right)$ connected
  to $x$ under $\omega_{|E^-_{h_k}}$ (respectively, $y_k (\omega)$ is the
  corresponding point for $y$) -- see Figure~\ref{fig-def-xyk} for an illustration. If $h_k = 0$ we let $x_k (\omega) = x$ and
  $y_k (\omega) = y$.

\begin{figure}[!h]
\begin{center}
 \includegraphics[width=10cm]{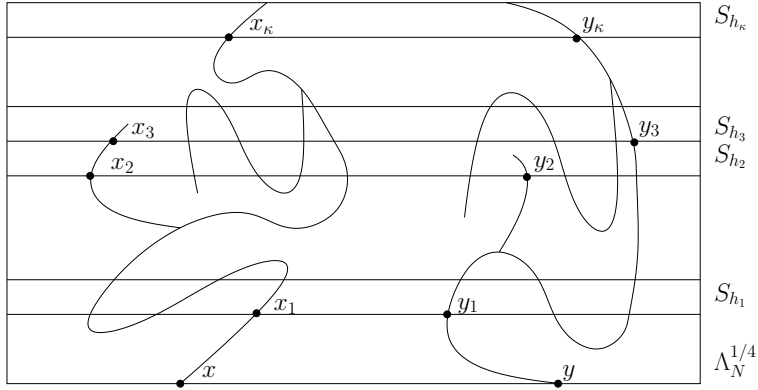}
\end{center}
  \caption{\label{fig-def-xyk} The $x_k$ and $y_k$ in Lemma~\ref{lemma-proba-vert-cl}.}
\end{figure}

Applying the DLR equation we get:
  \begin{equation}
    \Phi^{J, \pi}_{\Lambda^{1 / 4}_N} \left( D_{h_k} \right) = 
    \Phi^{J,\pi}_{\Lambda^{1 / 4}_N} \left( \mathbf{1}_{D_{h_k - 1}}
     \Phi^{J, \pi \vee \omega_{|E^w (S_{h_k})^c}}_{S_{h_k}} \left(
\begin{array}{l}
 x_k, y_k \stackrel{\omega_k \vee \xi}{\leftrightarrow} \Top(S_{h_k}) \\
\mbox{and } x_k \stackrel{\omega_k \vee \xi}{\nleftrightarrow} y_k
\end{array}
  \right) \right)
    \label{eq-DLR-D}
  \end{equation}
  where the variable $\omega$ (resp. $\omega_k$) corresponds to $\Phi^{J, \pi}_{\Lambda^{1 /
  4}_N}$ (resp. to $\Phi^{J, \pi \vee \omega_{|E^w
  (S_{h_k})^c}}_{S_{h_k}}$), $\xi = \xi (\omega) = \omega_{|E_{h_k}^-}$ is the
  restriction of $\omega$ to $E^-_{h_k}$ and $x_k$ and $y_k$ refer to $x_k
  (\omega)$ and $y_k (\omega)$. Here appears the reason for the
  introduction of $\xi$ in Lemma~\ref{lemma-mod-fpb} and in the definition of
  \eqref{eq-def-USP}: the cluster issued from $x$ under $\omega_k \vee \xi$ is
  the same as that issued from $x_k$ under $\omega_k \vee \xi$, while in
  general $x \stackrel{\omega_k \vee \xi}{\leftrightarrow} \Top(S_{h_k})$ does not imply $x_k \stackrel{\omega_k}{\leftrightarrow}
  \Top(S_{h_k})$.
  
  We use now the information that $S_{h_k}$ is a $J$-good slab. Given any $\pi
  \in \Omega_{E^w (S_{h_k})^c}$, $\xi \in \Omega_{E_{h_k}^-}$ and $z \in
  \Bottom(S_{h_k})$ we have, according to (\ref{eq-def-Sh-good}):
  \begin{equation}
    \Phi^{J, \pi}_{S_{h_k}} \left( z \stackrel{\omega_k \vee
    \xi}{\nleftrightarrow} \Top(S_{h_k}) \right) \geqslant
    \frac{\varepsilon}{2} \mbox{ \ or \ } \Phi^{J, \pi}_{S_{h_k}} \left( z
    \stackrel{\omega_k \vee \xi}{\leftrightarrow} o_{h_k} \right) \geqslant
    \frac{\varepsilon}{2} \label{eq-minzo-zT}
  \end{equation}
  Here we distinguish two cases. If 
\begin{equation*}\Phi^{J, \pi}_{S_{h_k}} \left( x_k
  \stackrel{\omega_k \vee \xi}{\nleftrightarrow} \Top(S_{h_k}) \right)
  \geqslant \frac{\varepsilon}{2} \mbox{ or } \Phi^{J, \pi}_{S_{h_k}} \left( y_k
  \stackrel{\omega_k \vee \xi}{\nleftrightarrow} \Top(S_{h_k}) \right)
  \geqslant \frac{\varepsilon}{2}\end{equation*} it is immediate that
  \begin{equation}
    \Phi^{J, \pi}_{S_{h_k}} \left( x_k \stackrel{\omega_k \vee
    \xi}{\leftrightarrow} \Top(S_{h_k}) \mbox{ and } y_k
    \stackrel{\omega_k \vee \xi}{\leftrightarrow} \Top(T_{h_k} \right)
    \leqslant 1 - \frac{\varepsilon}{2} . \label{eq-proba-xTyT}
  \end{equation}
  In the opposite case, (\ref{eq-minzo-zT}) implies that both 
\begin{equation*} \Phi^{J,
  \pi}_{S_{h_k}} (x_k \stackrel{\omega_k \vee \xi}{\leftrightarrow} o_{h_k})
  \geqslant \frac{\varepsilon}{2} \mbox{ \ and \ } \Phi^{J, \pi}_{S_{h_k}} (y_k
  \stackrel{\omega_k \vee \xi}{\leftrightarrow} o_{h_k}) \geqslant \frac{\varepsilon}{2} \end{equation*}
 and the FKG inequality tells us that
  \begin{equation}
    \Phi^{J, \pi}_{S_{h_k}} \left( x_k \stackrel{\omega_k \vee
    \xi}{\leftrightarrow} y_k \right) \geqslant \Phi^{J, \pi}_{S_{h_k}} \left(
    x_k \stackrel{\omega_k \vee \xi}{\leftrightarrow} o_{h_k} \right) \times
    \Phi^{J, \pi}_{S_{h_k}} \left( y_k \stackrel{\omega_k \vee
    \xi}{\leftrightarrow} o_{h_k} \right) \geqslant \frac{\varepsilon^2}{4}
    \label{eq-proba-xy} .
  \end{equation}
  Since either (\ref{eq-proba-xTyT}) or (\ref{eq-proba-xy}) occurs in a good
  slab, we see that
  \begin{equation*} \inf_{\pi \in \Omega_{E^w (S_{h_k})^c}} \inf_{\xi \in \Omega_{E_{h_k}^-}}
     \Phi^{J, \pi}_{S_{h_k}} \left( x_k, y_k \stackrel{\omega_k \vee
     \xi}{\leftrightarrow} \Top(S_{h_k}) \mbox{ and } x_k
     \stackrel{\omega_k \vee \xi}{\nleftrightarrow} y_k \right) \leqslant 1 -
     \frac{\varepsilon^2}{4} \end{equation*}
  and reporting in (\ref{eq-DLR-D}) we conclude that
  \begin{equation*} \Phi^{J, \pi}_{\Lambda^{1 / 4}_N} \left( D_{h_k} \right) \leqslant \left(
     1 - \frac{\varepsilon^2}{4} \right) \Phi^{J, \pi}_{\Lambda^{1 / 4}_N}
     \left( \mathbf{1}_{D_{h_k - 1}} \right), \end{equation*}
  which ends the induction step for the proof of (\ref{eq-ind-hyp}) as $D_{h_k - 1} \subset D_{h_{k - 1}}$.
 The proof of the Lemma follows
  combining (\ref{eq-proba-Shgood}) and (\ref{eq-ind-hyp}) with $k = \kappa =
  \left[ \varepsilon n / \left( 8 \log n \right) \right]$.
\end{pf}

We are now in a position to present a first version of the coarse graining:

\begin{proposition}
  \label{prop-weak-cg}Assume \eqref{eq-def-SP3}.
  Then for any $\varepsilon > 0$ there exists $N \in \mathbb{N}^{\star}$ such
  that
  \begin{equation}
    \mathbb{E} \inf_{\pi} \Phi^{J, \pi}_{\Lambda_N} \left( \begin{array}{l}
      \mbox{There exists a crossing cluster for $\omega$}\\
      \mbox{in $\Lambda_N$ and it is the only cluster of}\\
      \mbox{diameter larger or equal to N/4}
    \end{array} \right) \geqslant 1 - \varepsilon . \label{eq-weak-cg}
  \end{equation}
\end{proposition}

This estimate is clearly weaker than Theorem~\ref{thm-strong-cg}, yet it provides enough information to establish Theorem~\ref{thm-strong-cg}
with the help of renormalization techniques (Section~\ref{sec-renorm}). Note that at the price of little modifications in
  the proof below one could prove the
  following fact, assuming \eqref{eq-def-SP3}: there exist $L \subset \mathbb{N}^{\star}$ and $c > 0$
  such that, for any $N$ large enough multiple of $L$ and any function $g$
  such that $\left( \log N \right)^2 \ll g (N) \leqslant N$,
  \begin{equation*} \mathbb{E} \inf_{\pi} \Phi^{J, \pi}_{\Lambda_N} \left( \begin{array}{l}
       \mbox{There exists a crossing cluster for $\omega$}\\
       \mbox{in $\Lambda_N$ and it is the only cluster of}\\
       \mbox{diameter larger or equal to } g (N)
     \end{array} \right) \geqslant 1 - \exp \left( - \frac{c g (N)}{\log N}
     \right) \end{equation*}
  Yet, this formulation suffers from arbitrary restrictions: the
  logarithm in the denominator and the condition that $N$ be a multiple of
  $L$. This is the reason for our choice of establishing a simpler control in Proposition~\ref{prop-weak-cg}, that will be reinforced later on by the use of renormalization techniques.

\begin{pf}
  In Corollary~\ref{corollary-exists-cl} we have seen the existence of $L_1 \in
  \mathbb{N}^{\star}$ such that, for any $N$ large enough multiple of
  $L_1$,
  \begin{equation}
    \mathbb{E} \inf_{\pi} \Phi^{J, \pi}_{\Lambda_N} \left( \mbox{There exists
    a crossing cluster for $\omega$ in $\Lambda_N$} \right) \geqslant 1 -
    \varepsilon / 2. \label{eq-exists-cc}
  \end{equation}
  It remains to prove that it is the only large cluster. We fix $L_2 \in
  \mathbb{N}^{\star}$ and $c > 0$ according to Lemma~\ref{lemma-proba-vert-cl}, and assume that $N$ is a large enough multiple of
  $L_1$ and of $L_2$ so that both (\ref{eq-exists-cc}) and
  (\ref{eq-proba-vert-cl}) hold. We consider the event
  \begin{equation*} A = \left\{ \begin{array}{l}
       \mbox{There exists a crossing cluster $\mathcal{C}^{\star}$}\\
       \mbox{for } \omega \mbox{ and another } \mathcal{C}' \mbox{ of
       diameter}\\
       \mbox{larger or equal to } N / 4
     \end{array} \right\} . \end{equation*}
  For any $\omega \in \Omega_{E^w(\Lambda_N)} \cap A$, there exists some direction
  $k \in \{1, \ldots, d\}$ in which the extension of $\mathcal{C}'$ is at least
  $N / 4$. Since all directions are equivalent we assume that $k = d$. If we
  denote $h = \inf \{z \cdot \mathbf{e}_d, z \in \mathcal{C}' \}$ and
  $\Lambda_N^{1 / 4, h} = \Lambda_N^{1 / 4} + h\mathbf{e}_d$, there exist
  $x, y \in \Bottom\left( \Lambda_N^{1 / 4, h} \right)$ such that
  \begin{equation*} x, y \stackrel{\omega^r}{\leftrightarrow} \Top\left( \Lambda_N^{1 /
     4, h} \right) \mbox{ and } x \stackrel{\omega^r}{\nleftrightarrow} y \end{equation*}
  where $\omega^r = \omega_{|E^w (\Lambda_N^{1 / 4, h})}$. As a consequence,
  \begin{align}
    \mathbb{E} \sup_{\pi} \Phi^{J, \pi}_{\Lambda_N} \left( A \right) &
    \leqslant  d \sum_{\tmscript{\begin{array}{c}
      h = 0 \ldots \lceil 3 N / 4 \rceil\\
      x, y \in \Bottom(\Lambda_N^{1 / 4, h})
    \end{array}}} \mathbb{E} \sup_{\pi} \Phi^{J, \pi}_{\Lambda_N^{1 / 4, h}}
    \left( 
\begin{array}{l}
x, y \stackrel{\omega^r}{\leftrightarrow} \Top\left( \Lambda_N^{1 / 4, h} \right) \\
\mbox{and } x \stackrel{\omega^r}{\nleftrightarrow} y
\end{array} \right) \notag \\
    & \leqslant  d (3 N / 4 + 2) N^{2 (d - 1)} \exp \left( - c \frac{N}{\log
    N} \right) \notag
  \end{align}
  which goes to $0$ as $N \rightarrow \infty$ and the proof is over.
\end{pf}

\subsection{The two dimensional case}

\label{sec-dim-two}In the two dimensional case the adaptation of Proposition
\ref{prop-weak-cg} is an easy exercise: it is enough to realize a few
horizontal and vertical crossings in $\Lambda_N$ to ensure the existence of a
crossing cluster, together with the uniqueness of large clusters.

\begin{proposition}
  \label{prop-weak-cg-d2}Assume \eqref{eq-def-SP2}. Then
  for any $\varepsilon > 0$, for any $N \in \mathbb{N}^{\star}$ large enough:
  \begin{equation}
    \mathbb{E} \inf_{\pi} \Phi^{J, \pi}_{\Lambda_N} \left( \begin{array}{l}
      \mbox{There exists a crossing cluster for $\omega$}\\
      \mbox{in $\Lambda_N$ and it is the only cluster of}\\
      \mbox{diameter larger or equal to N/4}
    \end{array} \right) \geqslant 1 - \varepsilon . \label{eq-weak-cg-d2}
  \end{equation}
\end{proposition}

\begin{pf}
  We divide $\Lambda_N$ in eight horizontal parts: for $k \in \{0, \ldots, 7\}$ we let
  \begin{equation*} P_{N, k} = \left\{ 1, \ldots, N - 1 \right\} \times \{ \left[ N k / 8 \right] + 1, \ldots, \left[ N (k + 1) / 8 \right] - 1\} \end{equation*}
  and then we decompose each $P_{N, k}$ in slabs of height $\kappa (N)$ where
  $\kappa$ is the function appearing in the definition of \eqref{eq-def-SP2}:
  for all 
\[h \in \{0, \ldots, [[N / 8] / \kappa (N)] - 1\},\] we define
  \begin{equation*} S_{N, k, h} = \left\{ 1, \ldots, N - 1 \right\} \times \{ \left[ N k / 8
     \right] + h \kappa (N) + 1, \ldots, \left[ N k / 8 \right] + (h + 1)
     \kappa (N) - 1\}. \end{equation*}
  Given $k \in \{0, \ldots, 7\}$ we consider the measure $\Psi$ on
  $(\mathcal{J}_{P_{N, k}}, \Omega_{P_{N, k}})$ induced by $(J_1 \vee \ldots
  \vee J_{h_{\max}}, \omega_1 \vee \ldots \vee \omega_{h_{\max}})$ under the
  product measure
  \begin{equation*} \bigotimes_{\tmscript{\begin{array}{c}
       h = 0, \ldots, h_{\max}
     \end{array}}} \mathbb{E} \Phi^{J, f}_{S_{N, k, h}} \end{equation*}
  where $h_{\max} = [[N / 8] / \kappa (N)] - 1$. Thanks to Proposition~\ref{prop-comp-averaged-prod} we know that $\Psi$ is stochastically smaller
  than $\mathbb{E} \Phi^{J, f}_{\Lambda_N}$, and thus than $\mathbb{E}
  \Phi^{J, \tilde{\pi} (J)}_{\Lambda_N}$ if $\tilde{\pi}$ is a worst boundary condition for
\eqref{eq-weak-cg-d2}, cf.~(\ref{eq-meas-pit}). Consider now the event
\begin{equation*}
\mathcal{E}_k = \left\{ \omega \in \Omega : \mbox{\begin{tabular}{l}   there exists $h \in \{0, \ldots, h_{\max} \}$ such that $\omega$\\
       presents an horizontal crossing in $S_{N, k, h}$
     \end{tabular}} \right\},
\end{equation*}
thanks to \eqref{eq-def-SP2} and to the product structure of $\Psi$ there exists some $c > 0$ such that
\begin{equation*} \Psi \left( \mathcal{E}_k \right) \geqslant 1 - \exp \left( -
     \frac{c N}{\kappa (N)} \right) \end{equation*}
  for any $N$ large enough, and because of the stochastic domination (remark
  that $\mathcal{E}_k$ is an increasing event) it follows that
  \begin{equation*} \mathbb{E} \Phi^{J, \tilde{\pi} (J)}_{\Lambda_N} \left( \mathcal{E}_0
     \cap \ldots \cap \mathcal{E}_7 \right) \geqslant 1 - 8 \exp \left( -
     \frac{c N}{\kappa (N)} \right) . \end{equation*}
  We proceed similarly in the vertical direction and let $\mathcal{E}'_k$ the
  event that $\omega$ presents a vertical link between $\Bottom(\Lambda_N)$ and $\Top(\Lambda_N)$ 
in the region  $$\{k N / 8, \ldots, (k + 1) N / 8\}\times\{1,\ldots, N-1\}.$$ The event $\mathcal{E}_0 \cap \ldots \cap \mathcal{E}_7 \cap
  \mathcal{E}'_0 \cap \ldots \cap \mathcal{E}'_7$ has a large probability
  under $\mathbb{E} \Phi^{J, \tilde{\pi} (J)}_{\Lambda_N}$, on the other hand
  it implies the existence of a crossing cluster, as well as the uniqueness of
  clusters of diameter larger than~$N / 4$.
\end{pf}

\section{Renormalization and density estimates}

\label{sec-renorm-density}In this Section we introduce renormalization
techniques, following Pisztora~\cite{N09} and Liggett, Schonmann and Stacey~\cite{N43}. We then finish the proof of the coarse graining (Theorem
\ref{thm-strong-cg} and Proposition~\ref{prop-cg-dens}). We also adapt the arguments of
Lebowitz \cite{N33} and Grimmett~\cite{N03} to the random media case and prove that for all $q \geqslant 1$ and all $\rho$, for all except at most
countably many values of $\beta$, the two extremal infinite volume measures with parameters $p (J_e) = 1 - \exp (- \beta J_e)$, $q$ and $\rho$ are equal. We conclude on an adaptation of the coarse graining to the Ising model.

\subsection{Renormalization framework}

\label{sec-renorm}The renormalization framework is made of two parts. First we describe a
geometrical decomposition of a large domain $\Lambda$ into a double sequence
of smaller cubes, then we present an adaptation of the stochastic domination
Theorem from~\cite{N43}.

We begin with a geometrical covering of $\Lambda$ with some double sequence
$(\Delta_i, \Delta'_i)_{i \in I}$. Its properties are described in detail in
the next Lemma, for the moment we just point out what we expect of the
$\Delta_i$ and of the $\Delta'_i$ respectively:
\begin{itemize}
  \item The $\Delta_i$ are boxes of side-length $L - 1$, they cover all of
  $\Lambda$ and most of them are disjoint. In the applications of
  renormalization they will typically help to control the local density of
  clusters.
  
  \item The $\Delta'_i$ are boxes of side-length $L + 2 L' - 1$ such that
  $\Delta'_i$ and $\Delta'_j$ have an intersection of thickness at least $2
  L'$ whenever $i$ and $j$ are nearest neighbors. The role of the $\Delta'_i$
  is to permit the connection between the main clusters of
  two neighbor blocks $\Delta_i$ and $\Delta_j$.
\end{itemize}
\begin{definition} 
  \label{def-cov-delta-L}Consider some domain $\Lambda$ of the form $$\Lambda =
z+  \prod_{k = 1}^d \{1, \ldots, L_k \}$$ with $z=(z_1,\ldots,z_d)\in{\mathbb{Z}^d}$, $L_k \in \mathbb{N}^{\star}$,
  and $L, L' \in \mathbb{N}^{\star}$ with $L' \leqslant L$. Assume that $L +
  2 L' \leqslant \min_{k = 1 \ldots d} L_k$, denote
  \begin{equation*} I_{\Lambda, L} = \prod_{k = 1}^d \left\{ 0, \ldots, \lceil L_k / L \rceil
     - 1 \right\} \end{equation*}
  and for all $i \in I_{\Lambda, L}$, call $x_i$ the point of coordinates
  $z_k+\min (Li \cdot \mathbf{e}_k, L_k - L)$ ($k = 1 \ldots d$) and $x_i'$ that
  of coordinates $z_k+\min (\max (Li \cdot \mathbf{e}_k, L'), L_k - L - L')$.
  Consider at last:
  \begin{equation*} \Delta_i = x_i +\{1, \ldots, L\}^d \mbox{ \ and \ } \Delta_i' = x'_i
     +\{- L'+1, \ldots, L + L' \}^d . \end{equation*}
  We say that $(\Delta_i, \Delta'_i)_{i \in I_{\Lambda, L}}$ is the $(L,
  L')$-covering of $\Lambda$.
\end{definition}

Remark that $x_i$ and $x'_i$ are the closest points to $Li$, with respect to the $\|.\|_{1}$ distance, such that $\Delta_i$ and $\Delta_i'$ are subsets of $\Lambda$.

\begin{lemma}
  \label{lemma-cov-delta}The properties of the sequence $(\Delta_i,
  \Delta'_i)_{i \in I_{\Lambda, L}}$ are as follows: for any $\Lambda, L, L'$
  as in definition~\ref{def-cov-delta-L}, we have:
  \begin{enumerate}
    \item The union $\bigcup_{i \in I_{\Lambda, L}} \Delta_i$ equals
    $\Lambda$.
    
    \item For every $i \in I_{\Lambda, L}$, $\Delta_i \subset \Delta'_i$ and
    $d (\Delta_i, \Lambda \setminus \Delta'_i) \geqslant L' + 1$.
    
    \item If $i, j \in I_{\Lambda, L}$ and $k \in \{1, \ldots, d\}$ satisfy $j
    = i +\mathbf{e}_k$, then both $\Delta'_i$ and $\Delta'_j$ contain the
    slab
    \begin{equation*} \{x \in \Delta'_j : (x - x_j') \cdot \mathbf{e}_k \leqslant L' \}. \end{equation*}
    \item For any $x \in \Lambda$ such that $x \cdot \mathbf{e}_k \leqslant L_k - L$
    for all $k = 1 \ldots d$, there exists a unique $i \in I_{\Lambda, L}$
    such that $x \in \Delta_i$.
    
    \item Given any $x \in \Lambda$, there exist at most $6^d$ indices $i \in
    I_{\Lambda, L}$ such that $x \in \Delta'_i$.
  \end{enumerate}
\end{lemma}

\begin{pf}
  We begin with the first point. If we denote $\Lambda_L =\{1, \ldots,, L\}^d$, it is clear that the sequence $(Li + \Lambda_L)_{i \in I_{\Lambda, L}}$ covers all $\Lambda$ and that:
  \begin{equation*} \forall i \in I_{\Lambda, L}, \left( Li + \Lambda_L \right) \cap \Lambda
     \subset \Delta_i \subset \Lambda \end{equation*}
  thanks to the definition of $x_i$. The equality $\bigcup_{i \in I_{\Lambda,
  L}} \Delta_i = \Lambda$ follows. For (ii), the inclusion
  $\Delta_i \subset \Delta'_i$ is a trivial consequence of the remark that
  $\left\| x_i - x'_i \right\|_{\infty} \leqslant L'$. As for the distance
  between $\Delta_i$ and $\Lambda \setminus \Delta'_i$, we compute the
  distance between $\Delta_i$ and the outer faces of $\Delta'_i$ included in
  $\Lambda$. In a given direction $\mathbf{e}_k$ (for some $k \in \{1,
  \ldots, d\}$), it is exactly $L' + 1$ whenever $x_i \cdot \mathbf{e}_k =
  x'_i \cdot \mathbf{e}_k$. If $x_i \cdot \mathbf{e}_k < x'_i \cdot
  \mathbf{e}_k$, then the block $\Delta'_i$ touches the face of $\Lambda$ of
  $\mathbf{e}_k$-coordinate $1$ and the distance between $\Delta_i$ and the
  unique outer face of $\Delta'_i$ normal to $\mathbf{e}_k$ and included in
  $\Lambda$ is larger than $L' + 1$. The same occurs if $x_i \cdot
  \mathbf{e}_k > x'_i \cdot \mathbf{e}_k$ with the opposite face of
  $\Lambda$. For (iii), remark that $x_j' - x_i' = l\mathbf{e}_k$
  with $l \leqslant L$. For (iv), consider such an $x$ and let $i
  \in I_{\Lambda, L}$ such that $x \in \Delta_i$ (it exists thanks to (i)).
Since the coordinates of $x_i$ are strictly smaller than those of
  $x$, they do not exceed $L_k - L-1$. In view of the definition
  of $x_i$ this implies that $x_i = Li$ and hence that $x \in Li + \Lambda_L$,
  which determines $i$. Consider at last $x \in \Lambda$ and $i \in
  I_{\Lambda, L}$ such that $x \in \Delta'_i$. For each $k = 1 \ldots d$, at
  least one of the following inequalities must hold:
  \begin{equation*} Li \cdot \mathbf{e}_k < L' \mbox{ \ or \ } Li \cdot \mathbf{e}_k
     > L_k - L - L' \mbox{ \ or \ } Li \cdot \mathbf{e}_k - L' +1\leqslant
     x \cdot \mathbf{e}_k \leqslant Li \cdot \mathbf{e}_k + L + L' \end{equation*}
  since the $k$-coordinate of $x'_i$ is $Li \cdot \mathbf{e}_k$ whenever the
  first two inequalities are not satisfied. The first condition yields only
  one possible value for $i \cdot \mathbf{e}_k$ : $i \cdot \mathbf{e}_k =
  0$ since $L' \leqslant L$. For the second we consider candidates of the form
  $i \cdot \mathbf{e}_k = \lceil L_k / L \rceil - n$ with $n \geqslant 1$
  (recall that $i \in I_{\Lambda, L}$), and there are at most two
  possibilities corresponding to $n \in \{1, 2\}$. At last, the third
  condition yields not more than $3$ possibilities for $i \cdot
  \mathbf{e}_k$ and the bound in (v) follows.
\end{pf}

We now present the stochastic domination Theorem and its adaptation to the
averaged measure. Stochastic domination is a natural and useful concept for
renormalization, that was already present in the pioneer work~\cite{N09}.
It goes one step beyond the Peierls
estimates we use in the proof of Theorem~\ref{thm-strong-cg} 
and could have used in that of Corollary~\ref{corollary-exists-cl}. It is of much help for
example in the proof of (\ref{eq-dens-low}) in Proposition~\ref{prop-cg-dens}. 
Let us recall Theorem 1.3 of~\cite{N43}:

\begin{theorem}
  \label{thm-stochdom-LSS}Let $G = (S, E)$ be a graph with a countable vertex
  set in which every vertex has degree at most $K \geqslant 1$, and in which
  every finite connected component of G contains a vertex of degree strictly
  less than $K$. Let $p \in [0, 1]$ and suppose that $\mu$ is a Borel probability
  measure on $X \in \{0, 1\}^S$ such that almost surely,
  \begin{equation*}
    \mu (X_s = 1| \sigma (\{X_t : \{s, t\} \notin E\})) \geqslant p, \forall s
    \in S.
  \end{equation*}
  Then, if $p \geqslant 1 - (K - 1)^{K - 1} / K^K$ and
  \begin{equation*}
    r (K, p) = \left( 1 - \frac{(1 - p)^{1 / K}}{(K - 1)^{(K - 1) / K}}
    \right) (1 - ((1 - p) (K - 1))^{1 / K}), \label{eq-def-r-stochdom}
  \end{equation*}
  the measure $\mu$ stochastically dominates the Bernoulli product measure on
  $S$ of parameter $r (K, p)$. Note that as $p$ goes to $1$, $r (K, p)$ tends
  to $1$.
\end{theorem}

We provide then an adaptation of the former Theorem to the averaged measure:

\begin{proposition}
  \label{prop-renormalization}Consider some finite domain $\Lambda \subset
  \mathbb{Z}^d$, $(E_i)_{i = 1 \ldots n}$ a finite sequence of subsets of $E^w
  (\Lambda)$ and $\left( \mathcal{E}_i \right)_{i = 1 \ldots n}$ a family of
  events depending respectively on $\omega_{|E_i}$ only. If the intersection
  of any $K + 1$ distinct $E_i$ is empty and if
  \begin{equation*}
    p = \inf_{i = 1 \ldots n} \mathbb{E} \inf_{\pi \in \Omega} \Phi^{J,
    \pi}_{E_i} \left( \mathcal{E}_i \right)
  \end{equation*}
is close enough to $1$,
  then for any increasing function $f: \{0, 1\}^n \rightarrow \mathbb{R}$ we
  have:
  \begin{equation*}
    \mathbb{E} \inf_{\pi \in \Omega} \Phi^{J, \pi_{|E^w(\Lambda)}}_{\Lambda} \left( f \left(
    \mathbf{1}_{\mathcal{E}_1}, \ldots, \mathbf{1}_{\mathcal{E}_n} \right) 
    \left| \omega = \pi \mbox{ on } E^w (\Lambda) \setminus \bigcup_{i = 1}^n E_i \right.
    \right) \geqslant \mathcal{B}^n_{r' (K, p)} \left( f \right)
    \end{equation*}
  where $\mathcal{B}^n_r$ is the Bernoulli product measure on $\{0, 1\}^n$ of
  parameter $r$ and $r' (K, p) = r^2 \left( K, 1 - \sqrt{1 - p} \right)$
  (with $r(.,.)$ taken from Theorem~\ref{thm-stochdom-LSS}). In particular, $\lim_{p \rightarrow 1} r' (K, p) = 1$.
\end{proposition}

The conditional formulation for the stochastic domination is 
motivated by the need to control some region of the domain uniformly over constraints in the remaining region.
A good example of this necessity will be seen in the formulation of the lower bound for $L^1$ phase
coexistence in the Ising model~\cite{M02}.

\begin{pf}
  The proof is based on Markov's inequality. Consider
  \begin{equation*} \mathcal{G}_i = \left\{ J : \inf_{\pi} \Phi^{J, \pi}_{E_i} \left(
     \mathcal{E}_i \right) \geqslant 1 - \sqrt{1 - p} \right\} . \end{equation*}
  Clearly, the $\mathcal{G}_i$ are $\mathcal{B}_{E_i}$-measurable and hence
  any two $\mathcal{G}_i, \mathcal{G}_j$ are independent under $\mathbb{P}$
  if $E_i \cap E_j = \emptyset$. Thanks to Markov's inequality (\ref{eq-EPPhi}),
as  $\mathbb{E}(1 - \inf_{\pi} \Phi^{J, \pi}_{E_i})\leqslant 1-p $
  it follows that $\mathbb{P} \left( \mathcal{G}_i \right) \geqslant 1 - \sqrt{1 - p}$
  for all $i = 1 \ldots n$. Consider now the graph on $I =\{1, \ldots, n\}$
  with edge set $L =\{\{i, j\} \in I^2 : i \neq j \mbox{ and } E_i \cap E_j \neq
  \emptyset\}$. All vertices of the graph have degree at most $K - 1$, while
  almost surely
  \begin{equation*} \inf_{i \in I} \mathbb{P} \left( \mathcal{G}_i |\mathcal{G}_j : \{i, j\}
     \notin L \right) = \inf_{i \in I} \mathbb{P} \left( \mathcal{G}_i
     \right) \geqslant 1 - \sqrt{1 - p} . \end{equation*}
  Hence the assumptions of Theorem~\ref{thm-stochdom-LSS} are satisfied for
  $p$ large enough and it follows that the law of the $\mathcal{G}_i$
  dominates a Bernoulli product measure of parameter $r = r (K, 1 - \sqrt{1 -
  p})$. We keep this fact in mind for the end of the proof and now fix a
  realization of the media $J$. We call $I' =\{i \in I : J \in \mathcal{G}_i
  \}$. Let $(I', L')$ be the restriction of the graph $(I, L)$ to $I'$: again,
  the maximal degree of all vertices is at most $K - 1$. We consider now the
  sequence $(\mathcal{E}_i)_{i \in I'}$ under the conditional measure
  \begin{equation*} \mu_{\pi} = \Phi^{J, \pi_{|E^w(\Lambda)}}_{\Lambda} \left( . \left| \omega = \pi \mbox{
     on } E^w (\Lambda) \setminus \bigcup_{i = 1}^n E_i \right. \right) \end{equation*}
  where $\pi \in \Omega$. Thanks to the DLR equation for $\Phi^{J,
  \pi}_{\Lambda}$ and to the definition of $\mathcal{G}_i$, we have again:
  \begin{equation*} \inf_{i \in I'} \mu_{\pi} (\mathcal{E}_i |\mathcal{E}_j : \{i, j\} \notin
     L') \geqslant \inf_{i \in I'} \inf_{\pi} \Phi^{J, \pi}_{E_i} \left(
     \mathcal{E}_i \right) \geqslant 1 - \sqrt{1 - p} \end{equation*}
  almost surely. Thus, according to Theorem~\ref{thm-stochdom-LSS}, if
  $\mathcal{B}^n_r$ is a Bernoulli product measure of parameter $r = r (K, 1 -
  \sqrt{1 - p})$ as above, and if we denote its variable $(X_i)_{i \in I}$,
  then the family $(\mathbf{1}_{\mathcal{E}_i})_{i \in I'}$ stochastically
  dominates $(X_i)_{i \in I'}$. In other words, for any increasing function $f: \{0, 1\}^n \rightarrow \mathbb{R}$ we can write (notice that $i \in I
  \setminus I' \Rightarrow \mathbf{1}_{\mathcal{G}_i} = 0$):
  \begin{equation*} \mu_{\pi} (f (\mathbf{1}_{\mathcal{G}_1} \mathbf{1}_{\mathcal{E}_1},
     \ldots, \mathbf{1}_{\mathcal{G}_n} \mathbf{1}_{\mathcal{E}_n}))
     \geqslant \mathcal{B}^n_r (f (\mathbf{1}_{\mathcal{G}_1} X_1, \ldots,
     \mathbf{1}_{\mathcal{G}_n} X_n)) \end{equation*}
  and taking the infimum over $\pi$ we get (since
  $\mathbf{1}_{\mathcal{G}_i} \mathbf{1}_{\mathcal{E}_i} \leqslant
  \mathbf{1}_{\mathcal{E}_i}$)
  \begin{eqnarray}
  \lefteqn{\hspace*{-6 cm} \inf_{\pi \in \Omega} \Phi^{J, \pi_{|E^w(\Lambda)}}_{\Lambda} \left( f
    (\mathbf{1}_{\mathcal{E}_1}, \ldots, \mathbf{1}_{\mathcal{E}_n}) \left| \,
    \omega = \pi \mbox{ on } E^w (\Lambda) \setminus \bigcup_{i = 1}^n E_i
    \right. \right) \geqslant} & & \notag \\
\hspace*{5cm} & & \mathcal{B}^n_r (f (\mathbf{1}_{\mathcal{G}_1} X_1,
    \ldots, \mathbf{1}_{\mathcal{G}_n} X_n)) . \label{eq-stochdom-fixedJ}
  \end{eqnarray}
  At this point, we just need to exploit the stochastic minoration on the
  sequence $\left( \mathcal{G}_i \right)_{i \in I}$: let
  $\tilde{\mathcal{B}}^n_r$ another Bernoulli product measure of parameter $r$
  on $I$, and denote its variable $(Y_i)_{i \in I}$. Then,
  \begin{align}
    \mathcal{B}^n_r (f (\mathbf{1}_{\mathcal{G}_1} X_1, \ldots,
    \mathbf{1}_{\mathcal{G}_n} X_n)) & \geqslant  \tilde{\mathcal{B}}^n_r
    \left( \mathcal{B}^n_r (f (Y_1 X_1, \ldots, Y_n X_n)) \right) \notag \\
    & =  \mathcal{B}^n_{r^2} (f (X_1, \ldots, X_n)) \notag
  \end{align}
  and reporting in (\ref{eq-stochdom-fixedJ}) we prove the claim.
\end{pf}

\subsection{Structure of the main cluster}

\label{sec-strong-cg}Using the former geometrical decomposition, the weak form
of the coarse graining and the Peierls argument, we provide with Theorem
\ref{thm-strong-cg} the final version of the control on the structure of the
$\omega$-clusters under the averaged measure. Our result is, at last, entirely
similar to Theorem 3.1 of~\cite{N09}. We recall that a crossing
cluster in $\Lambda_N$ is a cluster that connects all outer faces of
$\Lambda_N$ (hence it lives on $E^w (\Lambda_N)$), and cite anew Theorem
\ref{thm-strong-cg}:

\begin{theorem}
Assumption {\rm(\textbf{SP})} implies the existence of $c > 0$ and $\kappa < \infty$ such that, for any $N
\in \mathbb{N}^{\star}$ large enough and for all $l \in [\kappa \log N, N]$,
\begin{equation*} \mathbb{E} \inf_{\pi} \Phi^{J, \pi}_{\Lambda_N} \left( \begin{array}{l}
     \mbox{There exists a crossing $\omega$-cluster $\mathcal{C}^{\star}$ in
     $\Lambda_N$}\\
     \mbox{and it is the unique cluster of diameter} \geqslant l
   \end{array} \right) \geqslant 1 - \exp \left( - c l \right)  \end{equation*}
where the infimum $\inf_{\pi}$ is taken over all boundary conditions $\pi \in
\Omega_{E(\mathbb{Z}^d)  \setminus E^w
(\Lambda_N)}$.
\end{theorem}

\begin{pf}
We begin with a geometrical covering of $\Lambda_N$: for $L \geqslant 2$ we let
$(\Delta_i, \Delta'_i)_{i \in I_{\Lambda_N, L}}$ the $(L, L - 1)$-covering of $\Lambda_N$
described at definition~\ref{def-cov-delta-L}. For each $i \in
  I_{\Lambda_N, L}$ we consider
  \begin{equation*}
    \mathcal{E}_i = \left\{ \omega \in \Omega : \begin{array}{l}
      \mbox{There exists a crossing cluster for $\omega$}\\
      \mbox{in $\Delta'_i$ and it is the only cluster of}\\
      \mbox{diameter larger or equal to } L \mbox{ in } \Delta'_i
    \end{array} \right\}  
  \end{equation*}
and denote by $A_l$ the event
\begin{equation*}
    A_l = \left\{ \omega \in \Omega : \mbox{\begin{tabular}{l}
There exists a crossing cluster $\mathcal{C}_{\mathcal{E}}$ for $\mathcal{E}_i$ in\\ $I_{\Lambda_N, L}$
such that the diameter of any connected \\ component
 of $I_{\Lambda_N, L}\setminus\mathcal{C}_{\mathcal{E}}$ 
is at most $\lceil l/L \rceil -1$
    \end{tabular}} \right\} . \end{equation*}
In a first time we prove the inclusion
  \begin{equation}
    A_l \subset \left\{ \omega \in \Omega : \mbox{\begin{tabular}{l}
      There exists a crossing $\omega$-cluster $\mathcal{C}^{\star}$ in
      $\Lambda_N$\\
      and it is the unique cluster of diameter $\geqslant l$
    \end{tabular}} \right\} . \label{eq-Al-subset-UCl} 
  \end{equation}
  To begin with, remark that if $i, j\in I_{\Lambda_N, L}$ are nearest neighbors, and if
  $\omega\in\mathcal{E}_i\cap\mathcal{E}_j$, then the
  corresponding $\omega$-crossing clusters in $\Delta'_i$ and $\Delta'_j$
are connected because the intersection $E^w (\Delta'_i) \cap E^w (\Delta_j')$ has a thickness at least $2 L -2$, cf.~Lemma~\ref{lemma-cov-delta} ({iii}). Hence we see that for every $\omega\in A_l$ there exists a crossing cluster $\mathcal{C}$ for
$\omega$ in $\Lambda_N$. Consider now $\omega \in A_l$ and some $\omega$-open path $c$ in $E^w(\Lambda_N)$ of diameter larger or equal to $l$.
  It has an extension at least $l$ in some direction $k$, thus we can find a
  connected path $i_1, \ldots, i_n$ in $I_{\Lambda_N, L}$ of extension at
  least $\lceil l/L \rceil$ in the same direction such that $c$ enters each
  $\Delta_{i_j}$. Because of the definition of $A_l$, at least one of the $i_j$ pertains to $\mathcal{C}_{\mathcal{E}}$.
 Yet in view of Lemma~\ref{lemma-cov-delta} ({ii}), $c$ has an incursion in $E^w(\Delta'_{i_j})$ 
of diameter at least $L$, hence $c$ touches the $\omega$-crossing cluster in $E^w (\Delta'_{i_j})$ which is a part of $\mathcal{C}$, thus $c=\mathcal{C}$ and \eqref{eq-Al-subset-UCl} is proved.

We need now a lower bound on the probability of $A_l$. If $\omega\in\Omega_{E^w(\Lambda_N)}$ 
is such that there exists no $\ast$-connected path $i_1,\ldots, i_n$ with $n=\lceil l/L \rceil$ and $\forall i\in\{1,\ldots, n\}, \omega\notin\mathcal{E}_i$, then
$\omega \in A_l$. This is a consequence of Lemma 2.1 in~\cite{N72} or of the (simpler) remark that the set of $\mathcal{E}_i$-good blocks
constitutes a connected interface in every slab of $I_{\Lambda_N, L}$ of height $\lceil l/L \rceil$, whatever is its orientation, 
hence the holes in $\mathcal{C}_{\mathcal{E}}$ have a diameter at most $\lceil l/L \rceil -1$.

Thanks to the stochastic domination (Proposition \ref{prop-renormalization}), and to the fact that $A_l$ is an increasing event, it follows that
\begin{equation*} \mathbb{E} \inf_{\pi} \Phi^{J, \pi}_{\Lambda_N} \left( A_l \right) \geqslant 
\mathcal{B}^{I_{\Lambda_N, L}}_{p_L}\left( 
\mbox{\begin{tabular}{l}
There is no $\ast$-connected path $i_1,..,i_n$ \\
in $I_{\Lambda_N, L}$ with $n=\lceil l/L \rceil$ and\\ $X_{i_k}=0$, for all $k\in\{1,..,n\}$
\end{tabular}}
\right) \end{equation*}
where $p_L$ can be chosen arbitrarily close to $1$ for an appropriate $L$ in view of Propositions~\ref{prop-weak-cg} and~\ref{prop-weak-cg-d2}.
We conclude using a Peierls estimate: there are no more than $|I_{\Lambda_N, L}| \times (3^d)^n$ $\ast$-connected paths of length $n$ in $I_{\Lambda_N, L}$, hence
\begin{equation*} \mathbb{E} \inf_{\pi} \Phi^{J, \pi}_{\Lambda_N} \left( A_l \right) \geqslant 
1-N^d (3^d)^n (1-p_L)^n . \end{equation*}
If we fix $L$ so that $p_L > 1 - 3^{-d}$, it follows that $(3^d)^n (1-p_L)^n = \exp(- c' n)$ for some $c'>0$, together with
\begin{equation*} \mathbb{E} \inf_{\pi} \Phi^{J, \pi}_{\Lambda_N} \left( A_l \right) \geqslant 
1-\exp \left( d \log N - c' l / L \right) \end{equation*}
hence the claim holds with $c = c' / (2 L)$ and $\kappa = d / c$.
\end{pf}

\subsection{Typical density of the main cluster}

In this Section we prove Proposition~\ref{prop-cg-dens} and provide estimates
on the averaged probability that the density of the main cluster be larger
than $\theta^w$ or smaller than $\theta^f$, where
\begin{equation}
  \theta^f = \lim_N \mathbb{E} \Phi_{\hat{\Lambda}_N}^{J, f} (0
  \stackrel{\omega}{\leftrightarrow} \partial \hat{\Lambda}_N) \mbox{ \ \ and \
  \ } \theta^w = \lim_N \mathbb{E} \Phi_{\hat{\Lambda}_N}^{J, w} (0
  \stackrel{\omega}{\leftrightarrow} \partial \hat{\Lambda}_N)
\end{equation}
(see after (\ref{eq-def-dL}) for the definitions of $\Lambda_N$ and $\hat{\Lambda}_N$). An important question is whether
these quantities are equal, and we will prove in Theorem~\ref{thm-unique-infvol}
this is the case for almost all values of $\beta$. We recall the contents of Proposition
\ref{prop-cg-dens}:

\begin{proposition}
For any $\varepsilon > 0$ and $d
\geqslant 1$,
\begin{equation*} \limsup_N  \frac{1}{N^d} \log \mathbb{E} \sup_{\pi} \Phi_{\Lambda_N}^{J,
   \pi} \left( \begin{array}{l}
     \mbox{Some crossing cluster $\mathcal{C}^{\star}$ has}\\
     \mbox{a density larger than $\theta^w + \varepsilon$}
   \end{array} \right) < 0 \end{equation*}
while assumption {\rm(\textbf{SP})} implies, for any $\varepsilon > 0$ and $d \geqslant 2$:
\begin{equation*} \limsup_N  \frac{1}{N^{d - 1}} \log \mathbb{E} \sup_{\pi}
   \Phi_{\Lambda_N}^{J, \pi} \left( \begin{array}{l}
     \mbox{There is no crossing cluster $\mathcal{C}^{\star}$}\\
     \mbox{of density larger than $\theta^f - \varepsilon$}
   \end{array}  \right) < 0. \end{equation*}
\end{proposition}

The proofs of these two estimates differ very little from the original ones in \cite{N09}, yet we state them as examples of applications of
the renormalization methods.

\begin{pf} (Upper deviations). Given $L \in \mathbb{N}^{\star}$ we consider $(\Delta_i, \Delta_i)_{i \in I_{\Lambda_N, L}}$ the $(L, 0)$-covering of $\Lambda_N$
and call $\tilde{I}_{\Lambda_N, L}=\{0,\ldots, [(N-1)/L]-1\}^d$, so that $\Delta_i$ and $\Delta_j$ are disjoint for any $i\neq j \in \tilde{I}_{\Lambda_N, L}$, cf.~Lemma \ref{lemma-cov-delta} ({iv}). We let furthermore
  \begin{equation*} Y_i = \frac{1}{L^d} \sum_{x \in \Delta_i} \mathbf{1}_{\left\{ x
     \stackrel{\omega}{\leftrightarrow} \partial^i \Delta_i \right\}} \;\; (i \in \tilde{I}_{\Lambda_N, L}), \end{equation*}
they are i.i.d. variables under the product measure
  $\bigotimes_{i \in  \tilde{I}_{\Lambda_N, L}} \mathbb{E} \Phi_{E^f(\Delta_i)}^{J, w}$ and their
expectation is not larger than $\theta^w+\varepsilon/4$ for $L$ large enough. Hence Cram\'er's Theorem yields:
  \begin{equation*} \limsup_N \frac{1}{|  \tilde{I}_{\Lambda_N, L} |} \log \bigotimes_{i \in
      \tilde{I}_{\Lambda_N, L}} \mathbb{E} \Phi_{E^f (\Delta_i)}^{J, w}
     \left( \frac{1}{|  \tilde{I}_{\Lambda_N, L} |} \sum_{i \in  \tilde{I}_{\Lambda_N, L}} Y_i
     \leqslant \theta^w + \frac{\varepsilon}{2} \right) < 0 \end{equation*}
for $L$ large enough. Thanks to the stochastic domination (Proposition~\ref{prop-comp-averaged-prod})
the same control holds under $\mathbb{E} \Phi_{\Lambda_N}^{J, w}$, and thanks to the remark that 
\begin{equation*}
    \sum_{x \in \Lambda_N} \mathbf{1}_{\{x \stackrel{\omega}{\leftrightarrow}
    \partial \Lambda_N \}} \leqslant L^d \sum_{i \in \tilde{I}_{\Lambda_N, L} } Y_i + d L N^{d - 1} \label{eq-comp-xLB} .
  \end{equation*} it follows that
  \begin{equation*} \limsup_N  \frac{1}{N^d} \log \mathbb{E} \sup_{\pi} \Phi_{\Lambda_N}^{J,
     \pi} \left( \frac{1}{| \Lambda_N |} \sum_{x \in \Lambda_N}
     \mathbf{1}_{\{x \stackrel{\omega}{\leftrightarrow} \partial \Lambda_N
     \}} \geqslant \theta^w + \varepsilon \right) < 0 \end{equation*}
  which implies the claim.
\end{pf}

The proof for the cost of lower deviations is more subtle as it relies on Theorem \ref{thm-strong-cg} and
Proposition~\ref{prop-renormalization}:

\begin{pf} (Lower deviations). Given $L \in \mathbb{N}^{\star}$ we call $(\Delta_i, \Delta'_i)_{i \in I_{\Lambda_N, L}}$
 the $(L, L - 1)$ covering of $\Lambda_N$. We use the same notation $\tilde{I}_{\Lambda_N, L}$
as in the previous proof and let
  \begin{equation*} Y_i = \frac{1}{L^d} \sum_{x \in \Delta_i} \mathbf{1}_{\left\{ \tmop{diam}(\mathcal{C}_x) \geqslant \sqrt{L} \right\}}  \;\; (i \in \tilde{I}_{\Lambda_N, L}) \end{equation*}
where $\mathcal{C}_x$ is the $\omega$-cluster containing $x$. One has $\liminf_{L\rightarrow\infty} \mathbb{E} \Phi_{E^f (\Delta_0)}^{J, f}(Y_0) \geqslant \theta^f$, hence Cram\'er's Theorem yields
  \begin{equation*} \limsup_N \frac{1}{|  \tilde{I}_{\Lambda_N, L} |} \log \bigotimes_{i \in
      \tilde{I}_{\Lambda_N, L}} \mathbb{E} \Phi_{E^f (\Delta_i)}^{J, f}
     \left( \frac{1}{|  \tilde{I}_{\Lambda_N, L} |} \sum_{i \in  \tilde{I}_{\Lambda_N, L}} Y_i
     \leqslant \theta^f - \frac{\varepsilon}{2} \right) < 0 \end{equation*}
for any $L$ large enough. Consider now $\tilde{\pi}_N: \mathcal{J}_{E^w (\Lambda_N)} \rightarrow \Omega_{E^w
  (\Lambda_N)^c}$ a measurable boundary condition as in~(\ref{eq-meas-pit}) that satisfies
  \begin{equation*}     \Phi_{\Lambda_N}^{J, \tilde{\pi}_N (J)} (\mathcal{A}_N^{\varepsilon}) =
    \sup_{\pi} \Phi_{\Lambda_N}^{J, \pi} (\mathcal{A}_N^{\varepsilon}) .  \end{equation*}
where $\mathcal{A}_N^{\varepsilon}$ is the event that there is no crossing cluster of density larger than $\theta^f - \varepsilon$ in $\Lambda_N$. Thanks to Proposition~\ref{prop-comp-averaged-prod} we infer that
  \begin{equation}
    \limsup_N \frac{1}{N^d} \log \mathbb{E} \Phi_{\Lambda_N}^{J,
    \tilde{\pi}_N (J)} \left( \frac{1}{|  \tilde{I}_{\Lambda_N, L} |} \sum_{i \in
     I_{\Lambda_N, L}} Y_i \leqslant \theta^f - \frac{\varepsilon}{2} \right) < 0
    \label{ineq-Yi-theta}
  \end{equation}
for any $L$ large enough. On the other hand, consider the collection of events
  \begin{equation*} \mathcal{E}_i =\left\{ \mbox{\begin{tabular}{l}
       There exists a crossing cluster for $\omega$ in $\Delta'_i$\\
       and it is the unique cluster of diameter $\geqslant \sqrt{L}$
     \end{tabular}} \right\} \end{equation*}
  for $i \in I_{\Lambda_N, L}$. Each $\mathcal{E}_i$ depends only on $\omega_{|E^w (\Delta'_i)}$ while Theorem~\ref{thm-strong-cg} implies:
  \begin{equation*} \lim_{L \rightarrow \infty} \mathbb{E} \inf_{\pi \in \Omega_{E^w
     (\Delta'_i)^c}} \Phi^{J, \pi}_{\Delta'_i} \left( \mathcal{E}_i \right) = 1 \end{equation*}
uniformly over $i\in I_{\Lambda_N, L}$.
Hence the assumptions of Proposition~\ref{prop-renormalization} are satisfied. Applying Theorem 1.1 of~\cite{N72} thus yields: for any $\delta > 0$, any $L$ large enough,
  \begin{equation}
    \limsup_N \frac{1}{N^{d - 1}} \log \mathbb{E} \Phi_{\Lambda_N}^{J,
    \tilde{\pi}_N (J)} \left( \mbox{\begin{tabular}{l}
      There exists no crossing cluster\\
      of density $\geqslant 1 - \delta$ for $(\mathcal{E}_i)_{i \in I_{\Lambda_N, L}}$
    \end{tabular}} \right) < 0 \label{eq-clust-high-dens} .
  \end{equation}
Assume now that $\omega \in \Omega_{E^w(\Lambda_N)}$ realizes neither of the events in~(\ref{ineq-Yi-theta}) and~(\ref{eq-clust-high-dens}) -- this is the typical behavior 
under $\mathbb{E} \Phi_{\Lambda_N}^{J, \tilde{\pi}_N (J)}$ up to surface order large deviations. Call $\mathcal{C}\subset I_{\Lambda_N, L}$ the
crossing cluster for $\mathcal{E}_i$. Because of the
  overlapping between the $\Delta'_i$ (Lemma~\ref{lemma-cov-delta}~(iii)), to $\mathcal{C}$ corresponds a crossing cluster $\mathcal{C}^{\star}$ 
for $\omega$ in $\Lambda_N$ that passes through every $\Delta'_i$ for $i \in \mathcal{C}$. Since $\mathcal{C}^{\star}$ is the
  only large cluster in each $\Delta'_i$ when $i \in \mathcal{C}$, we have
  \begin{align}
    \left| \mathcal{C}^{\star} \right| & \geqslant  \sum_{i \in \mathcal{C} \cap \tilde{I}_{\Lambda_N, L}} \left( L^d Y_i - 2 d \sqrt{L} L^{d - 1} \right) \notag \\
    & \geqslant  \left[ \frac{N-1}{L}\right]^d L^d  \left( \theta^f - \frac{\varepsilon}{2} - \frac{2
    d}{\sqrt{L}} \right) - \delta L^d  \left( \frac{N}{L} + 1 \right)^d, \notag
  \end{align}
which is not smaller than $N^d(\theta^f - \varepsilon)$ provided that $\delta = \varepsilon / 6$, $L > (12 d / \varepsilon)^2$ and $N$ is large enough.
\end{pf}

\subsection{Uniqueness of the infinite volume measure}

\label{sec-uniq-infvolEPhi}Adapting the arguments of Lebowitz~\cite{N33}
 and Grimmett~\cite{N03} to the random media case, we
prove that for all except at most countably many values of the inverse
temperature, the boundary condition does not influence the infinite volume
limit of joint FK measures.

To begin with, given the parameters $\rho, q, p (J) = 1 - \exp (- \beta J)$ with $\beta \geqslant 0$ we define two infinite volume measures on $\mathcal{J} \times \Omega$
 by
\begin{equation}
  \Theta_{\infty}^f = \lim_{N \rightarrow \infty} \mathbb{E}
  \Phi_{\hat{\Lambda}_N}^{J, f} \mbox{ \ and \ } \Theta_{\infty}^w = \lim_{N
  \rightarrow \infty} \mathbb{E} \Phi_{\hat{\Lambda}_N}^{J, w}.
\end{equation}
As in the uniform media case,
these limits exist and $\Theta_{\infty}^f$ is stochastically smaller than $\Theta_{\infty}^w$ thanks to the stochastic inequalities 
\begin{equation*} \mathbb{E} \Phi_{\hat{\Lambda}_N}^{J, f} \underset{stoch.}{\leqslant} \mathbb{E} \Phi_{\hat{\Lambda}_{N + 1}}^{J, f}
   \underset{stoch.}{\leqslant} \mathbb{E} \Phi_{\hat{\Lambda}_{N +
   1}}^{J, w} \underset{stoch.}{\leqslant} \mathbb{E}
   \Phi_{\hat{\Lambda}_N}^{J, w} \end{equation*}
regarding the law induced on $(J, \omega)_{|E^w ( \hat{\Lambda}_N)}$. Let us recall Theorem~\ref{thm-unique-infvol}:

\begin{theorem}
If the interaction equals $p (J_e) =
1 - \exp (- \beta J_e)$, for any Borel probability measure $\rho$ on $[0, 1]$, any $q \geqslant 1$
and any dimension $d \geqslant 1$, the set
\begin{equation*} \mathcal{D}_{\rho, q, d} = \left\{ \beta \geqslant 0 : \lim_{N \rightarrow
   \infty} \mathbb{E} \Phi^{J, f}_{\hat{\Lambda}_N} \neq \lim_{N \rightarrow
   \infty} \mathbb{E} \Phi^{J, w}_{\hat{\Lambda}_N} \right\} \end{equation*}
is at most countable.
\end{theorem}

We will present the proof of this Theorem after we state one Lemma. Given a
finite edge set $E$, a realization of the media $J \in \mathcal{J}_E$ and a
boundary condition $\pi \in \Omega_{E^c}$ we denote
\begin{equation}
  Y^{J, \pi}_E = \sum_{\omega \in \Omega_E} \prod_{e \in E} \left( \frac{p
  (J_e)}{1 - p (J_e)} \right)^{\omega_e} \times q^{C_E^{\pi}(\omega)} \label{eq-def-YpiE}
\end{equation}
the (adapted) partition function (see Section~\ref{sec-def-RCRM} for the definition of $C_E^{\pi}(\omega)$).

\begin{lemma}
  \label{lemma-conv-pressure}Let $(\pi_N)_{N \in \mathbb{N}^{\star}}$ such
  that $\forall N \in \mathbb{N}^{\star}, \pi_N \in \Omega_{E^w
  (\Lambda_N)^c}$. Then, the limit
  \begin{equation}
    y (\rho, q, \beta) = \lim_{N \rightarrow \infty} \frac{1}{(2 N + 1)^d}
    \mathbb{E} \log Y^{J, \pi_N}_{E^w ( \hat{\Lambda}_N)}
    \label{eq-def-pressure}
  \end{equation}
  exists and is independent of $(\pi_N)$. Furthermore, $y$ and $\mathbb{E}
  \log Y^{J, \pi}_E$ (for any $E \subset E(\mathbb{Z}^d) $ finite and $\pi \in
  \Omega_{E^c}$) are convex functions of $\log \beta$.
\end{lemma}

The parameter $\log \beta$ for the convexity appears naturally in the proof, see below after (\ref{ineq-d2logY}).

\begin{pf} As in the non-random case, the convergence in  (\ref{eq-def-pressure}) with $\pi_N=f$ follows from the
sub-additivity of the free energy. The influence of the boundary condition is 
negligible as $C_{E^w (\Lambda)}^{\pi}(\omega)$ fluctuates of at most $|\partial \Lambda |$ with $\pi$.

We address now the question of convexity. Let $I$ be an interval and $F: I \rightarrow \mathbb{R}_+$ a twice derivable function. 
We parametrize the inverse temperature letting $\beta = F(\lambda)$ and denote on the other hand
  $\lambda_e = \log (p (J_e) / (1 - p (J_e)))\in \mathbb{R}\cup\{-\infty\}$, thus
  \begin{equation}
    Y^{J, \pi}_E = \sum_{\omega \in \Omega_E} \exp \left( \sum_{e \in E}
    \omega_e \lambda_e \right) \times q^{C_E^{\pi} \left( \omega \right)}
  \end{equation}
with the convention that $\omega_e \lambda_e= \omega_e \frac{d^n \lambda_e}{d \lambda^n} = 0$ when $\omega_e=0$ and $\lambda_e = -\infty$. Using in particular the equality
  \begin{equation}
    \forall \omega \in \Omega_E, \Phi^{J, \pi}_E \left( \left\{ \omega
    \right\} \right) = \frac{1}{Y^{J, \pi}_E} \exp \left( \sum_{e \in E}
    \omega_e \lambda_e \right) \times q^{C_E^{\pi} \left( \omega \right)}
    \label{eq-link-Y-Phi}
  \end{equation}
  we get after standard calculations that:
  \begin{equation*} \frac{d^2}{d \lambda^2} \log Y^{J, \pi}_E = \Phi^{J, \pi}_E \left(
     \sum_{e \in E} \omega_e \frac{d^2 \lambda_e}{d \lambda^2} + \left(
     \sum_{e \in E} \omega_e  \frac{d \lambda_e}{d \lambda} \right)^2 \right)
     - \left( \Phi^{J, \pi}_E \left( \sum_{e \in E} \omega_e  \frac{d \lambda_e}{d
     \lambda} \right) \right)^2 \end{equation*}
  and Jensen's inequality implies:
  \begin{equation}
    \frac{d^2}{d \lambda^2} \log Y^{J, \pi}_E \geqslant \Phi^{J, \pi}_E \left(
    \sum_{e \in E} \omega_e \frac{d^2 \lambda_e}{d \lambda^2}  \right)
    \label{ineq-d2logY}
  \end{equation}
Here we recover the result of~\cite{N03}\footnote{
In the same direction we could prove the following: if $\forall e \in E, J_e = 0$ or $J_e \geqslant \varepsilon$, then
$\log Y^{J, \pi}_E$ is a convex function of $\log (p(\varepsilon)/(1-p(\varepsilon)))$ as $\beta$ varies, since
$f_{\alpha}: x \mapsto \log ( ( 1 + e^x )^{\alpha} - 1 )$ is convex for every
  $\alpha \geqslant 1$:
\begin{equation*}f'_{\alpha} (x) = \frac{\alpha e^x (1 + e^x)^{\alpha - 1}}{\left( 1 + e^x
     \right)^{\alpha} - 1}\mbox{ \ and \ }\left( \log \left( f'_{\alpha} (x) \right) \right)' = \frac{\left( 1 + e^x
     \right)^{\alpha} - 1 - \alpha e^x}{[1 + e^x] [ \left( 1 + e^x
     \right)^{\alpha} - 1]} \geqslant 0.\end{equation*}}: if $J\equiv1$ we have $\lambda_e=\log (p(1)/(1-p(1)))$, hence $\log Y^{1, \pi}_E$ is a convex function of $\lambda=\log (p(1)/(1-p(1)))$.
Now, let us develop the expression $\lambda_e = \log \left( e^{\beta J_e} - 1 \right)$ and calculate its second derivative in terms of
$\frac{d \beta}{d \lambda}$ and $\frac{d^2 \beta}{d \lambda^2}$:
  \begin{align}
    \frac{d^2 \lambda_e}{d \lambda^2} &= \frac{J_e  \frac{d^2 \beta}{d
    \lambda^2} }{1 - e^{- \beta J_e}} - \frac{\left( J_e  \frac{d \beta}{d
    \lambda} \right)^2 e^{- \beta J_e}}{\left( 1 - e^{- \beta J_e}
    \right)^2} \notag \\
    &= \frac{J_e}{\left( 1 - e^{- \beta J_e} \right)^2} \left[ \frac{d^2
    \beta}{d \lambda^2} - e^{- \beta J_e} \left( \frac{d^2 \beta}{d \lambda^2}
    + J_e \left( \frac{d \beta}{d \lambda} \right)^2  \right) \right] \notag
  \end{align}
  and fix at last $\beta = e^{\lambda}$, so that the former line simplifies to
  \begin{equation*} \frac{d^2 \lambda_e}{d \lambda^2} = \frac{J_e \beta}{\left( 1 - e^{-
     \beta J_e} \right)^2} \left[ 1 - e^{- \beta J_e} \left( 1 + \beta J_e 
     \right) \right] \end{equation*}
  which is non-negative since $J_e \geqslant 0$ and $1 + \beta J_e \leqslant
  e^{\beta J_e}$. In view of (\ref{ineq-d2logY}) this implies the convexity of
  $\log Y^{J, \pi}_E$ along $\lambda = \log \beta$, and the convexity of
  $\mathbb{E} \log Y^{J, \pi}_E$ and $y$ follows.
\end{pf}

\begin{pf}
  (Theorem~\ref{thm-unique-infvol}). We call again $\lambda = \log \beta$ and for any
  $N \in \mathbb{N}^{\star}$, $\pi \in \{f, w\}$ we denote
  \begin{equation*} y_N^{\pi} = \frac{1}{(2 N + 1)^d} \mathbb{E} \log Y^{J, \pi}_{E^w (
     \hat{\Lambda}_N)} \end{equation*}
  Consider some $q \geqslant 1$ and a Borel probability measure $\rho$ on $[0,
  1]$. Since $y$ is a convex function of $\lambda$ (Lemma~\ref{lemma-conv-pressure}), the set
  \begin{equation*} \mathcal{D}=\{\lambda \in \mathbb{R}: y \mbox{ is not derivable at }
     \lambda\} \end{equation*}
  is at most countable. Then, for any $\lambda \in \mathbb{R} \setminus
  \mathcal{D}$, $\pi \in \{f, w\}$ we have
  \begin{equation}
    \lim_N  \frac{d y^{\pi}_N}{d \lambda} =  \frac{d y}{d \lambda} \label{eq-conv-pN}
  \end{equation}
  thanks to the convexity of $y^{\pi}_N$ and to the pointwise convergence to $y$.
  Calculating the derivative we get:
  \begin{equation*} \frac{d y^{\pi}_N}{d \lambda} = \frac{1}{(2 N + 1)^d} \mathbb{E}
     \Phi^{J, \pi}_{\hat{\Lambda}_N} \left( \sum_{e \in E^w ( \hat{\Lambda}_N)}
     \frac{\beta J_e}{1 - \exp \left( - \beta J_e \right)} \omega_e \right) .
  \end{equation*}
  We fix now $e_0= \{0,\mathbf{e}_1\}$ the edge issued from $0$ that heads to $\mathbf{e}_1$ and denote
  \begin{equation*} r^f_L =\mathbb{E} \frac{\beta J_{e_0} \Phi_{\hat{\Lambda}_L}^{J, f} \left(
     \omega_{e_0} \right)}{1 - \exp \left( - \beta J_{e_0}
     \right)}  \mbox{ \ and \ } r^w_L =\mathbb{E} \frac{\beta J_{e_0}
     \Phi_{\hat{\Lambda}_L}^{J, w} \left( \omega_{e_0}
     \right)}{1 - \exp \left( - \beta J_{e_0} \right)}.  \end{equation*}
  For any $x \in \hat{\Lambda}_N$ and $e \in E^w ( \hat{\Lambda}_N)$ we have
  \begin{equation*} \mathbb{E} \Phi_{\hat{\Lambda}_N}^{J, f} \left( \omega_e \right)
     \leqslant \mathbb{E} \Phi_{x + \hat{\Lambda}_{2 N}}^{J, f} \left(
     \omega_e \right) \leqslant \mathbb{E} \Phi_{x + \hat{\Lambda}_{2 N}}^{J,
     w} \left( \omega_e \right) \leqslant \mathbb{E}
     \Phi_{\hat{\Lambda}_N}^{J, w} \left( \omega_e \right) \end{equation*}
  therefore, choosing $x = x_e$ such that $e =\{x_e, x_e \pm \mathbf{e}_k \}$ and summing over $e \in E^w ( \hat{\Lambda}_N)$ we obtain
  \begin{equation*} \frac{d y^f_N}{d \lambda} \leqslant \frac{|E^w (\Lambda_N) |}{(2 N + 1)^d}
     r^f_{2 N} \leqslant \frac{|E^w (\Lambda_N) |}{(2 N + 1)^d} r^w_{2 N}
     \leqslant \frac{d y^w_N}{d \lambda}  \end{equation*}
as the actual direction of $e_0$ in the definition of $r^w_L$ and $r^f_L$ does not influence their value.
  In view of (\ref{eq-conv-pN}) this implies that the limits of $r^f_{2 N}$
  and $r^w_{2 N}$ are equal, hence
\begin{equation*} \lim_{N\rightarrow\infty} \mathbb{E} \frac{\beta J_{e_0}}{1 - \exp \left( - \beta J_{e_0} \right)} \left( 
\Phi_{\hat{\Lambda}_{2 N}}^{J, w} \left(
     \omega_{e_0} \right)  - \Phi_{\hat{\Lambda}_{2 N}}^{J, f} \left(
     \omega_{e_0} \right) \right) = 0.  \end{equation*}
As $\beta J_{e_0} \geqslant 1 - \exp( - \beta J_{e_0})$ and $\Phi_{\hat{\Lambda}_{2 N}}^{J, w} ( \omega_{e_0} )  \geqslant \Phi_{\hat{\Lambda}_{2 N}}^{J, f} ( \omega_{e_0} )$, the equality  $\Theta^f ( \omega_{e_0} ) = \Theta^w ( \omega_{e_0} )$ follows. The stochastic domination $\Theta^f \leqslant  \Theta^w $ leads then to the conclusion: $\Theta^f = \Theta^w, \forall \lambda \in
  \mathbb{R} \setminus \mathcal{D}$.
\end{pf}

\subsection{Application to the Ising model}
\label{sec-cg-Ising}
In this last Section we adapt the coarse graining to the dilute Ising model (Theorem~\ref{thm-cg-Ising}). Applications include the study of equilibrium phase coexistence~\cite{M02} following~\cite{N06,N07,N11,N12,N13,N70}.

We start with a description of the Ising model with random ferromagnetic couplings. Given a domain $\Lambda \subset \mathbb{Z}^d$ we consider the set of spin configurations on $\Lambda$ with plus boundary condition
\begin{equation*} \Sigma^+_{\Lambda} = \left\{ \sigma : \mathbb{Z}^d \rightarrow \{-1,+1\}\mbox{ with }\sigma(x)=+1\mbox{ for all } x\notin \Lambda \right\} . \end{equation*}

The Ising measure on $\Lambda$ under the media $J\in\mathcal{J}_{E^w(\Lambda)}$, at inverse
temperature $\beta \geqslant 0$ and with plus boundary condition is defined by its weight on every spin configuration: $\forall \sigma \in \Sigma^{+}_{\Lambda}$,
\begin{equation}
  \mu_{\Lambda, \beta}^{J, +} (\{\sigma\}) = 
\frac{1}{Z_{\Lambda, \beta}^{J, +}} \exp \left( \beta \sum_{e =\{x, y\}
    \in E^w(\Lambda)} J_e \sigma_x \sigma_y
 \right) \label{eq-def-mu-Ising}
\end{equation}
where $Z_{\Lambda, \beta}^{J, +}$ is the partition function
\begin{equation*}
  Z_{\Lambda, \beta}^{J, +} = \sum_{\sigma \in \Sigma^{+}_{\Lambda}} \exp \left( \beta \sum_{e =\{x, y\} \in E^w (\Lambda)} J_e
\sigma_x \sigma_y \right) . \end{equation*}

The Ising model is closely related to the random-cluster model. To
begin with, we say that $\omega \in \Omega_{E^w(\Lambda)}$ and $\sigma \in \Sigma^{+}_{\Lambda}$ are
compatible, and we denote this by $\sigma \prec \omega$ if:
\begin{equation*}
  \forall e =\{x, y\} \in E^w(\Lambda), \; \; \omega_e = 1 \Rightarrow \sigma_x = \sigma_y . 
\end{equation*}
We consider then the joint measure $\Psi^{J,+}_{\Lambda,\beta}$ defined
again by its weight on each configuration $(\omega,\sigma) \in \Omega_{E^w(\Lambda)} \times \Sigma^{+}_{\Lambda}$:
\begin{equation}
  \Psi_{\Lambda, \beta}^{J,+} \left( \left\{ (\sigma, \omega) \right\} \right)
  = \frac{ \mathbf{1}_{\{\sigma \prec \omega\}}}{\tilde{Z}_{\Lambda, \beta}^{J,+}}
    \prod_{e \in E^w (\Lambda)} p(J_e)^{\omega_e} (1 - p(J_e))^{1 - \omega_e} 
\end{equation}
where $p(J_e) = 1 - \exp (- 2 \beta J_e)$ and $\tilde{Z}_{\Lambda, \beta}^{J,+}$ is the partition function that makes
$\Psi_{\Lambda, \beta}^{J,+}$ a probability measure. It is well known (see~\cite[Chapter 3]{N20} for a proof and for advanced remarks on the FK representation, including a random cluster representation for spin systems with non-ferromagnetic interactions) that:

\begin{proposition} \label{prop-FK-Ising}The marginals of $\Psi_{\Lambda, \beta}^{J,+}$ on $\sigma$ and $\omega$ are respectively
\begin{equation*} \mu_{\Lambda, \beta}^{J, +} \mbox{ \  and \ } \Phi^{J, p, 2, w}_{\Lambda}. \end{equation*}
Conditionally on $\omega$, the spin $\sigma$ is constant on each $\omega$-cluster,
equal to one on all clusters touching $\partial \Lambda$, independently and uniformly distributed on $\{-1,+1\}$
on all other clusters. Conditionally on $\sigma$, the $\omega_e$ are independent 
and $\omega_e=1$ with probability $\mathbf{1}_{\{\sigma_x=\sigma_y\}} \times p(J_e)$ if $e =\{x, y\}$.
\end{proposition}

Direct applications of the previous Proposition yield the following facts: first, the averaged magnetization
\begin{equation*} m_\beta = \lim_{N\rightarrow\infty} \mathbb{E} \mu_{\hat{\Lambda}_N, \beta}^{J, +}(\sigma_0) \end{equation*}
equals the cluster density $\theta^w$ defined at (\ref{eq-def-thetafw}). Second, assumption \eqref{eq-def-SP3}
can be reformulated as follows: there exists $H \in \mathbb{N}^{\star}$ such that
\begin{equation*} \inf_{N \in \mathbb{N}^{\star}} \inf_{x, y \in \overline{S}_{N, H}} \mathbb{E} \mu_{\overline{S}_{N, H}, \beta}^{J, f} \left( \sigma_x \sigma_y \right) > 0 \end{equation*}
where $\mu_{\Lambda, \beta}^{J, f}$ is the Ising measure with free boundary condition, that
one obtains considering $E^f(\Lambda)$ instead of $E^w(\Lambda)$ in (\ref{eq-def-mu-Ising}),
and $\overline{S}=S \cup \partial S$. On the other hand, a sufficient condition for \eqref{eq-def-SP2} is:
 there exists a function $\kappa (N): \mathbb{N}^{\star}
\rightarrow \mathbb{N}^{\star}$ with $\kappa (N) / N \longrightarrow 0$ as $N \rightarrow \infty$ and
\begin{equation*}
  \lim_{N \rightarrow \infty} 
\sup_{\tmscript{\begin{array}{c}
x\in\tmop{Left}(S_{N, \kappa (N)}) \\
y\in\tmop{Right}(S_{N, \kappa (N)})\end{array}}}
\mathbb{E} \mu_{\overline{S}_{N, \kappa (N)}, \beta}^{J, f} \left(
\sigma_x \sigma_y \right) > 0 \end{equation*}
where  $\tmop{Left}(S)$ and $\tmop{Right}(S)$ stand for the two vertical faces of $\partial S$.

We now present the adaptation of the coarse graining to the Ising model with random ferromagnetic couplings
(the adaptation to the Potts model would be similar). As in~\cite{N09}
it provides strong information on the structure of the local phase 
by the mean of \emph{phase labels}~$\phi$. 
Given $N,L\in{\mathbb{N}^\star}$ with $3 L \leqslant N + 1$, we denote
$(\Delta_i, \Delta'_i)_{i \in I_{\Lambda_N, L}}$ the $(L, L)$-covering of $\Lambda_N$
as in Definition \ref{def-cov-delta-L}. For any $i \in I_{\Lambda_N, L}$ we let
$\mathcal{M}_i^L(\sigma)$ the magnetization on  $\Delta_i$, that is
\begin{equation*} \mathcal{M}_i^L(\sigma) = \frac{1}{L^d} \sum_{x \in \Delta_i} \sigma_x. \end{equation*}

\begin{theorem} \label{thm-cg-Ising} Assume that $\beta \geqslant 0$ 
realizes {\rm(\textbf{SP})} and $\Theta^f=\Theta^w$. 
Let $N,L\in{\mathbb{N}^\star}$ with $3 L \leqslant N + 1$ and $\delta>0$. 
Then, there exists a sequence of variables
$(\phi_i)_{i\in I_{\Lambda_N, L}}$ taking values in $\{-1,0,1\}$,
with the following properties:

\begin{enumerate}
 \item For any $i \in I_{\Lambda_N, L}$, we have
$$ \phi_i\neq 0 \Rightarrow \left| \mathcal{M}_i^L(\sigma) - m_\beta \, \phi_i \right| \leqslant \delta.$$
The event $\phi_i\neq 0$ implies the existence of a $\omega$-crossing cluster 
and the uniqueness of $\omega$-clusters of diameter at least $L$ in $E^w(\Delta'_i)$.

\item If one extends $\phi$ letting $\phi_i= 1$ for 
$i \in \mathbb{Z}^d \setminus I_{\Lambda_N, L}$, then:
$$ \phi_i \; \phi_j  \geqslant 0,  \ \ \ \ \forall i,j\in \mathbb{Z}^d \mbox{ with } i\sim j. $$

\item For every $i \in I_{\Lambda_N, L}$,  $\phi_i$ is determined by $\sigma_{|\Delta_i}$ and $\omega_{|E^w(\Delta'_i)}$.

\item The sequence $(|\phi_i|)_{i\in I_{\Lambda_N, L}}$ stochastically dominates
a Bernoulli product measure with high density in the following sense: for every 
$p<1$, if $L$ is large enough, then for any 
$I\subset I_{\Lambda_N, L}$ and any increasing 
function $f: \{0, 1\}^I \rightarrow \mathbb{R}^+$, we have
\begin{equation}
 \mathbb{E} \inf_{\pi}  \Psi_{\Lambda_N, \beta}^{J,+} \left(
  f\left( \left( |\phi_i| \right)_{i\in I}\right) \left| \;
    \omega = \pi \mbox{ on } E^w (\Lambda_N) \setminus \bigcup_{i \in I} E^w(\Delta'_i) \right. \right) \geqslant \mathcal{B}^I_{p} \left( f \right) \label{eq-stochdom-phi} 
\end{equation}
where $\mathcal{B}^I_{p}$ is the Bernoulli product measure on $I$ of parameter $p$.
\end{enumerate}
\end{theorem}

\begin{pf}
We define the variable $\phi_i$ in two steps. First we 
let $\delta'>0$ and consider
\begin{equation*}
  \mathcal{E}_i = \left\{ \omega \in \Omega :
  \mbox{\begin{tabular}{l}
    In $E^w (\Delta'_i)$, there exists a crossing cluster for $\omega$,\\
    it is the unique cluster of diameter $\geqslant L^{1 / 3}$. \\
    In $E^w (\Delta_i)$, there exists a crossing cluster $\mathcal{C}_i$ for
    $\omega$,\\
    its relative density belongs to $[m_\beta (1 \pm \delta / 2)]$ and\\
    there are at least $\delta' L^d$ isolated $\omega$-clusters.
  \end{tabular}} \right\}
\end{equation*}
and 
\begin{equation*}
  \mathcal{G}_i = \left\{ (\sigma, \omega) : \mbox{
\begin{tabular}{l}
    $\omega \in \mathcal{E}_i$, $\sigma$ and $\omega$ are compatible\\
and $\left| \mathcal{M}_i^L(\sigma) - m_\beta \, \varepsilon_i(\sigma, \omega) \right| \leqslant \delta$
  \end{tabular}} \right\}
\end{equation*}
where $\varepsilon_i (\sigma, \omega)$ is the value of $\sigma$ on the main $\omega$-cluster in $E^w (\Delta_i)$. Then we let
\begin{equation*}
  \phi_i = \left\{ 
\begin{array}{ll}
\varepsilon_i (\sigma, \omega) & \mbox{if } (\sigma, \omega)\in \mathcal{G}_i \\
0 & \mbox{else.}
\end{array} \right.  \end{equation*}
Properties {(i)} to {(iii)} follow from the definition of $\mathcal{E}_i$ and $\mathcal{G}_i$, together with the
plus boundary condition imposed by $\Psi_{\Lambda_N, \beta}^{J,+}$ on $\sigma$.

We turn now to the proof of the stochastic domination and use the
 hypothesis \textbf{(SP)} and $\Theta^f=\Theta^w$. 
Combining Theorem~\ref{thm-strong-cg}, Proposition~\ref{prop-cg-dens} and the remark that for any $\delta' > 0$ small enough,
  \begin{equation*} \limsup_N \frac{1}{N^d} \log \mathbb{E} \sup_{\pi} \Phi^{J,\pi}_{\Lambda_N} \left( \mbox{
\begin{tabular}{l}
There are less than $\delta' N^d$ \\
clusters made of $1$ point in $\Lambda_N$
\end{tabular}} \right) < 0
    \label{eq-proba-isolated-clusters} \end{equation*}
 (remark that $\{x\}$ is a cluster for $\omega$ in $\Lambda_N$ if all the $\omega_e$
  with $x \in e$ are closed, which happens with probability at least $e^{- 2 d \beta}$ conditionally on the state of all other edges, uniformly over $J \in
  \mathcal{J}$),
we see that there exists $p_{L,\delta, \delta'}$ with $p_{L,\delta, \delta'} \rightarrow 1$
 as $L\rightarrow\infty$
(for small enough $\delta' >0$) such that, uniformly over $N$ and $i\in I_{\Lambda_N, L}$,
\begin{equation*} \mathbb{E} \inf_{\pi} \Phi^{J,\pi}_{\Delta'_i} (\mathcal{E}_i) \geqslant p_{L,\delta, \delta'}. \end{equation*}
Given $\omega \in \mathcal{E}_i$ we examine as in~\cite{N12} the conditional 
probability  for having $(\sigma, \omega)\in \mathcal{G}_i$.
The contribution of the main $\omega$-cluster $\mathcal{C}_i$
 to $\mathcal{M}_i^L(\sigma)$ belongs to $\varepsilon m_\beta (1 \pm \delta / 2)$
where $\varepsilon$ stands for the value of $\sigma$ on $\mathcal{C}_i$. Then, 
if $2 d L^{- 2 / 3} \leqslant \delta / 4$ the contribution of the small 
clusters connected to the boundary of $\Delta_i$ is not larger than $\delta / 4$ 
and it remains to control the contribution of the small clusters
  not connected to the boundary. Since the spin of these clusters are independent and
  uniformly distributed on $\{\pm 1\}$, Lemma 5.3~of~\cite{N09} tells us that:
\begin{equation*}
    \Psi_{\Lambda_N, \beta}^{J,+} \left( \left. \left| \frac{1}{
\left| \tmop{SC}_{\Delta_i}(\omega) \right|
} \sum_{x \in
    \tmop{SC}_{\Delta_i}(\omega)} \sigma_x \right| > \frac{\delta}{4} \right|
    \omega \right) \leqslant 2 \exp \left( - \left| \tmop{SC}_{\Delta_i}(\omega) \right| \Lambda^{\star}
    \left( \frac{\delta }{4 L^{d / 3}} \right) \right) \end{equation*}
  where $\tmop{SC}_{\Delta_i}(\omega)$ is the set of small clusters for
  $\omega$ in $\Delta_i$ not connected to the boundary, $L^{d/3}$ an upper bound
on the volume of any small cluster, and
  \begin{equation*} \Lambda^{\star} (x) = \frac{1 + x}{2} \log \left( 1 + x \right) + \frac{1
     - x}{2} \log \left( 1 - x \right), \forall x \in (- 1, 1) \end{equation*}
  is the Legendre transform of the logarithmic moment generating function of
  $X$ of law $\delta_{- 1} / 2 + \delta_1 / 2$.
Because of the assumption $\omega \in \mathcal{E}_i$, we have $|\tmop{SC}_{\Delta_i}(\omega)| \geqslant \delta' L^d$. Hence,
  \begin{equation*}
    \Psi_{\Lambda_N, \beta}^{J,+} \left( \left. \left| \frac{1}{L^d} \sum_{x \in
    \tmop{SC}_{\Delta_i}(\omega)} \sigma_x \right| > \frac{\delta}{4} \right|
    \omega \right) \leqslant 2 \exp \left( - \delta' L^d \Lambda^{\star}
    \left( \frac{\delta }{4 L^{d / 3}} \right) \right) \end{equation*}
As $\Lambda^{\star} (x) \geqslant x^2 / 2$ and $m_\beta\leqslant1$
we conclude that for $L$ large enough, for any $\omega \in \mathcal{E}_i$,
\begin{equation*}
\Psi_{\Lambda_N, \beta}^{J,+} \left( \left.
\mathcal{G}_i \right| \omega, \sigma_{|\Lambda\setminus\Delta_i} \right) \geqslant p'_{L,\delta, \delta'} = 1-2 \exp(-\delta' \delta^2 L^{d/3}/16).
\end{equation*}
We now conclude the proof of the stochastic domination for $|\phi_i|=\mathbf{1}_{\mathcal{G}_i}$ and
 consider $I\subset I_{\Lambda_N, L}$, together with an increasing function
 $f: \{0, 1\}^I \rightarrow \mathbb{R}^+$. We fix $\omega\in\Omega_{E^w(\Lambda_N)}$ and consider
\begin{equation*}I'=\{i\in I:\omega\in \mathcal{E}_i\} \mbox{ \ and \ } f': \{0, 1\}^{I'} \rightarrow \mathbb{R}^+ \end{equation*}
defined by
\begin{equation*} f'((x_i)_{i\in I'})=f((x_i)_{i\in I}), \; \forall (x_i)\in\{0,1\}^I \mbox{ with } x_i=0, \forall i\in I\setminus I'.\end{equation*}
 Since no more than $6^d$ distinct $\Delta_i$ can intersect,
 Theorem~\ref{thm-stochdom-LSS} tells us that
\begin{align}
\Psi_{\Lambda_N, \beta}^{J,+}\left( \left. f\left( \left( \mathbf{1}_{\mathcal{G}_i} \right)_{i\in I}\right) \right| \omega \right) &
 =  \Psi_{\Lambda_N, \beta}^{J,+}\left( \left. f'\left( \left( \mathbf{1}_{\mathcal{G}_i} \right)_{i\in I'}\right) \right| \omega \right) \notag \\
& \geqslant  \mathcal{B}^{I'}_{r(6^d,p'_{L,\delta,\delta'})} \left( f'\left( \left( X_i \right)_{i\in I'}\right) \right) \notag \\
& =  \mathcal{B}^I_{r(6^d,p'_{L,\delta,\delta'})} \left( f\left( \left( X_i \mathbf{1}_{\mathcal{E}_i} \right)_{i\in I}\right) \right) \label{eq-stocdom-Ising-Gi}.
\end{align}
Integrating \eqref{eq-stocdom-Ising-Gi} under the conditional measure 
\begin{equation*}\Phi_{\Lambda_N}^{J,w} \left( \, . \, \left| \omega = \pi \mbox{ on } E^w (\Lambda_N) \setminus \bigcup_{i \in I} E^w(\Delta'_i) \right. \right)\end{equation*}
and taking $\mathbb{E} \inf_\pi$ we obtain 
on the
left hand side, thanks to Proposition \ref{prop-FK-Ising},
the left-hand side of \eqref{eq-stochdom-phi}.
For the right-hand side, we remark that
$$y=(y_i)_{i\in I}\mapsto\mathcal{B}^I_{X,p} ( f (( X_i y_i)_{i\in I}))$$
is an increasing function, hence 
Proposition~\ref{prop-renormalization} gives the lower bound
\begin{eqnarray}
\lefteqn{\hspace*{-4cm} \mathbb{E} \inf_\pi \Phi_{\Lambda_N}^{J,w} \left( 
\mathcal{B}^I_{r(6^d,p'_{L,\delta,\delta'})} \left( f\left( \left( X_i \mathbf{1}_{\mathcal{E}_i} \right)_{i\in I}\right) \right)
 \left| \omega = \pi \mbox{ on } E^w (\Lambda_N) \setminus \bigcup_{i \in I} E^w(\Delta'_i) \right. \right) } \notag \\
\hspace*{4cm} & \geqslant & 
\mathcal{B}^I_{Y,r'(6^d,p_{L,\delta,\delta'})} \left(
\mathcal{B}^I_{X,r(6^d,p'_{L,\delta,\delta'})} \left( f\left( \left( X_i Y_i \right)_{i\in I}\right) \right) \right) \notag \\
& = & \mathcal{B}^I_{X,r'(6^d,p_{L,\delta,\delta'}) \times r(6^d,p'_{L,\delta,\delta'})} \left( f\left( \left( X_i \right)_{i\in I}\right) \right) \notag
\end{eqnarray}
and the claim follows as, for any $\delta'>0$ small enough,
\begin{equation*} \lim_{L\rightarrow\infty} r'(6^d,p_{L,\delta,\delta'})\times r(6^d,p'_{L,\delta,\delta'}) = 1. \end{equation*}
\end{pf}

\section{Conclusion}

These estimates for the Ising model with random ferromagnetic couplings conclude our construction
of a coarse graining under the assumption of slab percolation. It turns out 
that apart from being a strong obstacle to the shortness of the construction, the media
randomness does not change the typical aspect of clusters (or the behavior of phase labels for spin models) in the
regime of slab percolation.

This coarse graining is a first step towards a study of
phase coexistence in the dilute Ising model that we propose in a
separate paper~\cite{M02}. Following~\cite{N07,N70} we 
describe the phenomenon of phase coexistence in a $L^1$ setting, under both
quenched and averaged measures. The notion of surface tension and
the study of its fluctuations as a function of the media are another key point of~\cite{M02}.

Another fundamental application of the coarse graining, together with the
study of equilibrium phase coexistence, concerns the dynamics of such random
media models. In opposition with the previous phenomenon which
nature is hardly modified by the introduction of random media, the media randomness introduces an abrupt change in the dynamics and we confirm
in~\cite{M03} several predictions of~\cite{N05}, among which a lower bound on the average spin autocorrelation
at time $t$ of the form $t^{-\alpha}$.

\section*{Acknowledgments}

The author thanks warmly Prof. Thierry Bodineau for the countless stimulating discussions that led to the present work.

\bibliographystyle{abbrv}
\bibliography{biblio}

\begin{thebibliography}{10}

\bibitem{N233}
M.~Aizenman and D.~J. Barsky.
\newblock Sharpness of the phase transition in percolation models.
\newblock {\em Comm. Math. Phys.}, 108(3):489--526, 1987.

\bibitem{N21}
M.~Aizenman, J.~T. Chayes, L.~Chayes, and C.~M. Newman.
\newblock The phase boundary in dilute and random {I}sing and {P}otts
  ferromagnets.
\newblock {\em J. Phys. A}, 20(5):L313--L318, 1987.

\bibitem{N210}
M.~Aizenman, J.~T. Chayes, L.~Chayes, and C.~M. Newman.
\newblock Discontinuity of the magnetization in one-dimensional {$1/\vert
  x-y\vert \sp 2$} {I}sing and {P}otts models.
\newblock {\em J. Statist. Phys.}, 50(1-2):1--40, 1988.

\bibitem{N22}
M.~Aizenman and J.~Wehr.
\newblock Rounding of first-order phase transitions in systems with quenched
  disorder.
\newblock {\em Phys. Rev. Lett.}, 62(21):2503--2506, 1989.

\bibitem{N227}
K.~S. Alexander, F.~Cesi, L.~Chayes, C.~Maes, and F.~Martinelli.
\newblock Convergence to equilibrium of random {I}sing models in the
  {G}riffiths phase.
\newblock {\em J. Statist. Phys.}, 92(3-4):337--351, 1998.

\bibitem{N06}
T.~Bodineau.
\newblock The {W}ulff construction in three and more dimensions.
\newblock {\em Comm. Math. Phys.}, 207(1):197--229, 1999.

\bibitem{N46}
T.~Bodineau.
\newblock Slab percolation for the {I}sing model.
\newblock {\em Probab. Theory Related Fields}, 132(1):83--118, 2005.

\bibitem{N07}
T.~Bodineau, D.~Ioffe, and Y.~Velenik.
\newblock Rigorous probabilistic analysis of equilibrium crystal shapes.
\newblock {\em J. Math. Phys.}, 41(3):1033--1098, 2000.
\newblock Probabilistic techniques in equilibrium and nonequilibrium
  statistical physics.

\bibitem{N231}
A.~Bovier.
\newblock {\em Statistical mechanics of disordered systems}.
\newblock Cambridge Series in Statistical and Probabilistic Mathematics.
  Cambridge University Press, Cambridge, 2006.
\newblock A mathematical perspective.

\bibitem{N230}
J.~Bricmont and A.~Kupiainen.
\newblock Phase transition in the {$3$}d random field {I}sing model.
\newblock {\em Comm. Math. Phys.}, 116(4):539--572, 1988.

\bibitem{N11}
R.~Cerf.
\newblock Large deviations for three dimensional supercritical percolation.
\newblock {\em Ast\'erisque}, (267):vi+177, 2000.

\bibitem{N70}
R.~Cerf.
\newblock {\em The {W}ulff crystal in {I}sing and percolation models}, volume
  1878 of {\em Lecture Notes in Mathematics}.
\newblock Springer-Verlag, Berlin, 2006.
\newblock Lectures from the 34th Summer School on Probability Theory held in
  Saint-Flour, July 6--24, 2004, With a foreword by Jean Picard.

\bibitem{N12}
R.~Cerf and A.~Pisztora.
\newblock On the {W}ulff crystal in the {I}sing model.
\newblock {\em Ann. Probab.}, 28(3):947--1017, 2000.

\bibitem{N13}
R.~Cerf and A.~Pisztora.
\newblock Phase coexistence in {I}sing, {P}otts and percolation models.
\newblock {\em Ann. Inst. H. Poincar\'e Probab. Statist.}, 37(6):643--724,
  2001.

\bibitem{N27}
J.~T. Chayes, L.~Chayes, and J.~Fr{\"o}hlich.
\newblock The low-temperature behavior of disordered magnets.
\newblock {\em Comm. Math. Phys.}, 100(3):399--437, 1985.

\bibitem{N72}
J.-D. Deuschel and A.~Pisztora.
\newblock Surface order large deviations for high-density percolation.
\newblock {\em Probab. Theory Related Fields}, 104(4):467--482, 1996.

\bibitem{N14}
H.-O. Georgii, O.~H{\"a}ggstr{\"o}m, and C.~Maes.
\newblock The random geometry of equilibrium phases.
\newblock In {\em Phase transitions and critical phenomena}, volume~18, pages
  1--142. Academic Press, San Diego, CA, 2001.

\bibitem{N228}
R.~Griffiths.
\newblock Non-analytic behaviour above the critical point in a random {I}sing
  ferromagnet.
\newblock {\em Phys. Rev. Lett.}, 23(1):17--19, 1969.

\bibitem{N03}
G.~Grimmett.
\newblock The stochastic random-cluster process and the uniqueness of
  random-cluster measures.
\newblock {\em Ann. Probab.}, 23(4):1461--1510, 1995.

\bibitem{N222}
G.~Grimmett.
\newblock {\em The random-cluster model}, volume 333 of {\em Grundlehren der
  Mathematischen Wissenschaften [Fundamental Principles of Mathematical
  Sciences]}.
\newblock Springer-Verlag, Berlin, 2006.

\bibitem{N170}
G.~Grimmett and J.~M. Marstrand.
\newblock The supercritical phase of percolation is well behaved.
\newblock {\em Proc. Roy. Soc. London Ser. A}, 430(1879):439--457, 1990.

\bibitem{N05}
D.~A. Huse and D.~S. Fisher.
\newblock Dynamics of droplet fluctuations in pure and random {I}sing systems.
\newblock {\em Phys. Rev. B}, 35(13):6841--6846, 1987.

\bibitem{N234}
Y.~Imry and S.-K. Ma.
\newblock Random-field instability of the ordered state of continuous symmetry.
\newblock {\em Phys. Rev. Lett.}, 35(21):1399--1401, Nov 1975.

\bibitem{N33}
J.~L. Lebowitz.
\newblock Coexistence of phases in {I}sing ferromagnets.
\newblock {\em J. Statist. Phys.}, 16(6):463--476, 1977.

\bibitem{N43}
T.~M. Liggett, R.~H. Schonmann, and A.~M. Stacey.
\newblock Domination by product measures.
\newblock {\em Ann. Probab.}, 25(1):71--95, 1997.

\bibitem{N15}
F.~Martinelli.
\newblock Lectures on {G}lauber dynamics for discrete spin models.
\newblock In {\em Lectures on probability theory and statistics (Saint-Flour,
  1997)}, volume 1717 of {\em Lecture Notes in Math.}, pages 93--191. Springer,
  Berlin, 1999.

\bibitem{N232}
M.~V. Menshikov.
\newblock Coincidence of critical points in percolation problems.
\newblock {\em Dokl. Akad. Nauk SSSR}, 288(6):1308--1311, 1986.

\bibitem{N20}
C.~M. Newman.
\newblock {\em Topics in disordered systems}.
\newblock Lectures in Mathematics ETH Z\"urich. Birkh\"auser Verlag, Basel,
  1997.

\bibitem{N09}
A.~Pisztora.
\newblock Surface order large deviations for {I}sing, {P}otts and percolation
  models.
\newblock {\em Probab. Theory Related Fields}, 104(4):427--466, 1996.

\bibitem{N236}
A.~C.~D. van Enter and C.~K{\"u}lske.
\newblock Two connections between random systems and non-{G}ibbsian measures.
\newblock {\em J. Stat. Phys.}, 126(4-5):1007--1024, 2006.

\bibitem{M03}
M.~Wouts.
\newblock Glauber dynamics in the dilute {I}sing model below ${T}_c$.
\newblock {\em In preparation}.

\bibitem{M02}
M.~Wouts.
\newblock Surface tension in the dilute {I}sing model. {T}he {W}ulff
  construction.
\newblock {\em In preparation}.

\end{thebibliography}

\end{document}